\documentclass[%
 preprint,
showpacs,preprintnumbers,
nofootinbib,
 amsmath,amssymb,
 aps,
prd,
floatfix,
]{revtex4-1}

\usepackage{graphicx}
\usepackage{dcolumn}
\usepackage{bm}

\usepackage{hyperref}
\usepackage{graphicx}
\usepackage{amsfonts}
\usepackage{amssymb}
\usepackage{cancel}
\usepackage{hepunits}
\usepackage{multirow}
\usepackage{pbox}
\usepackage{cleveref}
\usepackage{adjustbox}
\usepackage{enumitem}

\numberwithin{equation}{section}

\def\beq{\begin{equation}}
\def\eeq{\end{equation}}

\def\bea{\begin{eqnarray}}
\def\eea{\end{eqnarray}}

\def\beq{\begin{equation}}
\def\eeq{\end{equation}}
\def\bea{\begin{eqnarray}}
\def\eea{\end{eqnarray}}

\DeclareRobustCommand{\SkipTocEntry}[4]{}

\usepackage{color}

\newcommand{\dif}{\mathrm{d}}
\newcommand{\pt}{\partial}
\newcommand{\abs}[1]{\lvert #1 \rvert}

\newcommand{\lag}{\mathcal{L}}

\newcommand{\ga}{\gamma}
\newcommand{\mgamgam}{\ensuremath{m_{\ga\ga}}}

\begin{document}

\title{Exploring peaks and valleys in the diphoton spectrum}
\author{Nathaniel Craig$^{1,4}$}
\author{Sophie Renner$^{2,4}$}
\author{Dave Sutherland$^{3,4}$} 

\affiliation{$^1$ Department of Physics, University of California, Santa Barbara, CA 93106, USA}
\affiliation{$^2$  DAMTP, University of Cambridge, Cambridge CB3 0WA, UK}
\affiliation{$^3$  Cavendish Laboratory, University of Cambridge, Cambridge CB3 0HE, UK}
\affiliation{$^4$  KITP, University of California, Santa Barbara, CA  93106, USA}

\date{\today}

\begin{abstract}
We investigate the effects of resonance-continuum interference on the diphoton spectrum in the presence of a new spin-0 or spin-2 state produced via gluons or quarks and decaying to pairs of photons. Interference effects can significantly influence the extraction of resonance masses and widths from the diphoton spectrum, particularly in the case of a spin-2 resonance produced via quarks. We illustrate these effects via a binned likelihood analysis of LHC diphoton data at both 8 and 13 TeV.
\end{abstract}


\maketitle

\section{Introduction}

The diphoton spectrum provides a powerful probe for resonantly-produced states at the LHC,
playing an instrumental role in the discovery of the Standard Model-like Higgs boson \cite{Aad:2012tfa, Chatrchyan:2012xdj} and constraining a variety of new physics scenarios such as extended Higgs sectors in the alignment limit \cite{Craig:2013hca} and a plethora of beyond-the-Standard Model scenarios at higher masses (see e.g. \cite{Strumia:2016wys} for a recent review). Although Standard Model backgrounds to the diphoton final state are considerable, the smoothness of these backgrounds as a function of diphoton invariant mass (in conjunction with excellent diphoton mass resolution at ATLAS and CMS) permits the search for bumps in the diphoton invariant mass spectrum. The appreciable size of the continuum diphoton background also implies that resonance-continuum interference may have a significant impact on the shape and size of resonant signals. The precise impact of resonance-continuum interference depends sensitively on the spin and production mode of the resonance, and may allow for discrimination between different signal hypotheses in the event of an excess.

The effects of resonance-continuum interference in the diphoton final state at the LHC have been considered for the Standard Model Higgs boson \cite{Dicus:1987fk, Dixon:2003yb, Martin:2012xc, deFlorian:2013psa, Martin:2013ula, Dixon:2013haa, Coradeschi:2015tna} and for a now-disfavored 750 GeV state  \cite{Jung:2015etr, Fabbrichesi:2016jlo, Djouadi:2016ack, Martin:2016bgw}. Here we build on previous work by considering six possible combinations of initial state ($q \bar q$ or $gg$) and $J^{PC}$ quantum numbers ($0^{++}, 0^{-+}$, or $2^{++}$). We directly study the impact of resonance-continuum interference on signal interpretation by performing a binned likelihood analysis of diphoton data provided by the ATLAS and CMS collaborations. For the sake of concreteness, we illustrate the effects of interference on the interpretation of several distinctive statistical fluctuations in the data, including fluctuations near 750 GeV found in 8 TeV and early 13 TeV data as well as other features in the diphoton spectrum.\footnote{For recent fits near 750 GeV neglecting interference effects, see \cite{Buckley:2016mbr}.} This analysis demonstrates the impact of resonance-continuum interference on the extraction of resonance masses and widths. In addition to modifying the apparent peak shape of a resonance, in some cases resonance-continuum interference may lead to complete deficits in the diphoton spectrum, as in \cite{Jung:2015gta}. We illustrate these effects via a shared deficit in the ATLAS and CMS diphoton spectra near 400 GeV and a common peak-dip structure near 550 GeV.

The paper is organized as follows: In Section \ref{sec:signal} we establish notation and convention for scalar, pseudoscalar, and spin-2 resonances produced through $q \bar q$ or $gg$ and decaying to photon pairs.  We then revisit leading-order calculations of resonance-continuum interference of the six possible signal hypotheses. We translate these effects into the diphoton spectra measured by the ATLAS and CMS collaborations at $\sqrt{s} = 8, 13$ TeV in Section \ref{sec:data}. In Section \ref{sec:peak} we employ our results to perform a binned likelihood analysis of various signal hypotheses in the 700 -- 800 GeV region, for which statistical fluctuations in both 8 and 13 TeV data provide useful test cases. We apply the same techniques to search for {\it deficits} in the data resulting from resonance-continuum interference in Section \ref{sec:valley}. We conclude in Section \ref{sec:conc} with a summary of results and recommendations for future experimental analyses of the diphoton spectrum at the LHC.

\section{Signals and Backgrounds}\label{sec:signal}

\subsection{Signal Models}

We consider three possible resonance candidates -- two spin-0 candidates (scalar and pseudoscalar) and a spin-2 candidate, consistent with the Lee-Yang theorem. While such states may be produced in a variety of ways at the LHC, in order to interfere appreciably with the continuum background they should be predominantly produced via $q\bar q$ or $gg$ initial states. For each spin hypothesis we therefore consider couplings both to gluons and quarks, with signal interactions of the form
\begin{eqnarray}
\lag_{0^{++}} & \supset & -\frac{1}{\Lambda_g} \phi G^{\mu \nu} G_{\mu \nu} - \frac{1}{\Lambda_\gamma} \phi F^{\mu \nu} F_{\mu \nu} - c_q \phi \overline{q} q; \label{eq:lagS} \\
\lag_{0^{-+}} & \supset & -\frac{1}{\Lambda_g} \phi G^{\mu \nu} \tilde G_{\mu \nu} - \frac{1}{\Lambda_\gamma} \phi F^{\mu \nu} \tilde F_{\mu \nu} - i c_q \phi \overline{q} \gamma^5 q; \label{eq:lagPS} \\
\lag_{2^{++}} & \supset & -\frac{1}{\Lambda_T} \phi_{\mu \nu} \left[ -F^{\mu \lambda} F_{\lambda}^\nu + \frac{1}{4} g^{\mu \nu} F^2 -G^{\mu \lambda} G_{\lambda}^\nu + \frac{1}{4} g^{\mu \nu} G^2  \right. \nonumber \\
&& \left. + \frac{1}{2} i \overline{q}(\gamma^\mu \pt^\nu + \gamma^\nu \pt^\mu) q - i g^{\mu \nu} \overline{q} \cancel{\pt} q \right] \label{eq:lagT} .
\end{eqnarray}
Here we have assumed the spin-2 candidate to couple via the SM stress-energy tensor (the term in square brackets). In all three cases we use $q$ to denote a $u$ or $d$ quark, and assume our models couple to both with equal strength.

\subsection{Theory-level Diphoton Spectrum}

For a given signal hypothesis, we determine the theory-level diphoton spectrum including resonance-continuum interference by computing the corresponding helicity amplitudes for both signal and background. We compute all helicity amplitudes at tree level, save for the $gg \to \ga\ga$ background, which is performed using 5 flavours of massless quarks in~\cite{Dicus:1987fk} with care for the appropriate sign convention. The resulting lowest order helicity amplitudes $M_{\lambda_1 \lambda_2 \lambda_3 \lambda_4}$ are shown in Table~\ref{tab:helamp} for the background (BG), spin 0 scalar signal (S), spin 0 pseudoscalar signal (PS) and spin 2 signal (T), in both the $q\bar{q}$ and $gg$ initiated cases. 

\begin{table}
\begin{tabular}{c c c c | c c c c | c c c c}
\multicolumn{4}{c|}{Helicities} & \multicolumn{4}{|c|}{$q \bar{q} \to \ga \ga$} & \multicolumn{4}{|c}{$gg \to \ga\ga$} \\
$\lambda_1$ & $\lambda_2$ & $\lambda_3$ & $\lambda_4$ & BG/$4\pi \alpha Q^2$ & S & PS & T & BG/$\frac{44}{9} \alpha \alpha_s$ & S & PS & T \\ \hline
$+$&$+$&$+$&$+$  &
 $0$&$-\hat s^\frac{3}{2} P$&$-\hat s^\frac{3}{2} P$&$0$  &
 $M_1$&$- \hat s^2 P$&$-\hat s^2 P$&$0$  \\
 
$+$&$+$&$+$&$-$  &
 $0$&$0$&$0$&$0$  &
 $1$&$0$&$0$&$0$  \\
 
$+$&$+$&$-$&$+$  &
 $0$&$0$&$0$&$0$  &
 $1$&$0$&$0$&$0$  \\
 
$+$&$+$&$-$&$-$  &
 $0$&$-\hat s^\frac{3}{2} P$&$\hat s^\frac{3}{2} P$&$0$  &
 $1$&$- \hat s^2 P$&$\hat s^2 P$&$0$  \\ \hline
 
$+$&$-$&$+$&$+$  &
 $0$&$0$&$0$&$0$  &
 $1$&$0$&$0$&$0$  \\
 
$+$&$-$&$+$&$-$  &
 $2\sqrt{\frac{\hat u}{\hat t}}$&$0$&$0$&$\frac{1}{2} \hat u \sqrt{\hat u \hat t} P$  &
 $M_3$&$0$&$0$&$-\frac{1}{4} \hat u^2 P$  \\
 
$+$&$-$&$-$&$+$  &
 $-2\sqrt{\frac{\hat t}{\hat u}}$&$0$&$0$&$-\frac{1}{2} \hat t \sqrt{\hat u \hat t} P$  &
 $M_2$&$0$&$0$&$-\frac{1}{4} \hat t^2 P$  \\
 
$+$&$-$&$-$&$-$  &
 $0$&$0$&$0$&$0$  &
 $1$&$0$&$0$&$0$  \\ \hline
 
$-$&$+$&$+$&$+$  &
 $0$&$0$&$0$&$0$  &
 $1$&$0$&$0$&$0$  \\
 
$-$&$+$&$+$&$-$  &
 $-2\sqrt{\frac{\hat t}{\hat u}}$&$0$&$0$&$-\frac{1}{2} \hat t \sqrt{\hat u \hat t} P$  &
 $M_2$&$0$&$0$&$-\frac{1}{4} \hat t^2 P$  \\
 
$-$&$+$&$-$&$+$  &
 $2\sqrt{\frac{\hat u}{\hat t}}$&$0$&$0$&$\frac{1}{2} \hat u \sqrt{\hat u \hat t} P$  &
 $M_3$&$0$&$0$&$-\frac{1}{4} \hat u^2 P$  \\
 
$-$&$+$&$-$&$-$  &
 $0$&$0$&$0$&$0$  &
 $1$&$0$&$0$&$0$  \\ \hline

$-$&$-$&$+$&$+$  &
 $0$&$\hat s^\frac{3}{2} P$&$-\hat s^\frac{3}{2} P$&$0$  &
 $1$&$- \hat s^2 P$&$\hat s^2 P$&$0$  \\
 
$-$&$-$&$+$&$-$  &
 $0$&$0$&$0$&$0$  &
 $1$&$0$&$0$&$0$  \\
 
$-$&$-$&$-$&$+$  &
 $0$&$0$&$0$&$0$  &
 $1$&$0$&$0$&$0$  \\
 
$-$&$-$&$-$&$-$  &
 $0$&$\hat s^\frac{3}{2} P$&$\hat s^\frac{3}{2} P$&$0$  &
 $M_1$&$- \hat s^2 P$&$-\hat s^2 P$&$0$  \\
\end{tabular}
\caption{Helicity amplitudes for  background (BG) and scalar (S), pseudoscalar (PS), and spin-2 (T) signals for both $q \bar q$- and $gg$-initiated cases. Here the propagator factor $P$ is given in (\ref{eq:prop}), while the amplitude factors are given in (\ref{eq:m1})-(\ref{eq:m3}).  \label{tab:helamp}}
\end{table}

The amplitudes in Table~\ref{tab:helamp} are expressed in terms of the propagator factor 
\begin{equation} \label{eq:prop}
P = \frac{A}{\hat s - M^2 + i \Gamma \sqrt{\hat s}}
\end{equation}
and the amplitude factors $M_1, M_2, M_3$,
\begin{eqnarray} 
M_1 &=& M_1(\hat s,\hat t,\hat u) = -1 - \frac{\hat t-\hat u}{\hat s} \ln \left (\abs{\frac{\hat t}{\hat u}} \right) - \frac{\hat t^2 + \hat u^2}{2 \hat s^2} \left[ \ln \left(\abs{\frac{\hat t}{\hat u}}\right)^2 + \pi^2 \Theta\left(\frac{\hat t}{\hat u}\right) \right] \nonumber \\
&&+ i \pi (\Theta(\hat t) - \Theta(\hat u)) \left(\frac{\hat t-\hat u}{\hat s} + \frac{\hat t^2 + \hat u^2}{\hat s^2} \ln \left(\abs{\frac{\hat t}{\hat u}}\right) \right); \label{eq:m1} \\
M_2 &=& M_1(\hat t, \hat s,\hat u); \label{eq:m2}\\
M_3 &=& M_1(\hat u, \hat t,\hat s). \label{eq:m3}
\end{eqnarray}
In terms of the coefficients of the Lagrangians~(\ref{eq:lagS})-(\ref{eq:lagT}) at tree-level,
\begin{equation}
A =
\begin{cases}
\frac{4}{\Lambda_g \Lambda_\gamma} & \text{for the spin 0 $gg$ initiated process,} \\
\frac{4}{\Lambda_T^2} & \text{for the spin 2 $gg$ initiated process,} \\
\frac{2 c_q}{\Lambda_\gamma} & \text{for the spin 0 $q\bar{q}$ initiated process,} \\
\frac{2 c_q}{\Lambda_T} & \text{for the spin 2 $q\bar{q}$ initiated process.} \\
\end{cases}
\end{equation}
The dominant resonance-continuum interference effects are already apparent at the level of the helicity amplitudes. In particular, interference arises only for $gg$-initiated, not $q \bar q$-initiated, spin-0 resonances, while interference is possible for either $q\bar q$- or $gg$-initiated spin-2 signals. Given that the continuum background is dominated by $q \bar q$-initiated photon pair production at the high invariant masses that we consider, this implies that the strongest interference effects will arise for a spin-2 resonance produced via $q \bar q$.

Given these helicity amplitudes, the corresponding spectrum in $m_{\gamma \gamma} \equiv \sqrt{\hat s}$ is obtained via
\begin{align}
\frac{\dif \sigma}{\dif m_{\gamma\gamma}} = \frac{1}{8 \pi s m_{\gamma\gamma}^3 } \int_{-5}^5 \dif Y \int^0_{- 0.5 \hat s} \dif \hat t \, g(x_1) g(x_2) \overline{\abs{M}^2} \mathcal{A}(\hat s, \hat t, Y) ,
\end{align}
for the $gg$ case, and
\begin{align}
\frac{\dif \sigma}{\dif m_{\gamma\gamma}} = \frac{1}{8 \pi s m_{\gamma\gamma}^3 } \int_{-5}^5 \dif Y \int^0_{- 0.5 \hat s} \dif \hat t \, \left[ q(x_1) \overline{q}(x_2) \overline{\abs{M}^2}(\hat s,\hat t,\hat u) + \overline{q}(x_1) q(x_2) \overline{\abs{M}^2}(\hat s,\hat u,\hat t) \right] \mathcal{A}(\hat s, \hat t, Y),
\end{align}
for the $q\bar q$ case, where we integrate over the average rapidity of the two final state particles $Y = \frac{1}{2} (y_3 + y_4)$ up to an arbitrary cutoff $\abs{Y} < 5$, to which we hope to have no sensitivity on account of the acceptance function $\mathcal{A}(\hat s, \hat t, Y)$. In this notation $\hat s = x_1 x_2 s$ and $x_{1/2} = \sqrt{\frac{\hat s}{s}} e^{\pm Y}$, and we define $\overline{\abs{M}^2} = \frac{1}{4 N_c} \sum_{\lambda_1 \lambda_2 \lambda_3 \lambda_4} \abs{M_{\lambda_1 \lambda_2 \lambda_3 \lambda_4}}^2$, where $\lambda_i = \pm$ labels the helicity of each particle and $N_c = 3,8$ is the number of colour degrees of freedom of a particle in the initial state. Note the lower integration limit of $\hat t > -0.5 \hat s$ (\emph{i.e.}, $\theta < \frac{1}{2} \pi$) on account of the indistinguishability of the two final state photons. In what follows, when evaluating the PDFs we use the central values of the NNPDF 3.0 NNLO set with $\alpha_s(m_Z) = 0.118$ and  $Q^2 = \frac{1}{2} \hat s$ \cite{Ball:2014uwa}.

\subsection{Interference effects}

\begin{figure}[t] 
   \centering
\includegraphics[width=0.5\textwidth]{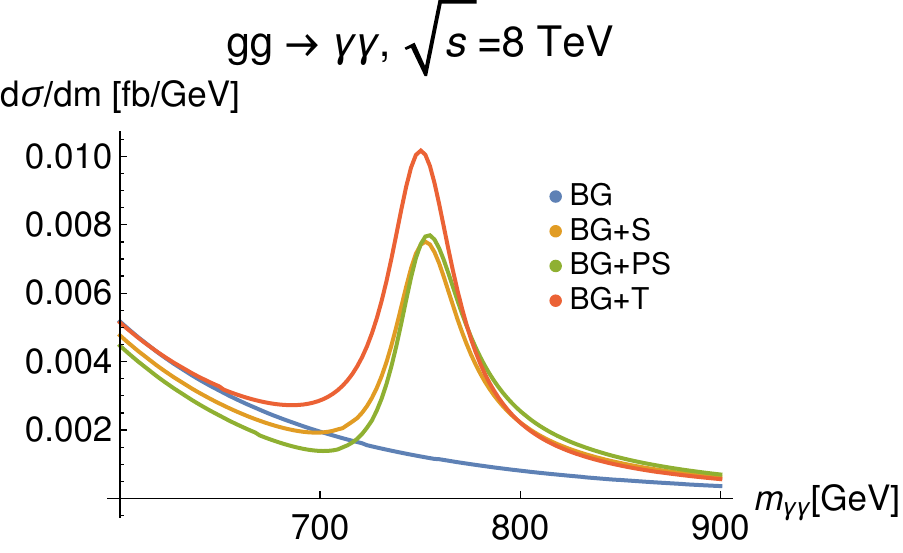}\includegraphics[width=0.5\textwidth]{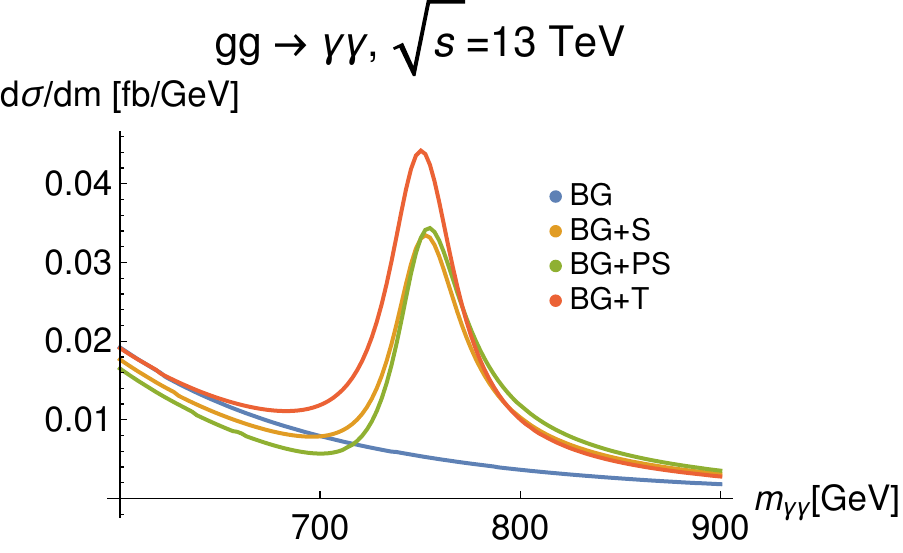}
\includegraphics[width=0.5\textwidth]{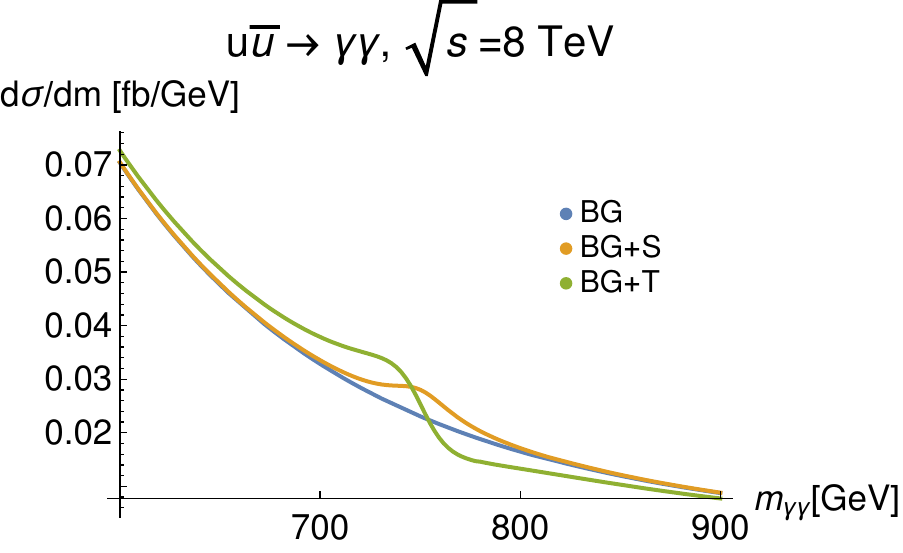}\includegraphics[width=0.5\textwidth]{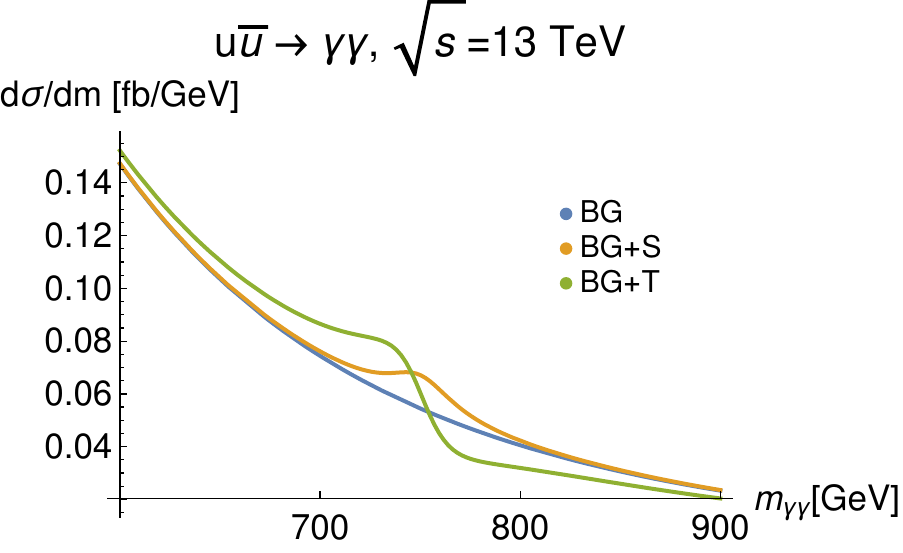}
   \caption{Scalar (S), pseudoscalar (PS), spin-2 (T) signals plus background at $\sqrt{s} =8, 13$ TeV. Upper left: $gg$ initiated signal at $\sqrt{s} = 8$ TeV, showing only the $gg$ initiated background for clarity. Upper right: $gg$ initiated signal at $\sqrt{s} = 13$ TeV, showing only the $gg$ initiated background. Lower left: $q \bar q$ initiated signal at $\sqrt{s} = 8$ TeV, showing only the $q \bar q$ initiated background for clarity. Lower right: $q \bar q$ initiated signal at $\sqrt{s} = 13$ TeV, showing only the $q \bar q$ initiated background. When $q \bar q$-initiated, the S and PS spectra are identical, and we omit the latter. In each case we have taken the resonance mass $M = 750\, \GeV$, resonance width $\Gamma = 40\, \GeV$, and amplitudes $A = (15\, \TeV)^{-2}$, for $gg$-initiated spin-0 resonances, $A = (200\, \TeV)^{-1}$ for $q \bar q$-initiated spin-0 resonances, and $A = (5\, \TeV)^{-2}$ for spin-2 resonances of either initial state. In each case we assume $A$ is real. We require $\abs{\eta} < 2.5$ for both photons.}
 \label{fig:interfere}
\end{figure}

To illustrate the effects of resonance-continuum interference, the scalar (S), pseudoscalar (PS), and spin-2 (T) signals (both $q \bar q$ and $gg$-initiated) plus background at $\sqrt{s} =8, 13$ TeV are shown in Figure \ref{fig:interfere}. In each case we show only the background contribution that potentially interferes with the signal (e.g., only the $q \bar q$-initiated background for a $q \bar q$-initiated signal); at a collider, this background would be summed incoherently with additional background contributions that do not interfere. 

\begin{figure}[h] 
   \centering
\includegraphics[width=0.5\textwidth]{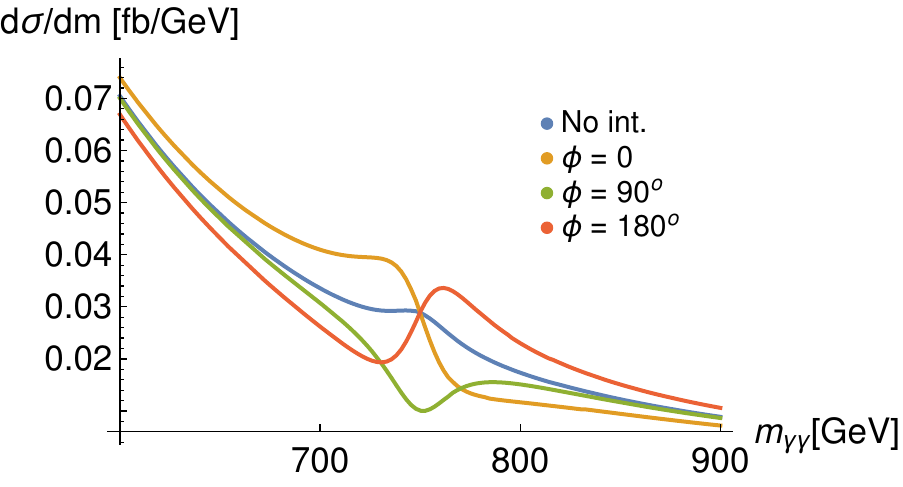}\includegraphics[width=0.5\textwidth]{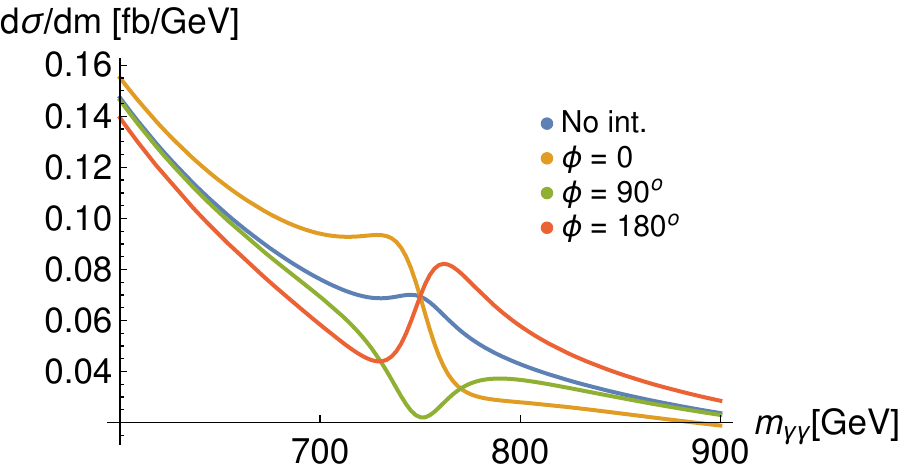}
   \caption{Spin-2 $q \bar q$-initiated signal plus $q \bar q$ background, varying the phase $\phi$ of $A$ at $\sqrt{s} =$ 8 TeV (left) and $\sqrt{s} =$ 13 TeV (right). The case of incoherently-summed signal and background is shown for comparison.}
 \label{fig:phase}
\end{figure}

In general, interference effects are modest at high invariant masses for $gg$-initiated signals, since the $gg$-initiated background is relatively small here. In contrast, for $q \bar q$-initiated signals the large $q \bar q$-initiated backgrounds at high invariant mass raise the prospect of considerable interference, but the size of these effects depends sensitively on the spin of the resonance. In the spin-0 case the signal and background do not interfere, while there are dramatic effects in the spin-2 case that lead to a characteristic peak-dip structure. 

Although we have chosen real $A$ for illustration in Figure \ref{fig:interfere}, it is in principle possible for signal amplitudes to carry a phase relative to the background. Such a phase may arise, for example, when part or all of the coupling of the spin-2 resonance to photons is induced by loops of particles light enough to be produced on-shell, in which case the phase is associated with a branch cut (see e.g. \cite{Geng:2016xin}). The effects of a phase in $A$ are again seen most clearly in the case of a spin-2 resonance, as illustrated in Figure \ref{fig:phase}, which shows the effect (at 8 and 13 \TeV ~respectively) of either turning off the interference or varying the phase $\phi$ of $A$, keeping its magnitude constant, for the dramatic spin 2 $q \bar q$ case.

While flipping the sign of the signal amplitude ($180^\circ$ phase) unsurprisingly transitions from a dip-peak structure to a peak-dip structure, note that in the case of a $90^\circ$ phase between signal and background, resonance-continuum interference leads to a strict deficit at the location of the resonance.

\section{Data and selections\label{sec:data}}

We now compare our theoretical predictions for signal and background to the diphoton spectrum at $\sqrt{s} = 8, 13$ TeV as measured by the ATLAS and CMS collaborations. To do so, we adopt a hybrid approach that employs a fit to the data for the continuum-only contribution and a suitably-normalized theoretical calculation for the resonance-continuum interference and the resonance-only contribution. 

To model the background-only component, we follow the collaborations in fitting one of two curves, $f_\text{ATL}^{\rm fit}$ or $f_\text{CMS}^{\rm fit}$, to each dataset, where the curves are given by \cite{ATLAS-CONF-2016-018,CMS-PAS-EXO-16-018}
\begin{eqnarray}
f_\text{ATL}^{\rm fit}(\mgamgam;N,b,a_0;\sqrt{s}) &=& N \left(1-\left(\frac{\mgamgam}{\sqrt{s}}\right)^\frac{1}{3}\right)^b \left(\frac{\mgamgam}{\sqrt{s}}\right)^{a_0} ,
\label{eq:ATLASfunc} \\
f_\text{CMS}^{\rm fit}(\mgamgam; N,a,b) &=& N \mgamgam^{ (a + b \ln \mgamgam) } .
\label{eq:CMSfunc}
\end{eqnarray}
The best fit parameters for each dataset, along with the relevant analysis cuts, are enumerated in Table \ref{tab:data}. Note that as we are fitting to the binned diphoton spectra provided by the collaborations, the best fit parameters are expected to differ modestly from those used by ATLAS and CMS.

\begin{table}
\begin{adjustbox}{width=\textwidth}
\begin{tabular}{ c | c | c | c | c | c | c}
Ref. & Dataset & Fit curve & Acceptance & $\sigma_\text{res}/\mgamgam$ & $L_\text{int} / \mathrm{fb}^{-1}$ & $C$ \\ \hline
\multirow{2}{*}{\cite{ATLAS-CONF-2016-018} } &
ATLAS13SPIN0 &$f_\text{ATL}^{\rm fit}(;2.68,14.9,-2.63;)$ &\pbox{4cm}{$\abs{\eta_{1/2}} \in [0,2.37]$ \\ $E_{T,1} > 0.4 \mgamgam$ \\ $E_{T,2} > 0.3 \mgamgam$} &
\multirow{2}{*}{ 0.01 } & \multirow{2}{*}{ $3.2$ } & \multirow{2}{*}{ $0.75$ }  \\ \cline{2-4}

 &
ATLAS13SPIN2 & $f_\text{ATL}^{\rm fit}(;4.03,11.2,-2.15;)$ & \pbox{4cm}{$\abs{\eta_{1/2}} \in [0,2.37]$ \\ $E_{T,1/2} > 55 \GeV$ \\ $\mgamgam > 200 \GeV$} & & &
 \\ \hline
 
\multirow{2}{*}{\cite{CMS-PAS-EXO-16-018} } &
CMS13EBEB & $f_\text{CMS}^{\rm fit}(;0.069,6.41,-0.89)$ &\pbox{4cm}{$\abs{\eta_{1/2}} \in [0,1.44]$ \\ $p_{T,1/2} > 75 \GeV$ \\ $\mgamgam > 230 \GeV$} & 0.01 & \multirow{2}{*}{ $2.7$ } & 0.81 \\ \cline{2-5} \cline{7-7}

 &
CMS13EBEE & $f_\text{CMS}^{\rm fit}(;0.013,6.28,-0.81)$ &\pbox{4cm}{$\abs{\eta_{1}} \in [0,1.44]$ \\ $\abs{\eta_{2}} \in [1.57,2.5]$\\ $p_{T,1/2} > 75 \GeV$ \\ $\mgamgam > 320 \GeV$} & 0.015 & & 0.73
 \\ \hline
 
\pbox{5cm}{ Data \cite{ATLAS-CONF-2016-018} \\ Cuts \cite{Aad:2015mna}}  &
ATLAS8SPIN2 & $f_\text{ATL}^{\rm fit}(;4792,14.5,-1.43;)$ & \pbox{4cm}{$\abs{\eta_{1/2}} \in [0,2.37]$ \\ $E_{T,1/2} > 50 \GeV$} &
0.01 & 20.3 & 0.75  \\ \hline 

\pbox{5cm}{ Data \cite{ATLAS-CONF-2016-018} \\ Cuts \cite{Aad:2014ioa}}  &
ATLAS8SPIN0 & $f_\text{ATL}^{\rm fit}(;4.43,11.5,-2.89;)$ &\pbox{4cm}{$\abs{\eta_{1/2}} \in [0,2.37]$ \\ $E_{T,1} > 0.4 \mgamgam$ \\ $E_{T,2} > 0.3 \mgamgam$} &
0.01 & 20.3 & 0.75  \\ \hline 

\pbox{5cm}{\cite{Khachatryan:2015qba}} &
CMS8HIGGS & $f_\text{ATL}^{\rm fit}(;126,14.2,-2.22;)$ &\pbox{4cm}{$\abs{\eta_{1/2}} \in [0,2.5]$ \\ $p_{T,1} > \frac{1}{3} \mgamgam$ \\ $p_{T,2} > \frac{1}{4} \mgamgam$} & 0.017 & 19.7 & 0.86 \\
\end{tabular}
\end{adjustbox}
\caption{ATLAS and CMS diphoton spectrum measurements at $\sqrt{s} = 8, 13$ TeV used in this analysis, including the best-fit values for the background curves $f_{\rm ATL}^{\rm fit}$ and $f_{\rm CMS}^{\rm fit}$; the geometric acceptance; the diphoton invariant mass resolution $\sigma_\text{res}/\mgamgam$; the integrated luminosity $L_{\rm int}$ in fb$^{-1}$; and the efficiency factor $C$ for each data set. In the ``Fit curve'' column, the entries are of the form $f_{\rm ATL}^{\rm fit}(;N,b,a_0;)$ and $f_{\rm CMS}^{\rm fit}(;N,a,b)$, corresponding to the parameters appearing in (\ref{eq:ATLASfunc}) and (\ref{eq:CMSfunc}), respectively. In the ``Acceptance'' column, the subscripts `1' and `2' respectively refer to the leading and subleading photon in $p_T$.   \label{tab:data}}
\end{table}

To model the resonance-continuum interference term and the pure resonance term, we suitably adapt a theoretical calculation of these contributions to account for the acceptance times efficiency in each analysis; higher-order corrections to signal and background; and potentially significant reducible backgrounds from $\gamma j$ and $jj$ processes in which one or more jets fake a photon. We first obtain a theoretical prediction for the continuum background $f_{\rm cont}^{\rm theory}(m_{\gamma \gamma})$ and compare it to the fitted curve $f^{\rm fit}(\mgamgam)$ obtained from data. This allows us to suitably normalize our theory calculation of the genuine $\gamma \gamma$ contribution to the continuum background. We then apply the same normalization factors to a theoretical prediction for the resonance-continuum interference term and the pure resonance term. 

In particular, we obtain a theoretical prediction for the continuum background as follows: Using gamma2MC~\cite{Bern:2002jx} at $\sqrt{s}=8,13 \, \TeV$, and considering respectively the $\bar{q}q$ and $gg$ initiated processes, we calculate $\mgamgam$-dependent scale factors, $K_q(\mgamgam,\sqrt{s})$ and $K_g(\mgamgam,\sqrt{s})$, from the ratio of the NLO to LO diphoton spectrum. Near $\mgamgam = 750$ GeV, these scale factors are $K_q \sim 1.3$ and $K_g \sim 1.45$ at $\sqrt{s} = 13$ TeV. We take the geometric acceptance $\mathcal{A}$ to be a step function, equal to 1 in the kinematic region defined for each dataset in Table~\ref{tab:data}, and 0 elsewhere. We also multiply by a constant efficiency $C$, which we estimate for ATLAS from the auxiliary material of~\cite{Aad:2012tba}, and for CMS using the text accompanying the individual datasets. The value of $C$ for each dataset is given in Table \ref{tab:data}. 

We then account for the contribution of $\gamma j$ and $jj$ fakes by extracting the quoted fractional fake rate $\epsilon(\mgamgam)$ for each $13 \, \TeV$ dataset. At invariant masses around $750$ GeV, the quoted central values of $\epsilon(\mgamgam)$ are 0.15 for CMS13EBEB and CMS13EBEE, and 0.05 for ATLAS13SPIN0 and ATLASSPIN2, all with large error bars. We obtain the combined theoretical estimate of the diphoton continuum background $f_{\rm cont}^{\rm theory}(m_{\gamma \gamma})$ by summing our calculated continuum $\gamma \gamma$ background rate (accounting for the bin size $B$ and integrated luminosity $L_\text{int}$) and the $\gamma j$ and $jj$ fake rate (computed as a fraction of the {\it fitted} background):
\begin{equation}
f_{\rm cont}^{\rm theory}(m_{\gamma \gamma}) = L_\text{int} B C \left[ K_q(\mgamgam)  \frac{\dif \sigma}{\dif m_{\gamma\gamma}} \Big|_\text{qq} + K_g(\mgamgam) \frac{\dif \sigma}{\dif m_{\gamma\gamma}} \Big|_\text{gg} \right] + \epsilon(\mgamgam) f^{\rm fit}(\mgamgam) ,\label{eq:bkgPred}
\end{equation}
where the fake rate $\epsilon \cdot f^{\rm fit}$ is determined using the appropriate ATLAS or CMS data set. 

We then compare this theory prediction for the diphoton continuum  background $f_{\rm cont}^{\rm theory}(m_{\gamma \gamma})$ to the fitted spectrum $f^{\rm fit}(\mgamgam)$  and compute the ratio $\mathcal{F} \equiv f^{\rm fit}(\mgamgam)/f_{\rm cont}^{\rm theory}(m_{\gamma \gamma})$ in the range $700 \, {\rm GeV} \leq \mgamgam \leq 800 \, {\rm GeV}$ for each data set. Thus computed, we find that $\mathcal{F} \simeq 1$ within error bars for the ATLAS13Spin0 and CMS13EBEB datasets, whereas for both ATLAS13Spin2 and CMS13EBEE $\mathcal{F} \simeq 1.25$. We use the appropriate $\mathcal{F}$ obtained for each data set as a final normalization factor for the resonance-continuum and pure resonance terms. For the $\sqrt{s} = 8$ TeV data sets, due to the large uncertainty on $\gamma j$ and $jj$ fakes at high invariant mass, we do not account for fake contributions and set $\mathcal{F} = 1$. Given that the high invariant mass signals we consider would be significantly more prominent in $\sqrt{s} = 13$ TeV data, this choice has negligible impact on our fits.

Finally, in all cases we convolve the signal features of our spectrum with a Gaussian line shape, $G(x)$, to simulate the detector response:
\begin{equation}
G(x;\sigma_{\rm res}) = \frac{1}{\sigma_{\rm res} \sqrt{2\pi}} \exp \left[ - \frac{x^2}{2 \sigma_{\rm res}^2}\right],
\end{equation}
where $\sigma_{\rm res}$ is calculated using the fractional uncertainties in the invariant mass, which are quoted in the experimental papers and tabulated in Table~\ref{tab:data}.

Combining the fitted background-only contribution with our suitably normalized resonance-continuum and pure resonance contributions, our prediction for the total number of events in a bin of width $B$ around a central value $\mgamgam$ is therefore
\begin{equation}
f_{\rm tot}^{\rm theory}(\mgamgam) = f^{\rm fit}(\mgamgam) + L_\text{int} C \mathcal{F} \int_{\mgamgam - \frac{1}{2}B}^{\mgamgam + \frac{1}{2}B} \dif \mgamgam^\prime \, \, G * \left( K \frac{\dif \sigma}{\dif m_{\gamma\gamma}^\prime} \Big|_\text{signal and interference} \right)
\label{eq:binSpec}
\end{equation}
where $*$ indicates convolution.

\section{Diphoton Peaks} \label{sec:peak}

Given our prediction for the total number of events in a binned $\mgamgam$ spectrum for a given signal hypothesis, we now investigate the implications of resonance-continuum interference for the extraction of model parameters from peaks and valleys in the measured diphoton spectrum. We begin by considering the effect of resonance-continuum interference on the interpretation of excesses in the diphoton spectrum, taking the statistical fluctuations around 750 GeV in the pre-2016 datasets as an example.

For our 6 signal hypotheses (S, PS and T, each either $\bar{q}q$ or $gg$ initiated), we scan over a grid of the parameters mass $M$, width $\Gamma$, and amplitude $A$ in the propagator factor (\ref{eq:prop}); at each point, we compute the likelihood of the datasets in Table~\ref{tab:data} given our prediction (\ref{eq:binSpec}) for the binned spectrum. We choose common points in mass and width of:
\begin{eqnarray}
\Gamma / \GeV &\in& \{5,10,20,30,40,50,60,70\}, \\
M / \GeV &\in& \{700,710,720,730,740,750,760,770,780,790\},
\end{eqnarray}
whereas the amplitude (which we here assume to be real) takes values
\begin{eqnarray}
\Lambda / \TeV &\equiv& A^{-\frac{1}{2}} /\TeV \in \{8,10,15,20,30\} \text{  for S gg}, \\
\Lambda / \TeV &\equiv& A^{-1} /\TeV \in \{50,100,150,200,250\} \text{  for S qq}, \\
\Lambda / \TeV &\equiv& A^{-\frac{1}{2}} /\TeV \in \{8,10,15,20,30\}  \text{  for PS gg}, \\
\Lambda / \TeV &\equiv& A^{-1} /\TeV \in \{50,100,150,200,250\} \text{  for PS qq}, \\
\Lambda / \TeV &\equiv& A^{-\frac{1}{2}} /\TeV \in \{2,4,6,8\} \text{  for T gg}, \\
\Lambda / \TeV &\equiv& A^{-\frac{1}{2}} /\TeV \in \{2,4,6,8\} \text{  for T qq}.
\end{eqnarray}
For each bin of the datasets, we compute the Poisson log likelihood of the signal hypothesis. We sum the log likelihoods of the datasets ATLAS13Spin0, CMS13EBEB, CMS13EBEE to calculate the likelihood of a `Combined13' dataset, as well as summing those of ATLAS8Spin0 and CMS8Higgs to make `Combined8'. Fits of the `Combined8+13' dataset are then the sum of likelihoods of the `Combined8' and `Combined13' datasets.

\begin{figure}
   \centering
\includegraphics[width=0.24\textwidth]{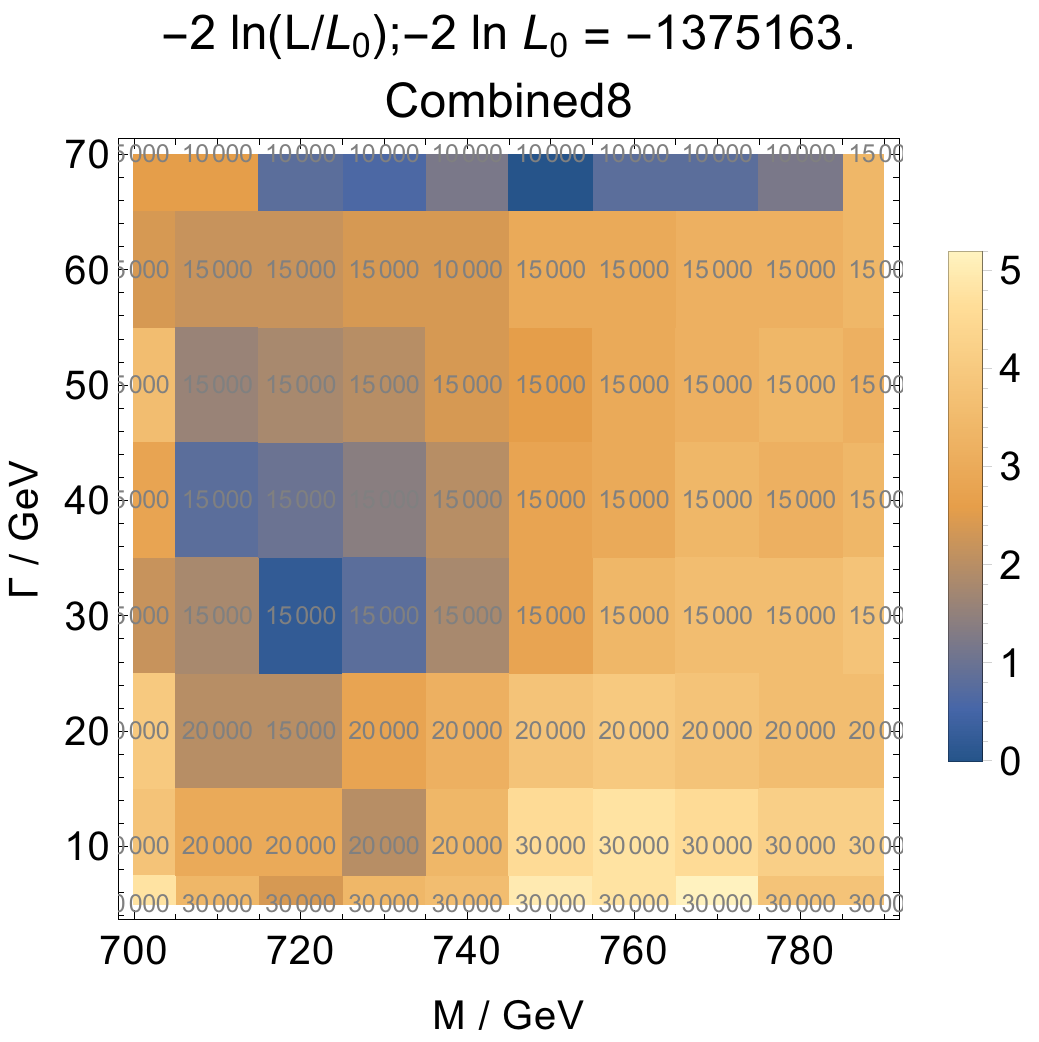}
\includegraphics[width=0.24\textwidth]{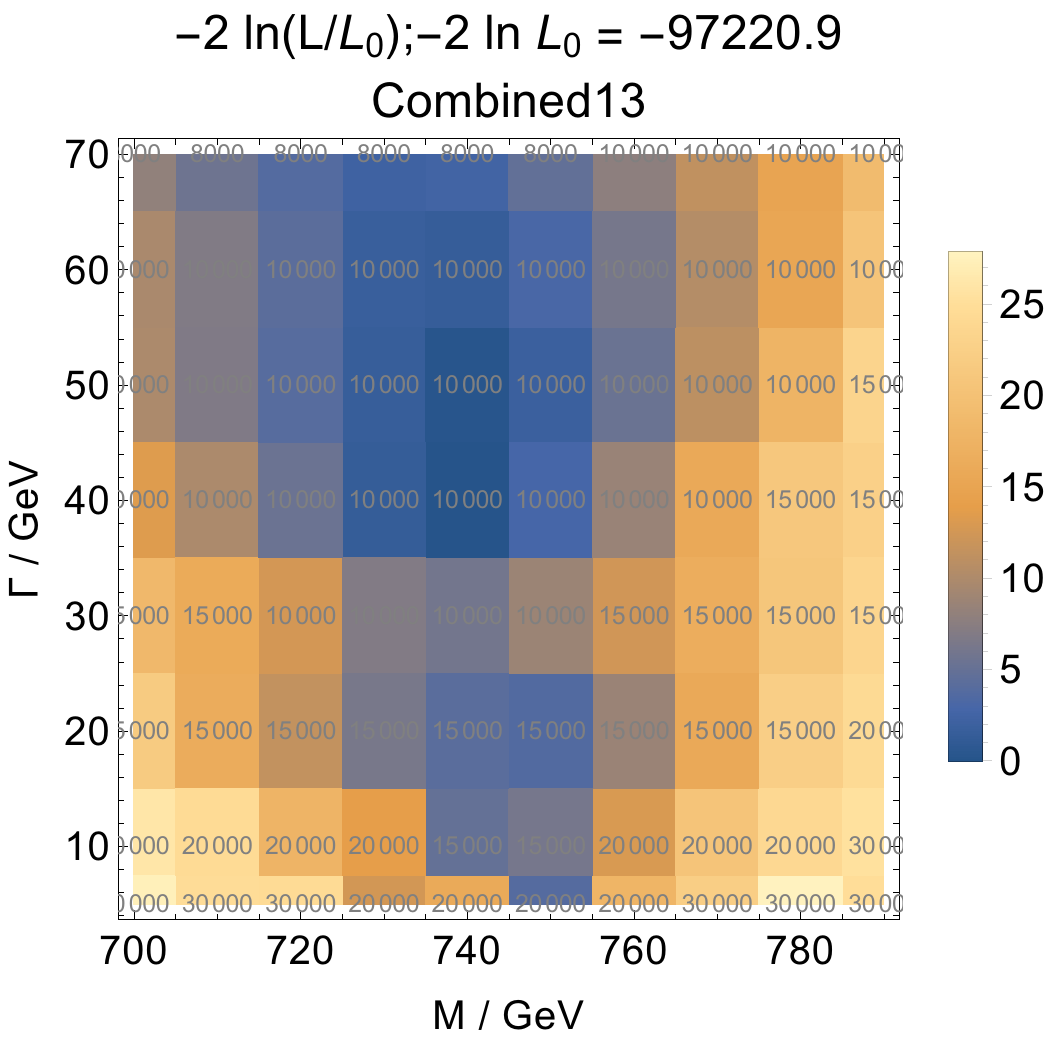}
\includegraphics[width=0.24\textwidth]{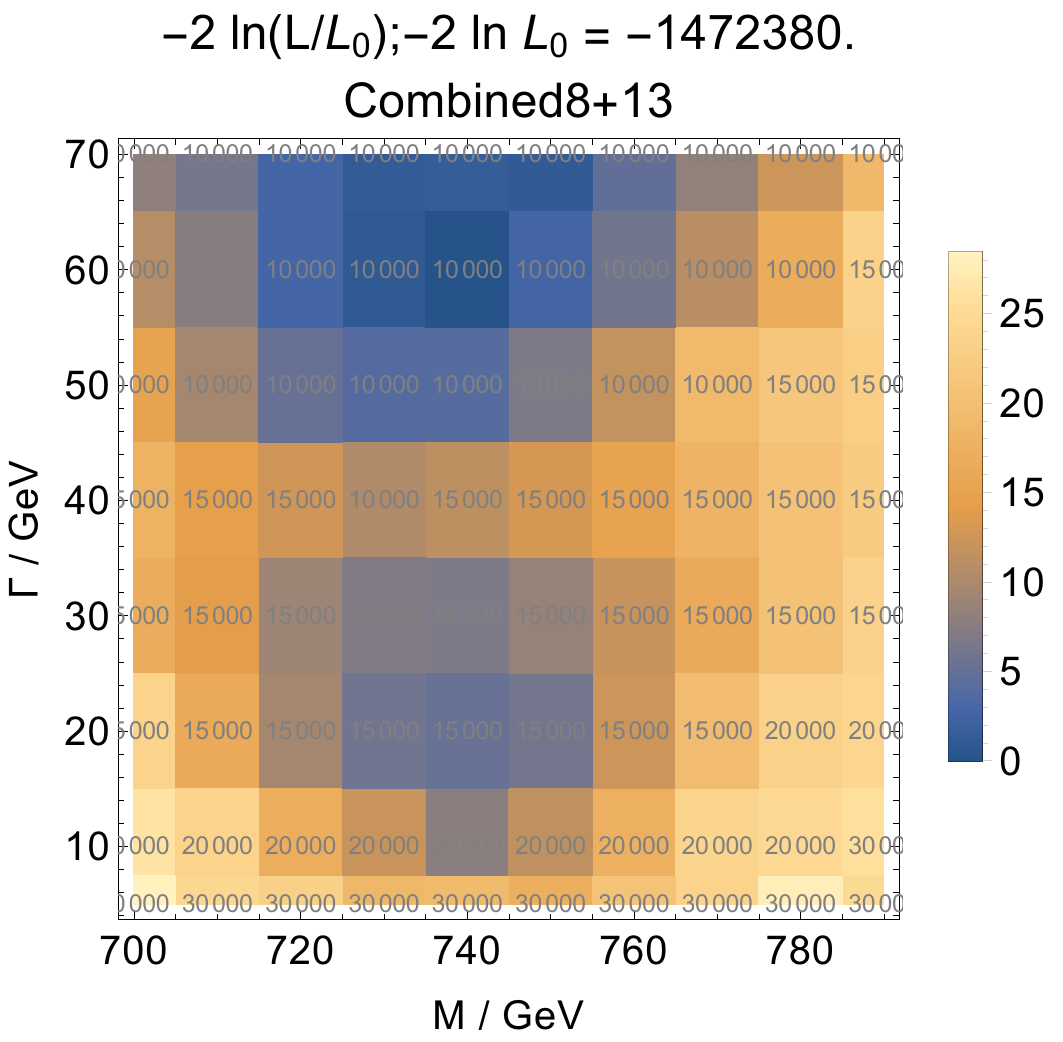}
\includegraphics[width=0.24\textwidth]{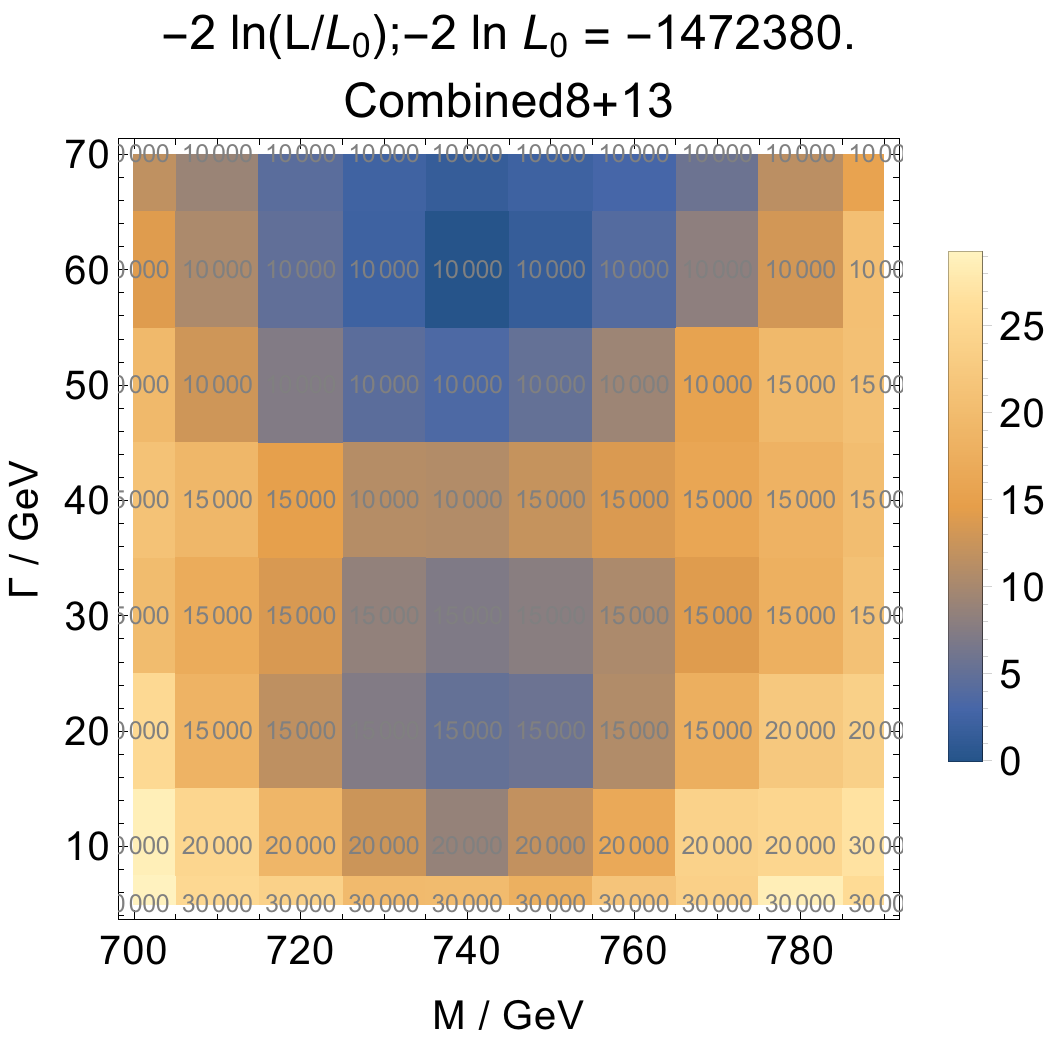}

\includegraphics[width=0.24\textwidth]{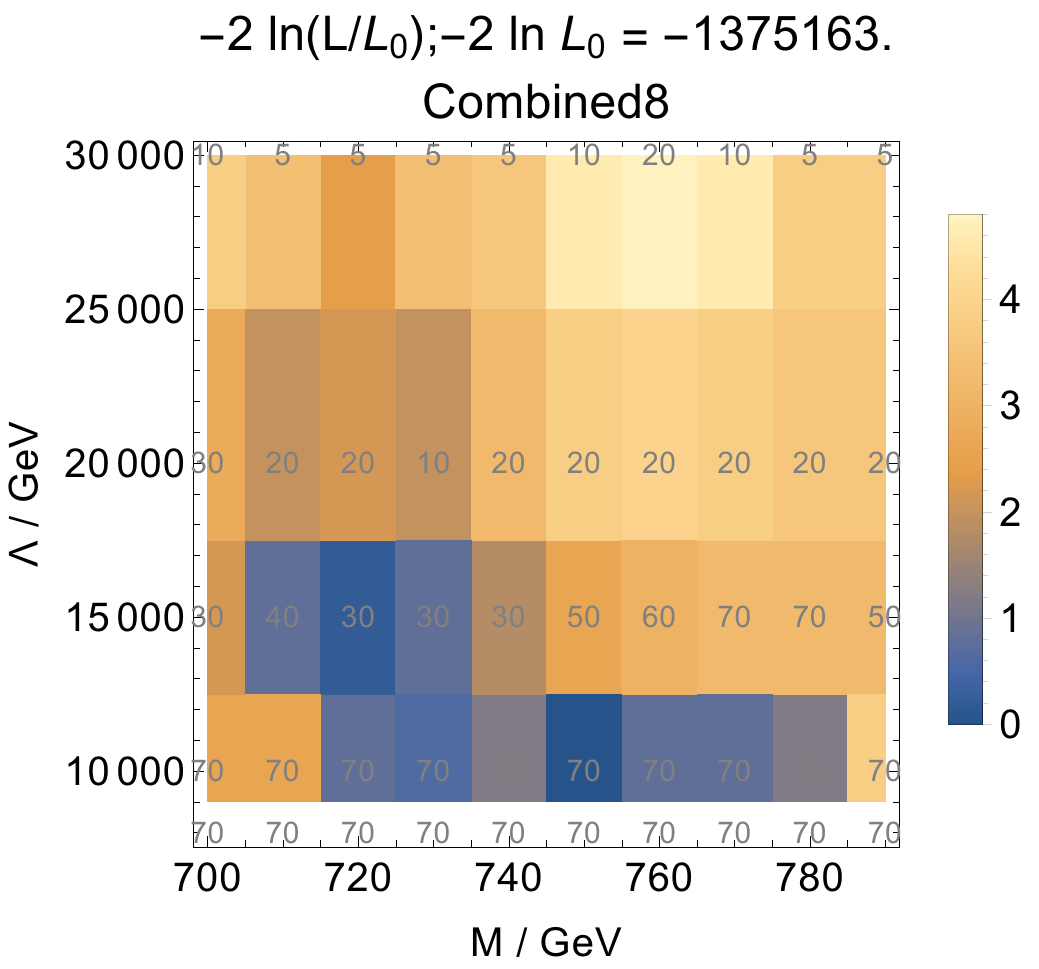}
\includegraphics[width=0.24\textwidth]{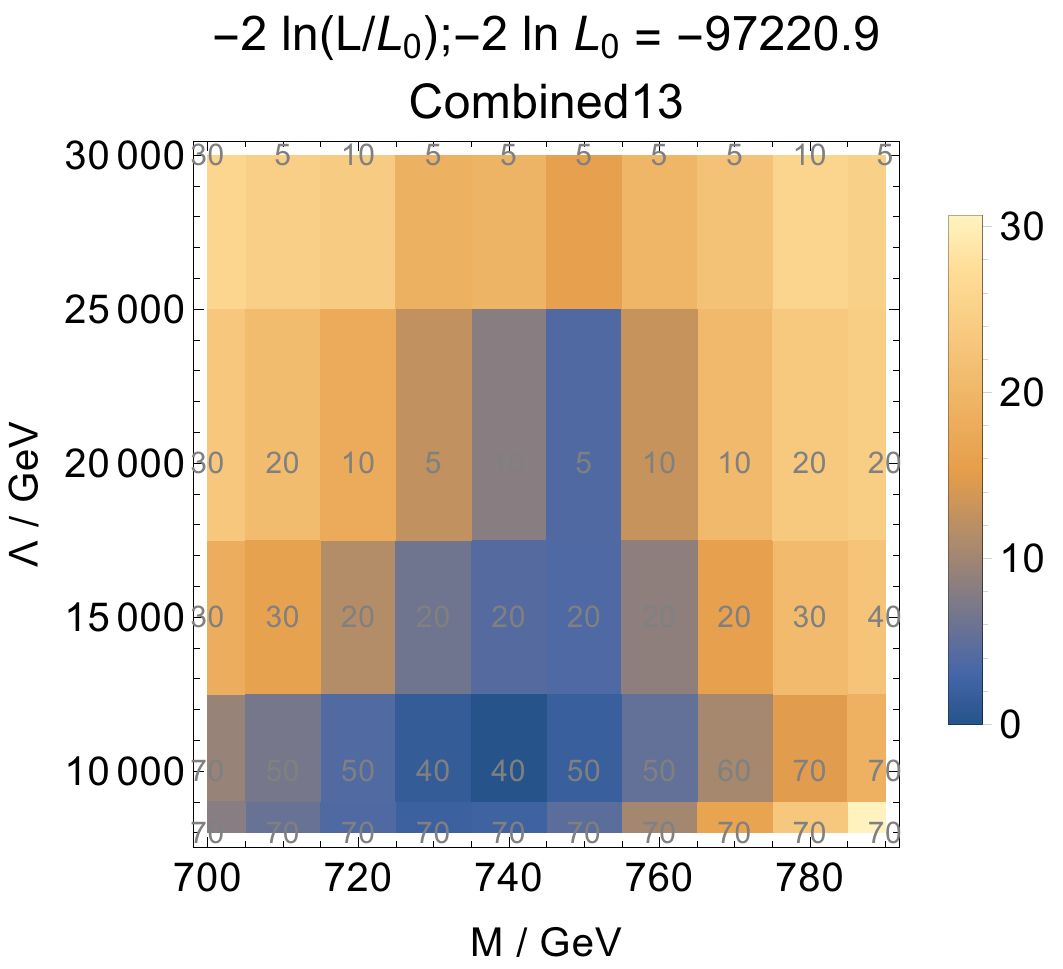}
\includegraphics[width=0.24\textwidth]{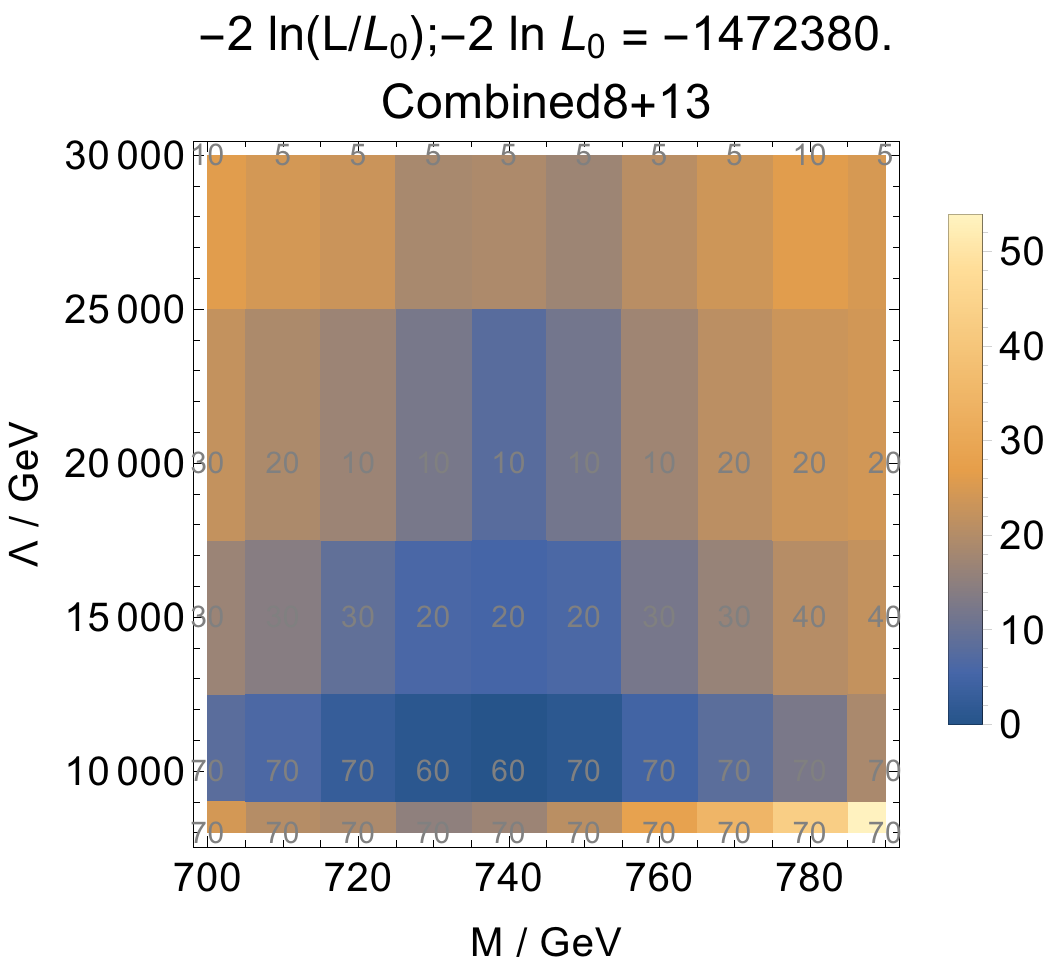}
\includegraphics[width=0.24\textwidth]{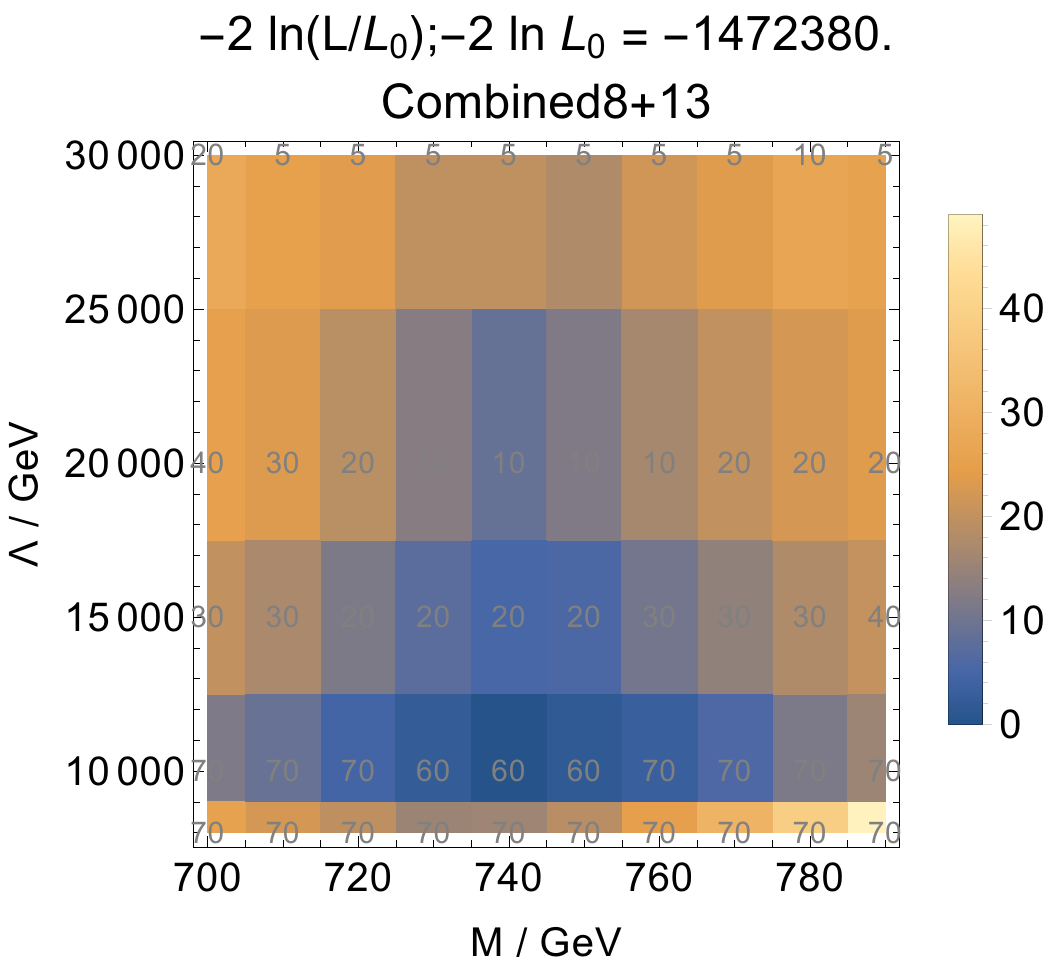}
\caption{Log likelihoods relative to the best fit points for the $gg$-initiated spin-0 scalar signal hypothesis. Top row: Likelihoods as a function of signal mass $M$ and width $\Gamma$, profiling over the signal amplitude parameter $\Lambda$. The profiled value of $\Lambda$ is shown at each grid point. From left to right, the likelihoods are shown for the combined $\sqrt{s} = 8$ TeV data, combined $\sqrt{s} = 13$ TeV data, combined $\sqrt{s} = 8+13$ TeV data, and combined $\sqrt{s} = 8+13$ TeV data {\it neglecting interference effects}. Bottom row: Likelihoods as a function of signal mass $M$ and signal amplitude parameter $\Lambda$, profiling over the signal width $\Gamma$ (with the profiled value of $\Gamma$ shown at each grid point). \label{fig:sgg}}
\end{figure}

\begin{figure}
   \centering
\includegraphics[width=0.24\textwidth]{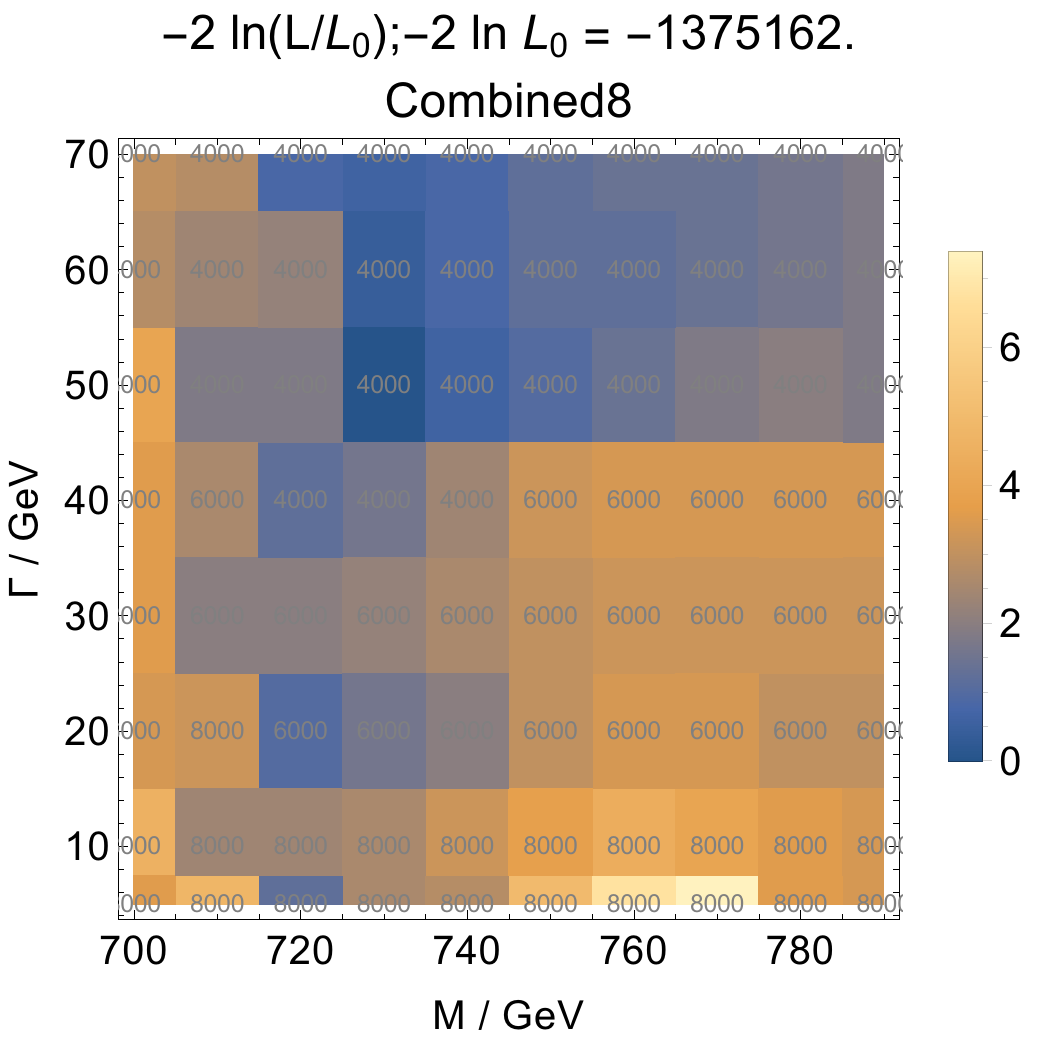}
\includegraphics[width=0.24\textwidth]{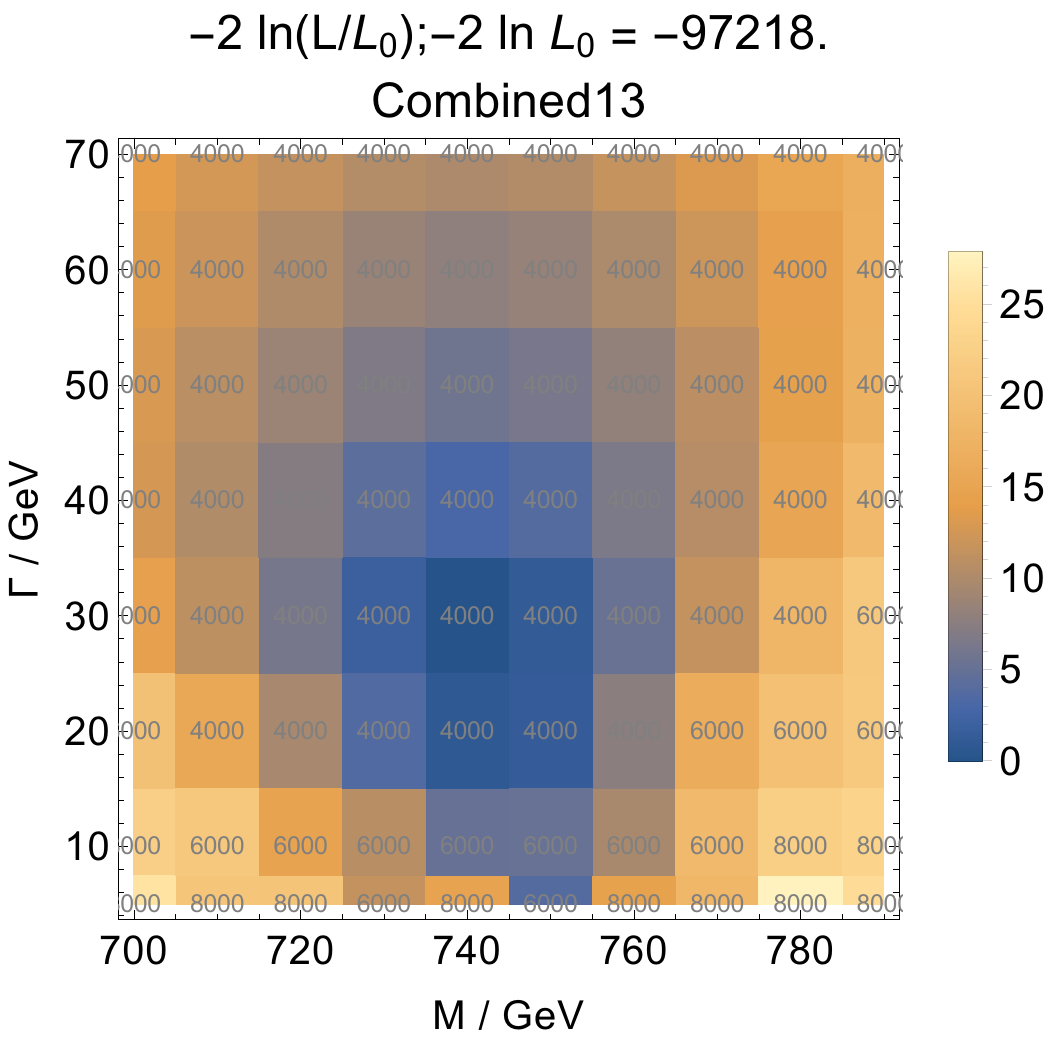}
\includegraphics[width=0.24\textwidth]{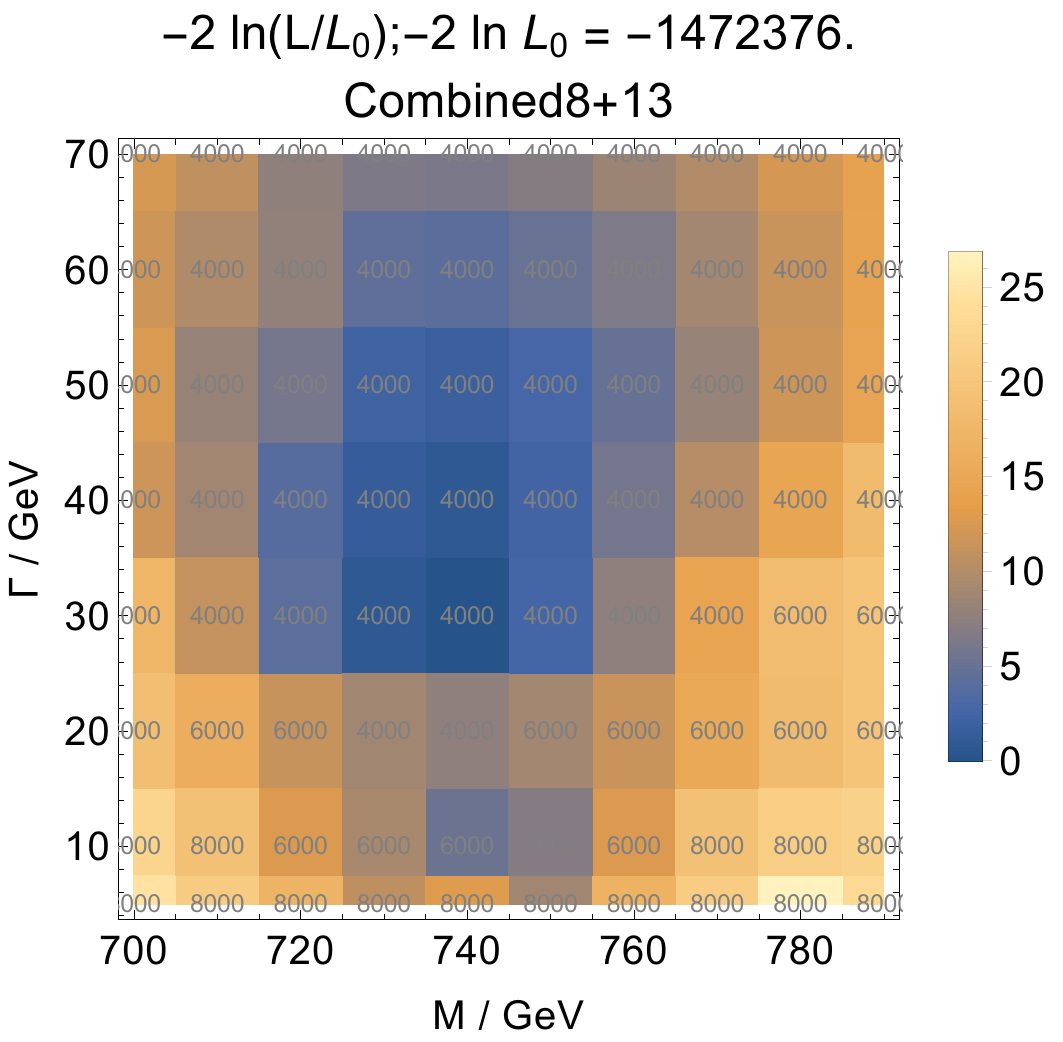}
\includegraphics[width=0.24\textwidth]{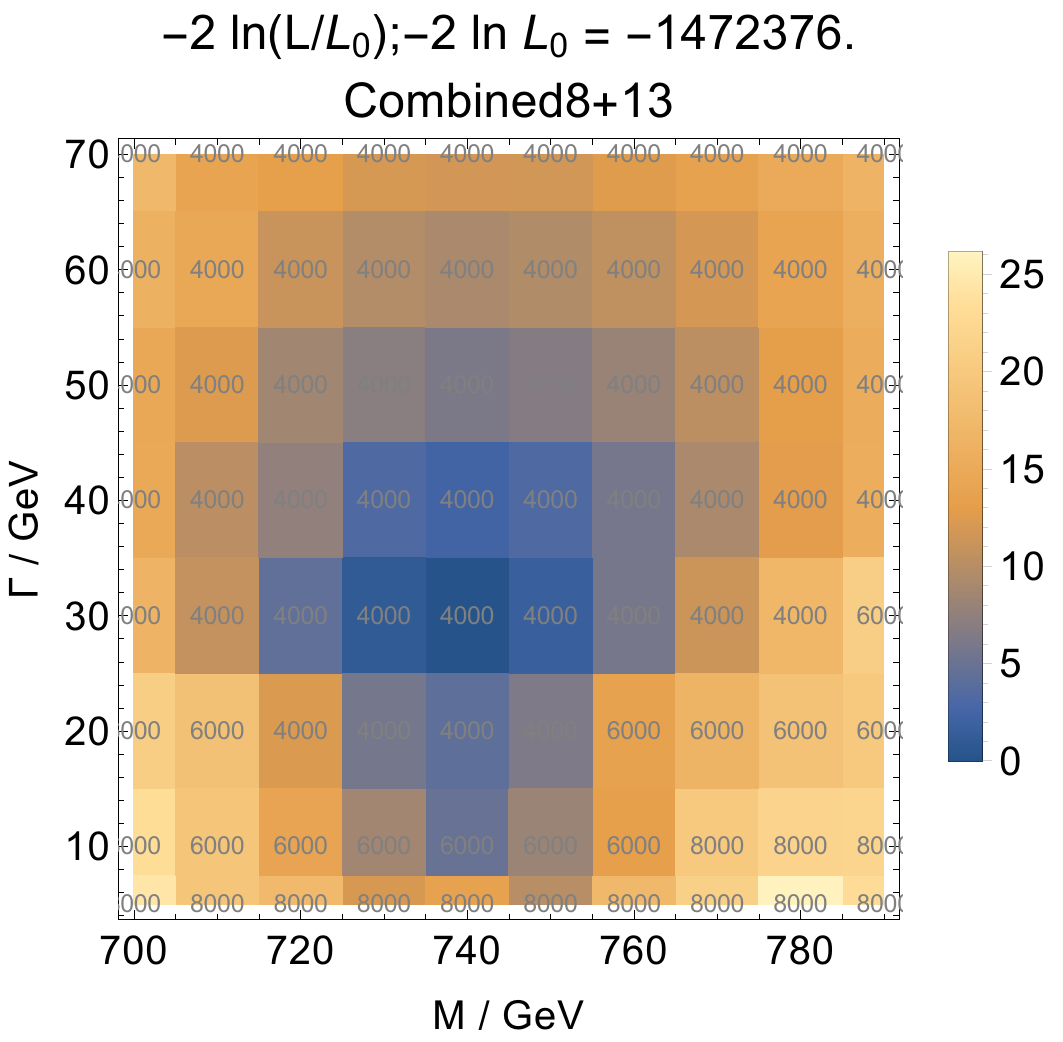}

\includegraphics[width=0.24\textwidth]{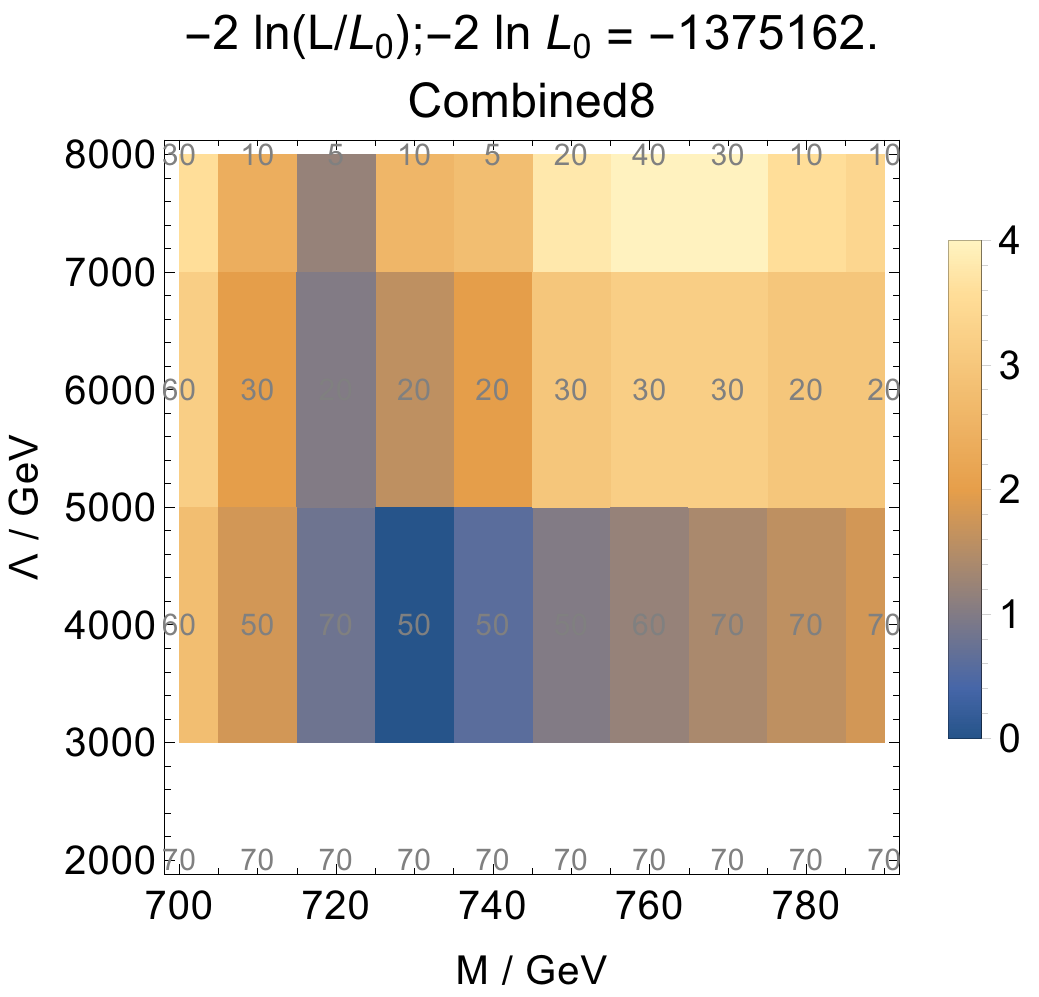}
\includegraphics[width=0.24\textwidth]{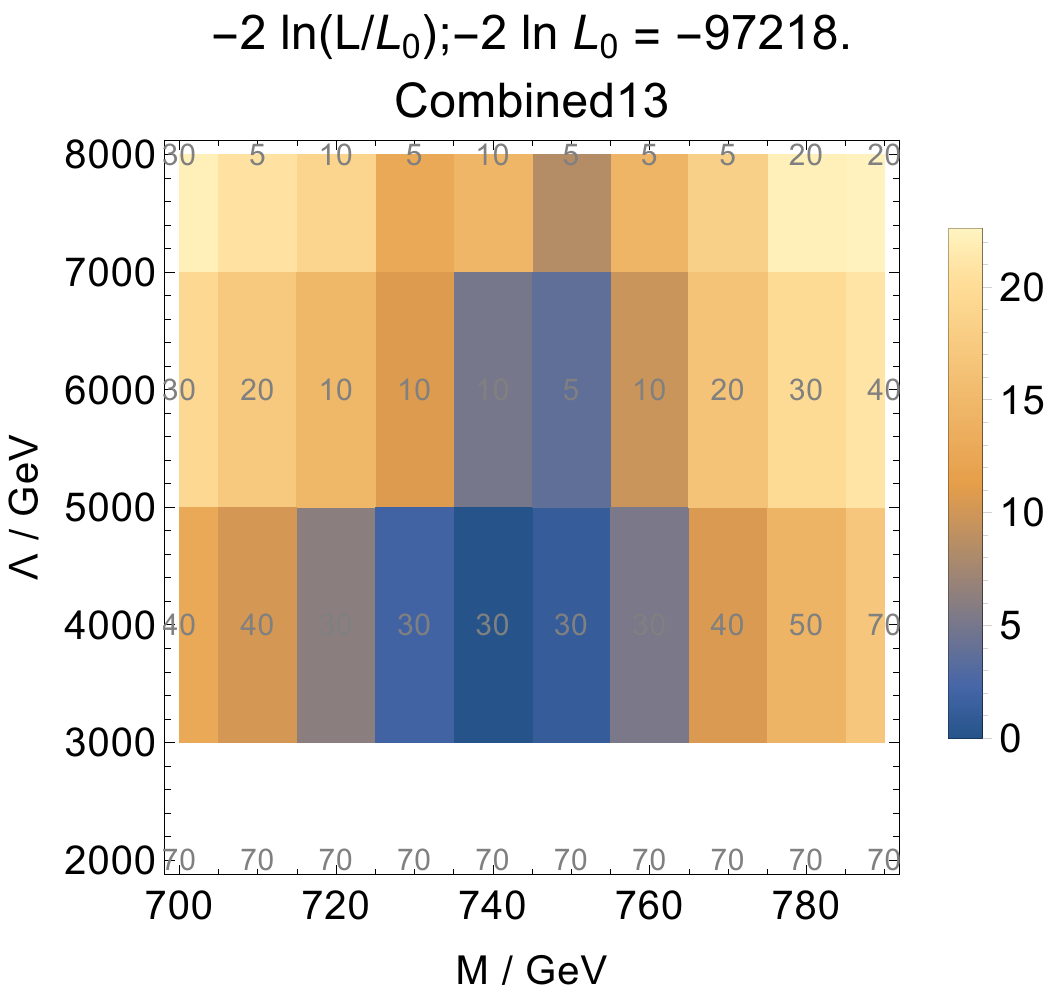}
\includegraphics[width=0.24\textwidth]{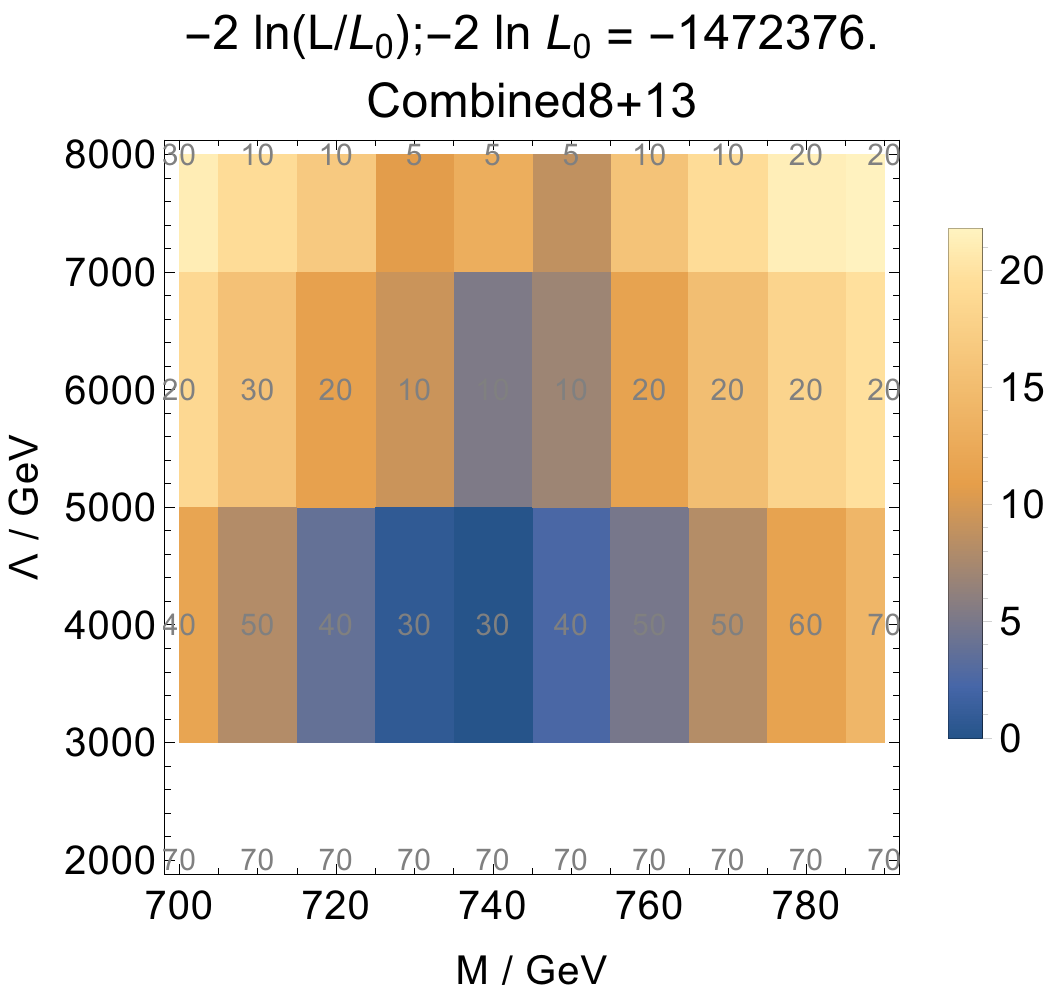}
\includegraphics[width=0.24\textwidth]{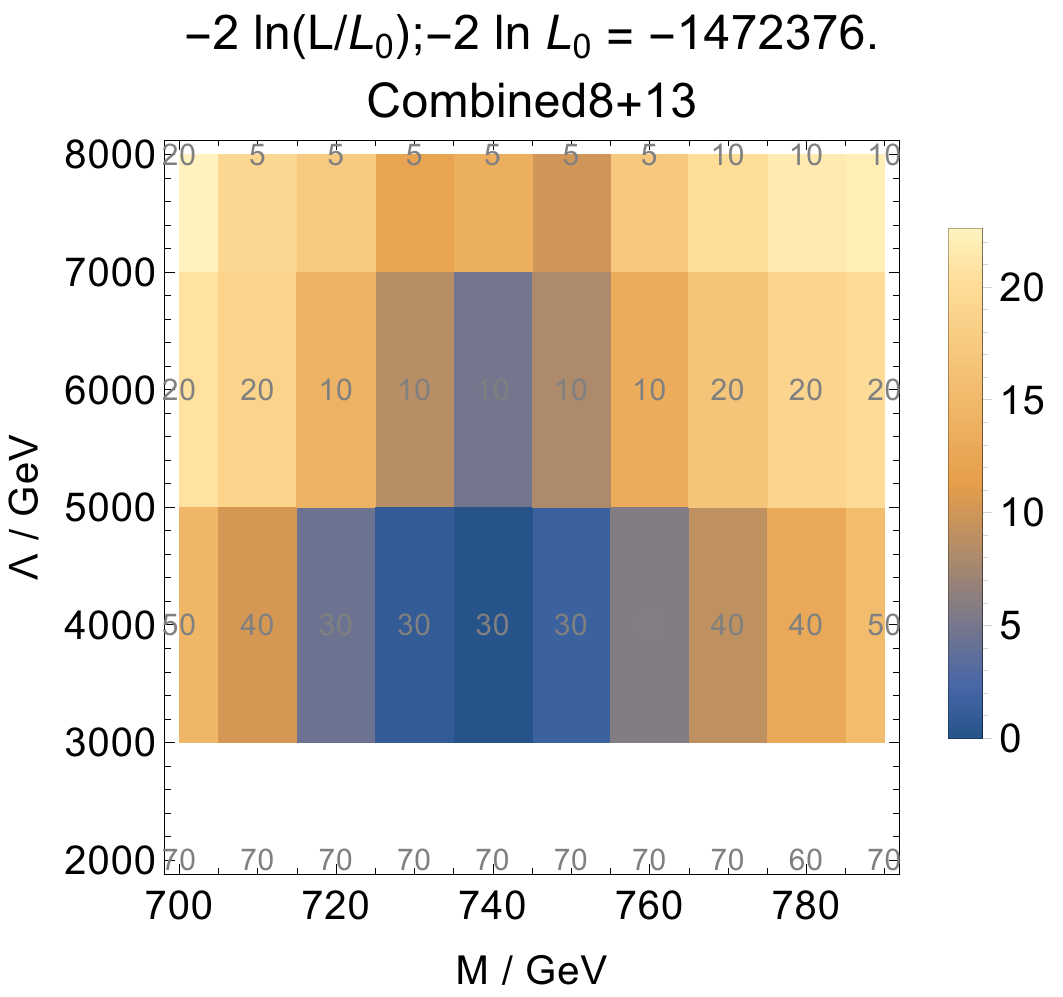}
\caption{Log likelihoods relative to the best fit points for the $gg$-initiated spin-2 signal hypothesis. Top row: Likelihoods as a function of signal mass $M$ and width $\Gamma$, profiling over the signal amplitude parameter $\Lambda$. The profiled value of $\Lambda$ is shown at each grid point. From left to right, the likelihoods are shown for the combined $\sqrt{s} = 8$ TeV data, combined $\sqrt{s} = 13$ TeV data, combined $\sqrt{s} = 8+13$ TeV data, and combined $\sqrt{s} = 8+13$ TeV data {\it neglecting interference effects}. Bottom row: Likelihoods as a function of signal mass $M$ and signal amplitude parameter $\Lambda$, profiling over the signal width $\Gamma$ (with the profiled value of $\Gamma$ shown at each grid point).\label{fig:tgg}}
\end{figure}

\begin{figure}
   \centering
\includegraphics[width=0.24\textwidth]{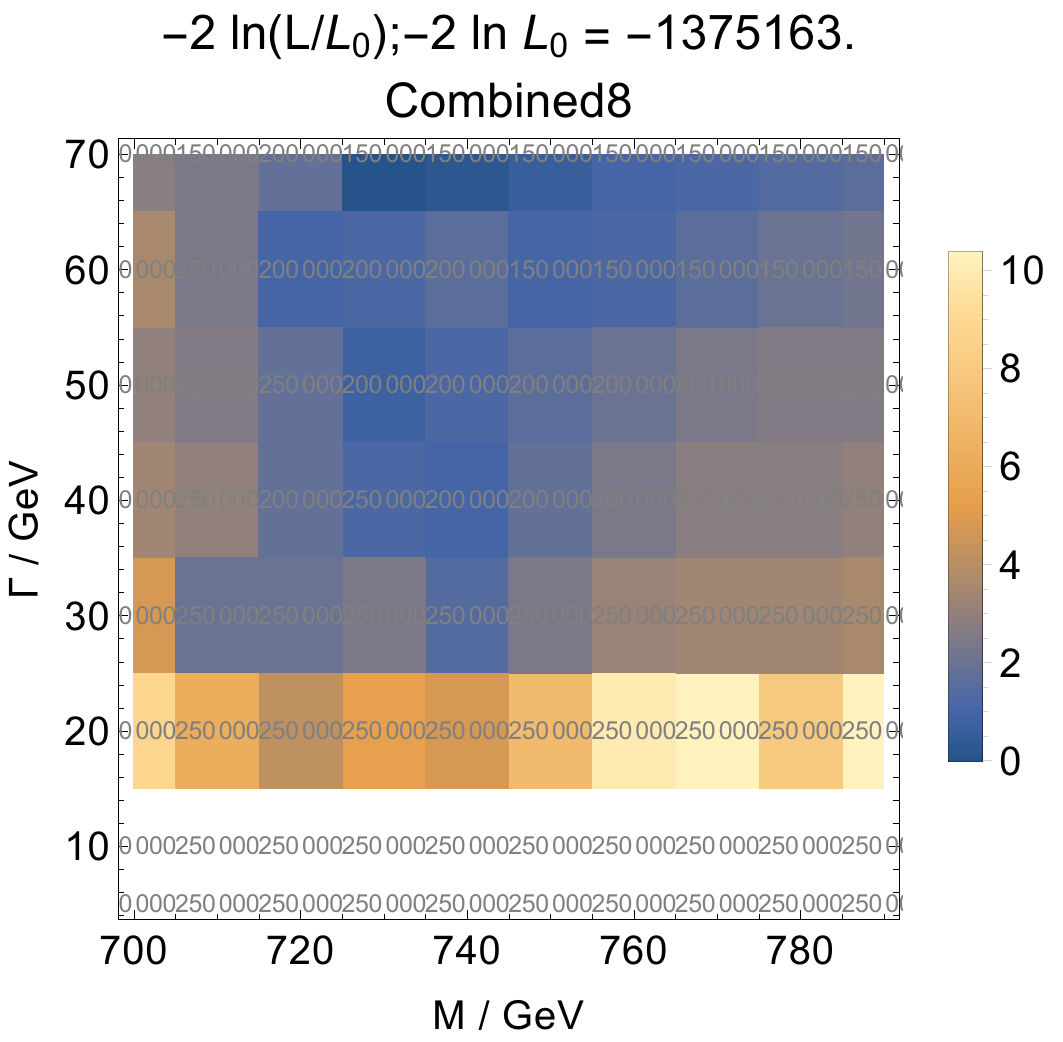}
\includegraphics[width=0.24\textwidth]{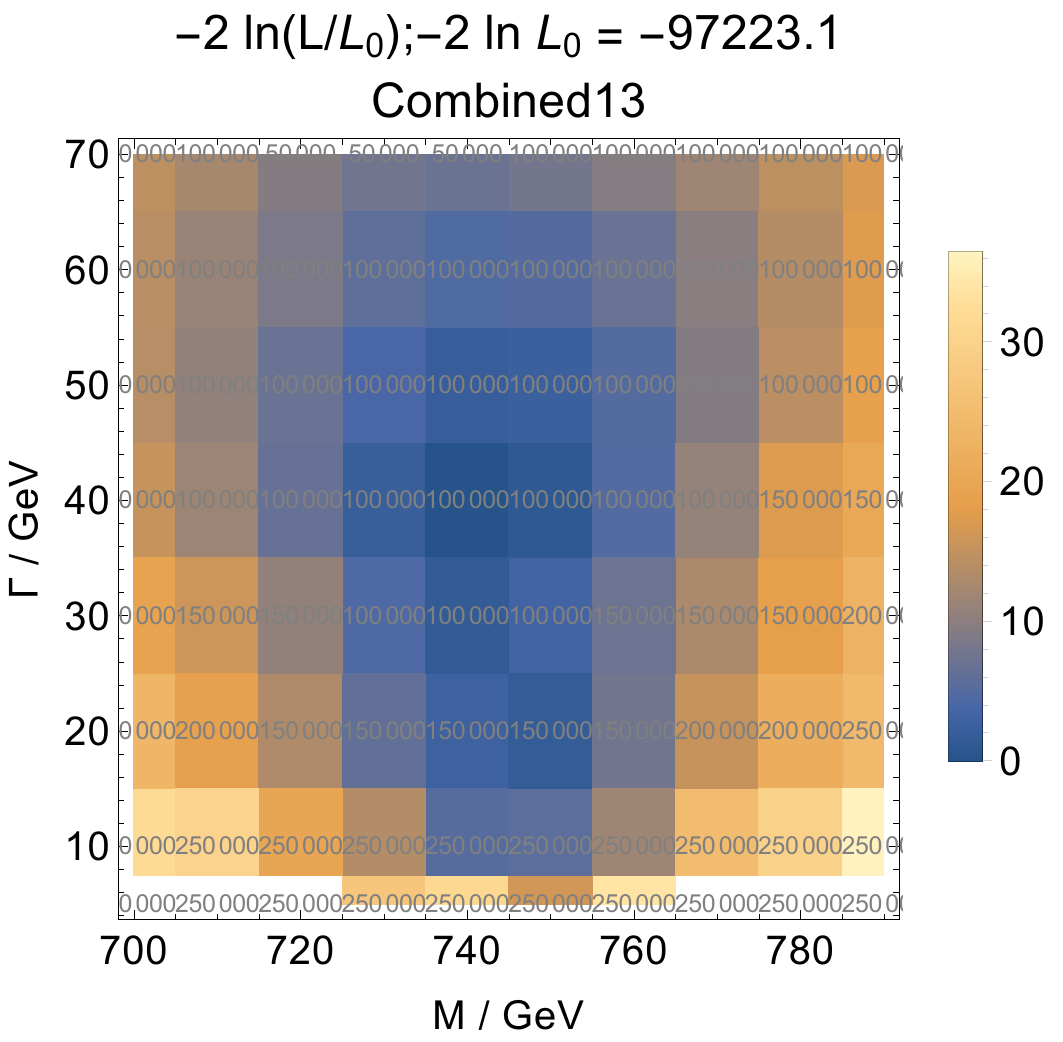}
\includegraphics[width=0.24\textwidth]{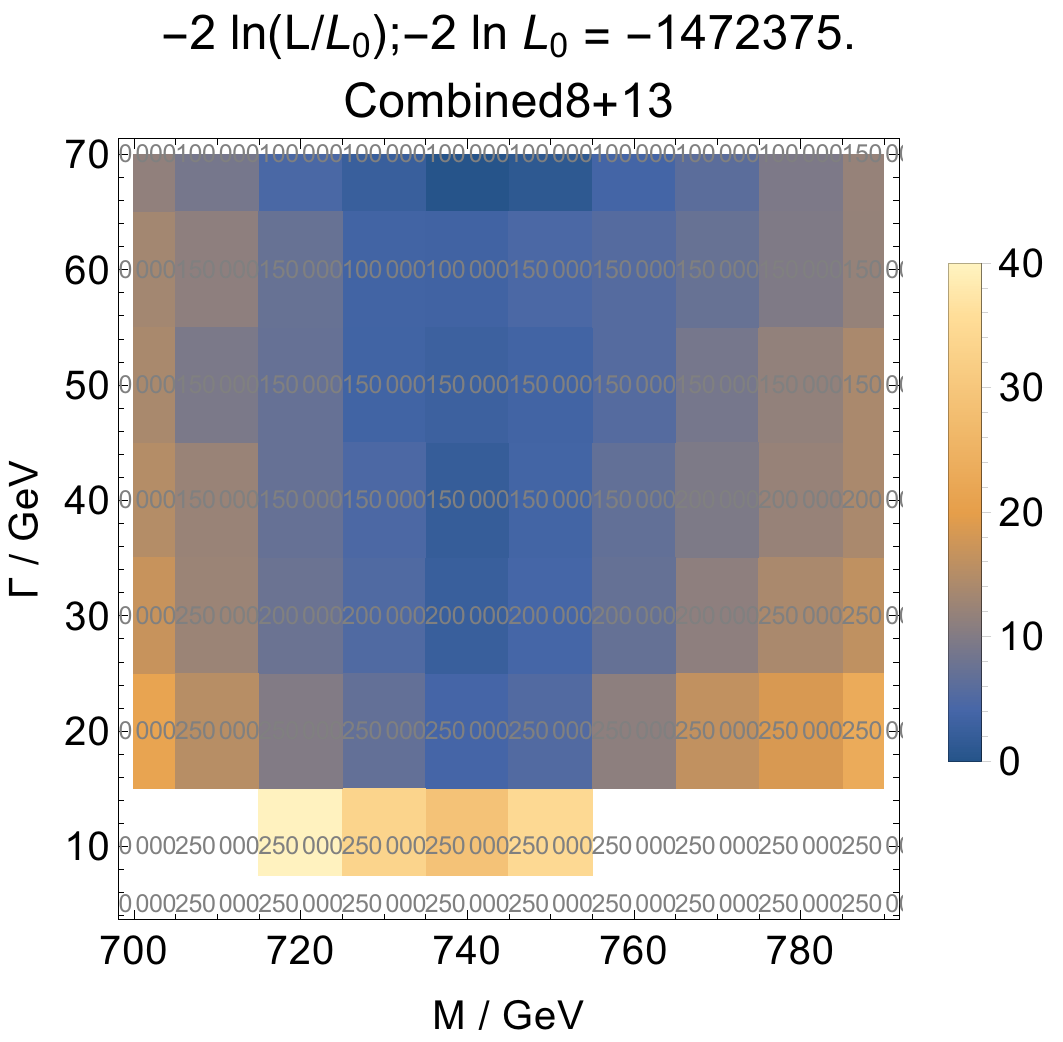}
\includegraphics[width=0.24\textwidth]{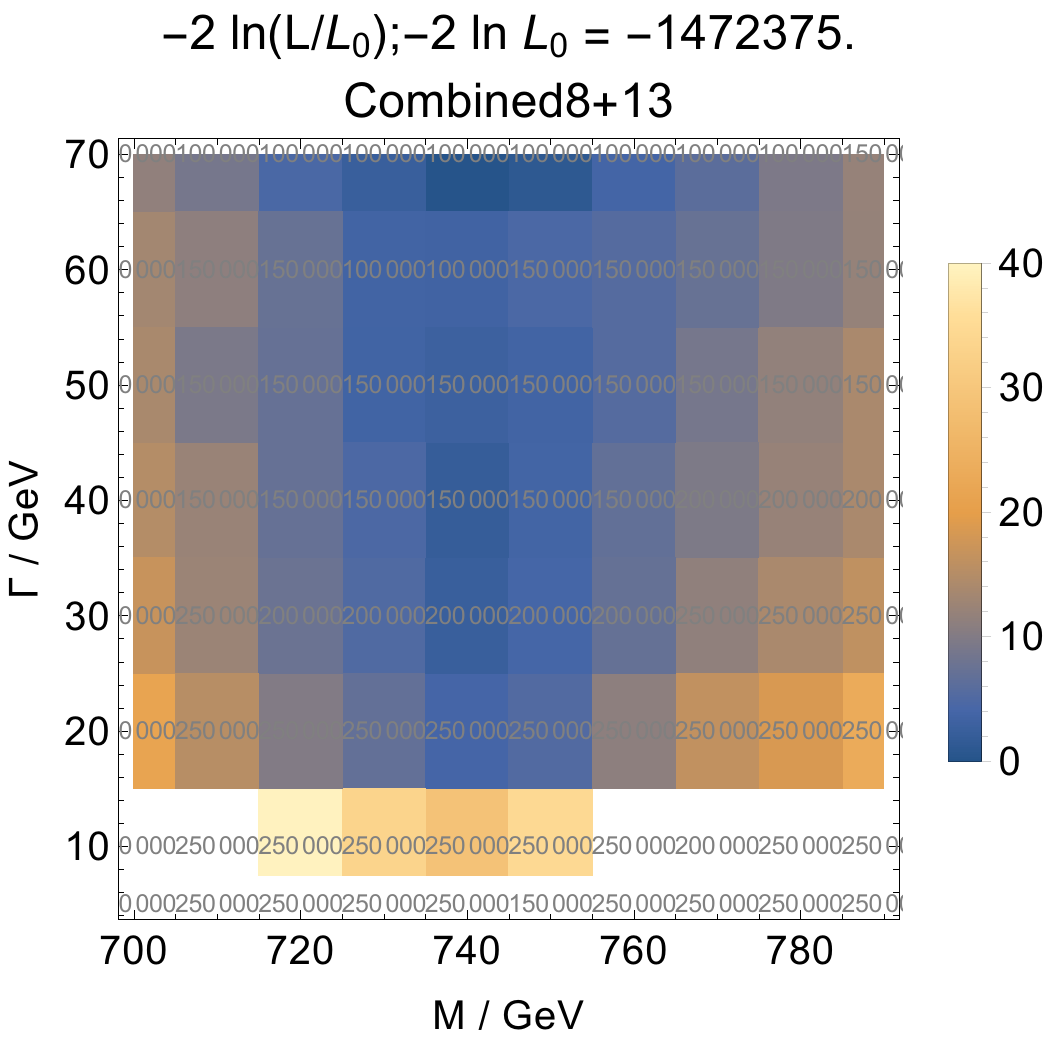}

\includegraphics[width=0.24\textwidth]{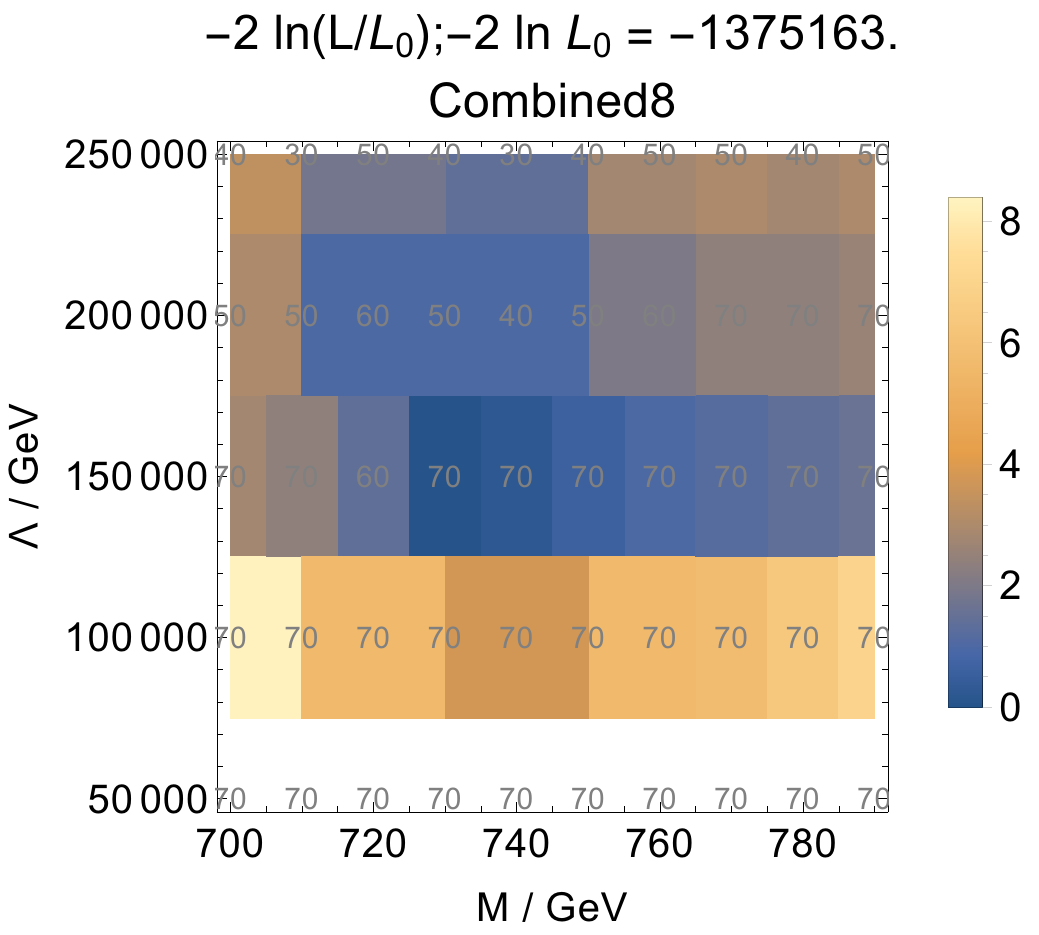}
\includegraphics[width=0.24\textwidth]{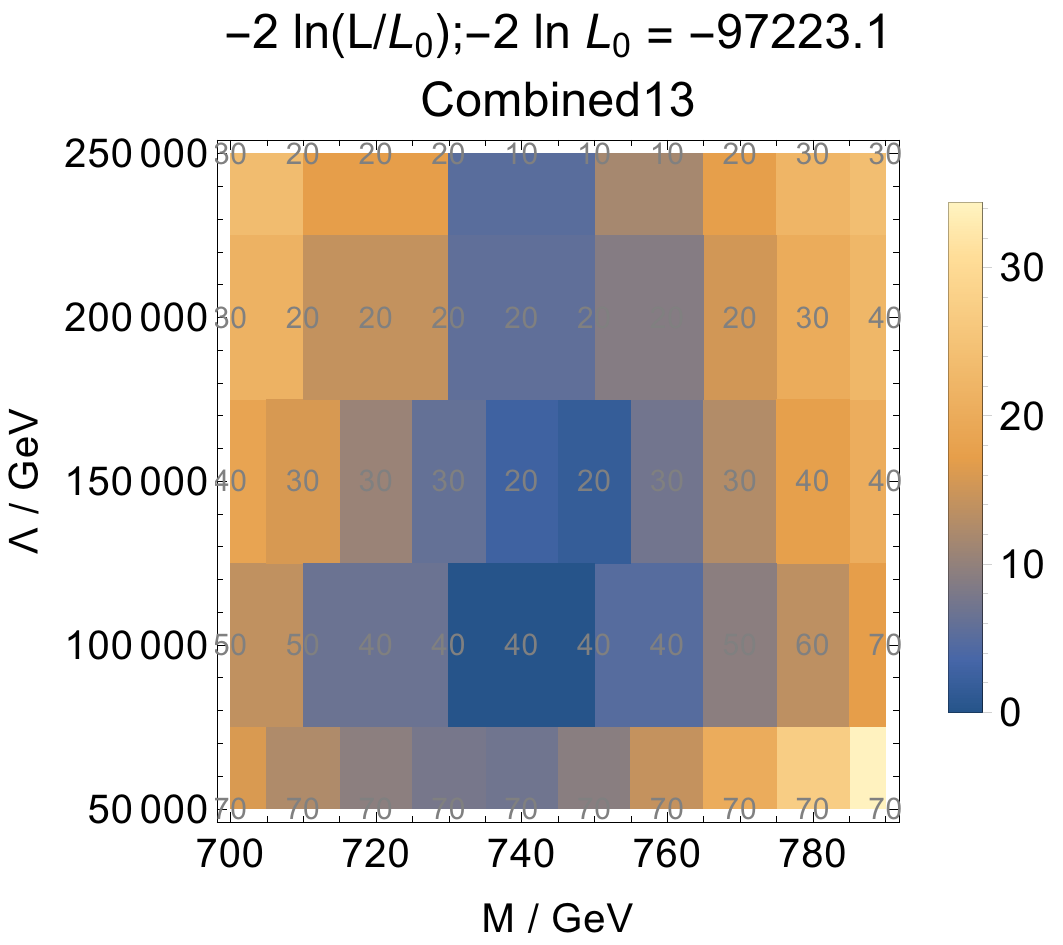}
\includegraphics[width=0.24\textwidth]{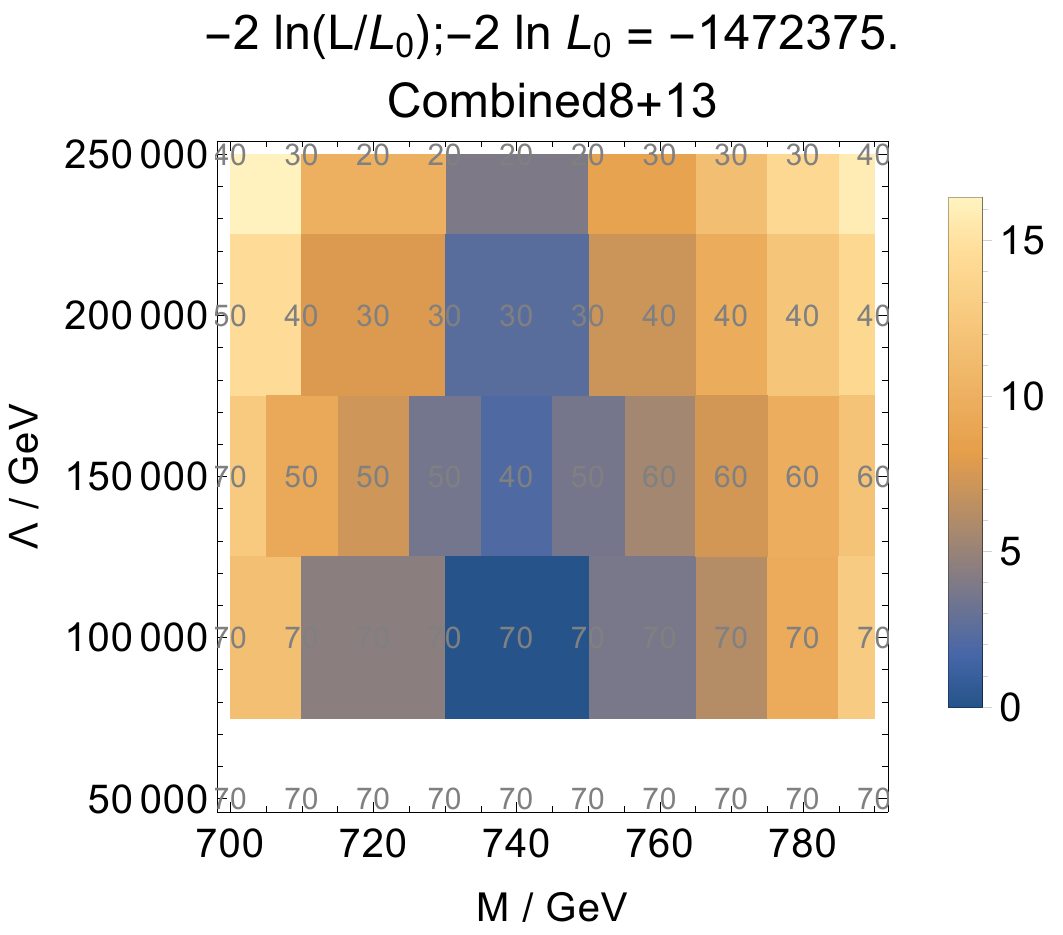}
\includegraphics[width=0.24\textwidth]{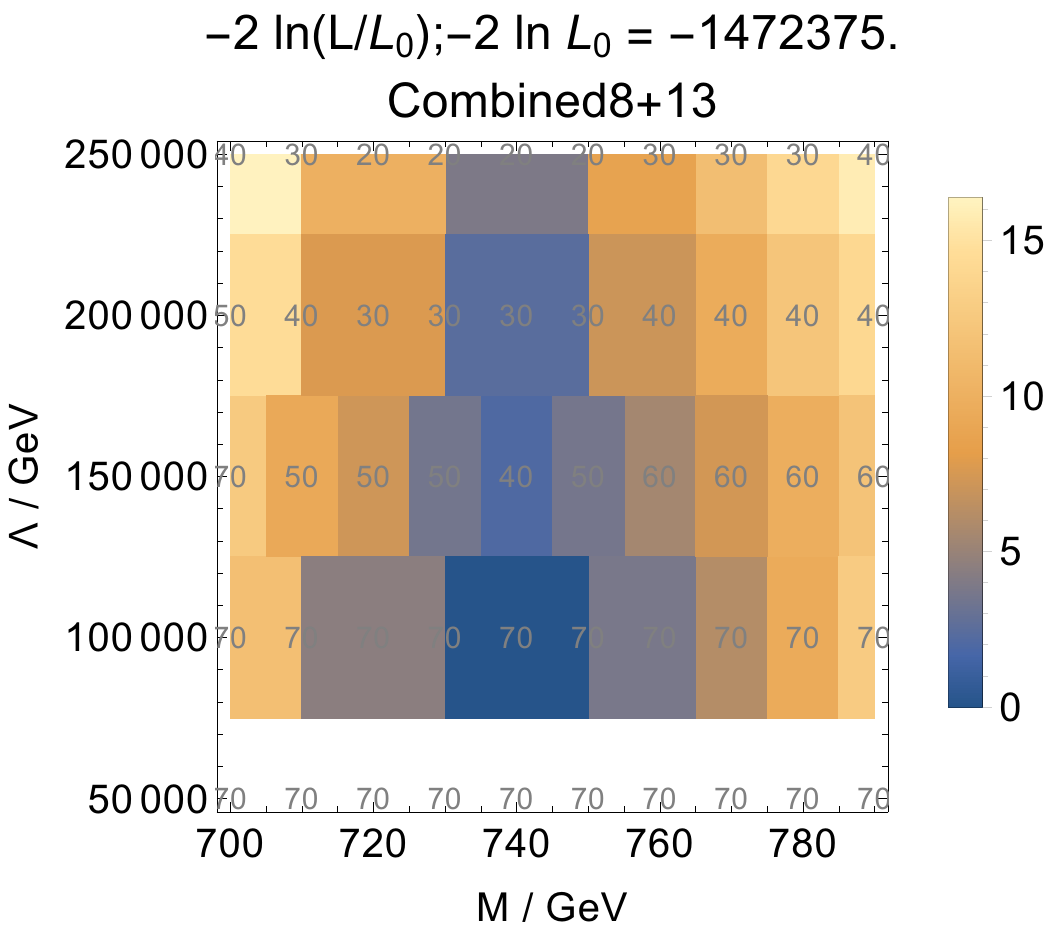}
\caption{Log likelihoods relative to the best fit points for the $qq$-initiated spin-0 scalar signal hypothesis. Top row: Likelihoods as a function of signal mass $M$ and width $\Gamma$, profiling over the signal amplitude parameter $\Lambda$. The profiled value of $\Lambda$ is shown at each grid point. From left to right, the likelihoods are shown for the combined $\sqrt{s} = 8$ TeV data, combined $\sqrt{s} = 13$ TeV data, combined $\sqrt{s} = 8+13$ TeV data, and combined $\sqrt{s} = 8+13$ TeV data {\it neglecting interference effects}. Bottom row: Likelihoods as a function of signal mass $M$ and signal amplitude parameter $\Lambda$, profiling over the signal width $\Gamma$ (with the profiled value of $\Gamma$ shown at each grid point). \label{fig:sqq}}
\end{figure}

\begin{figure}
   \centering
\includegraphics[width=0.24\textwidth]{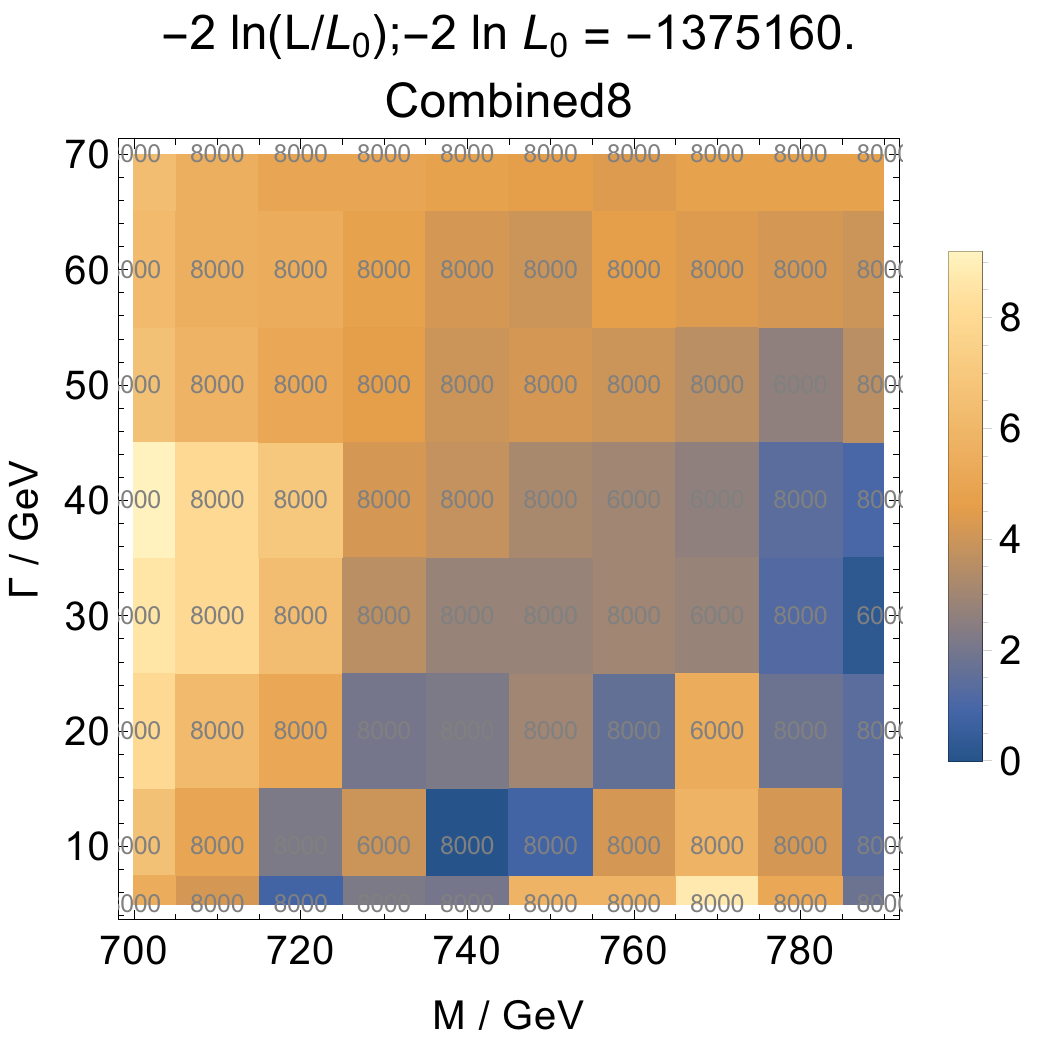}
\includegraphics[width=0.24\textwidth]{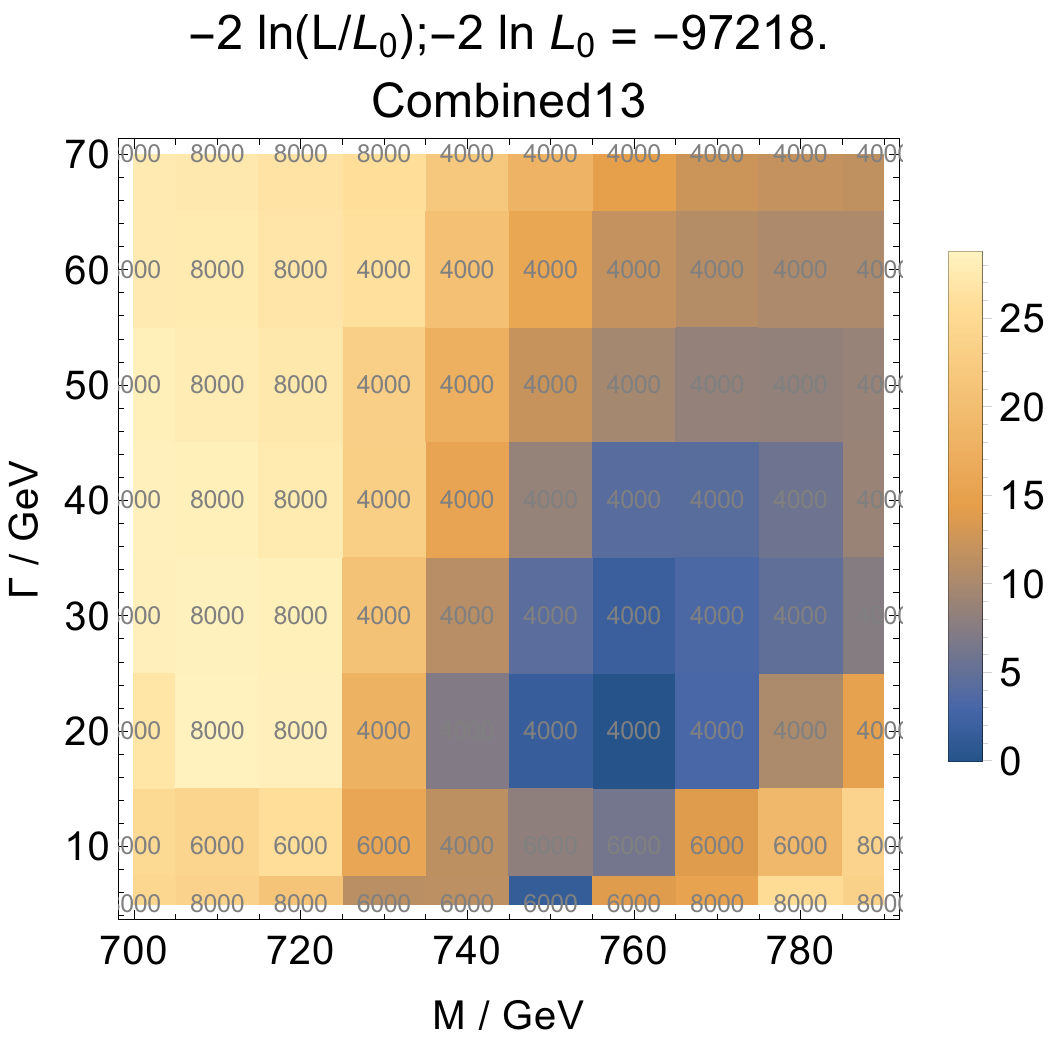}
\includegraphics[width=0.24\textwidth]{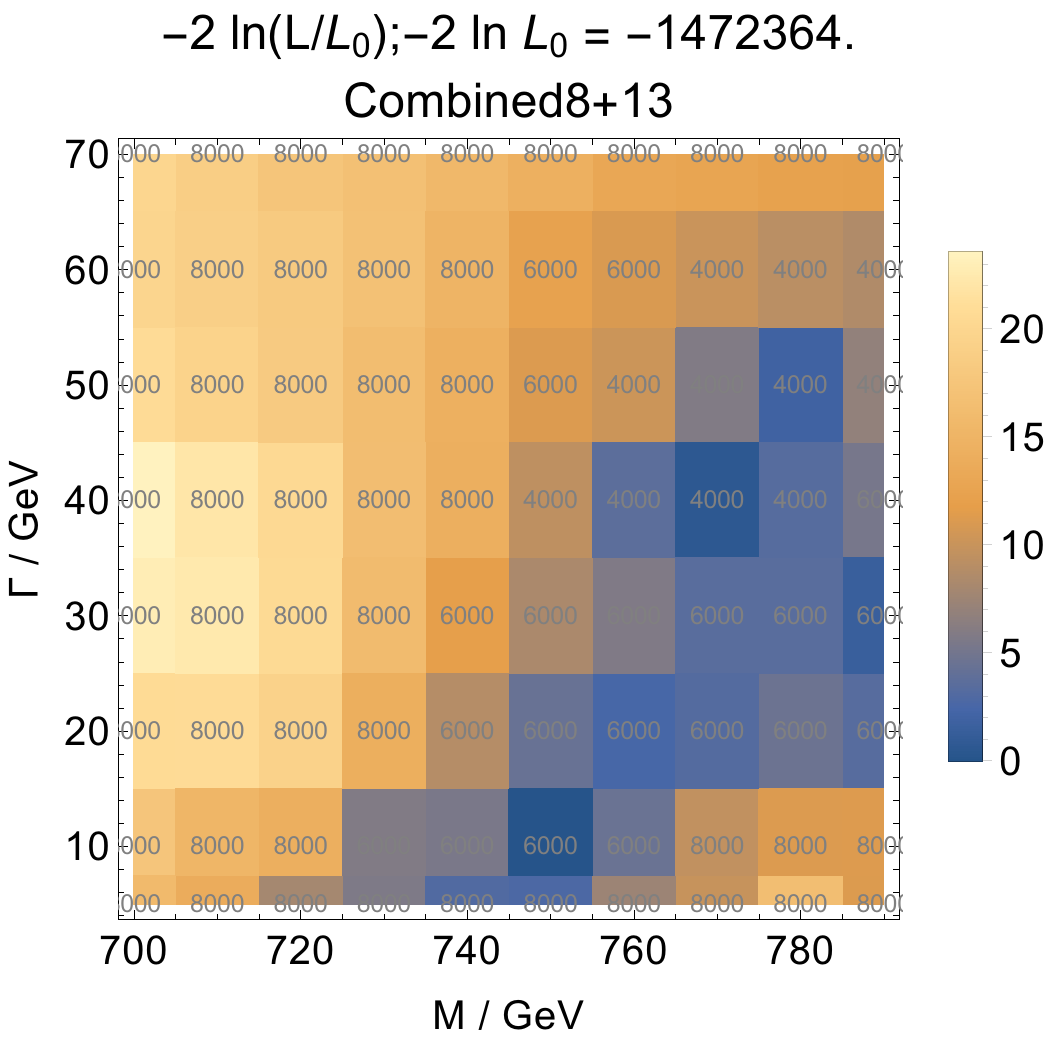}
\includegraphics[width=0.24\textwidth]{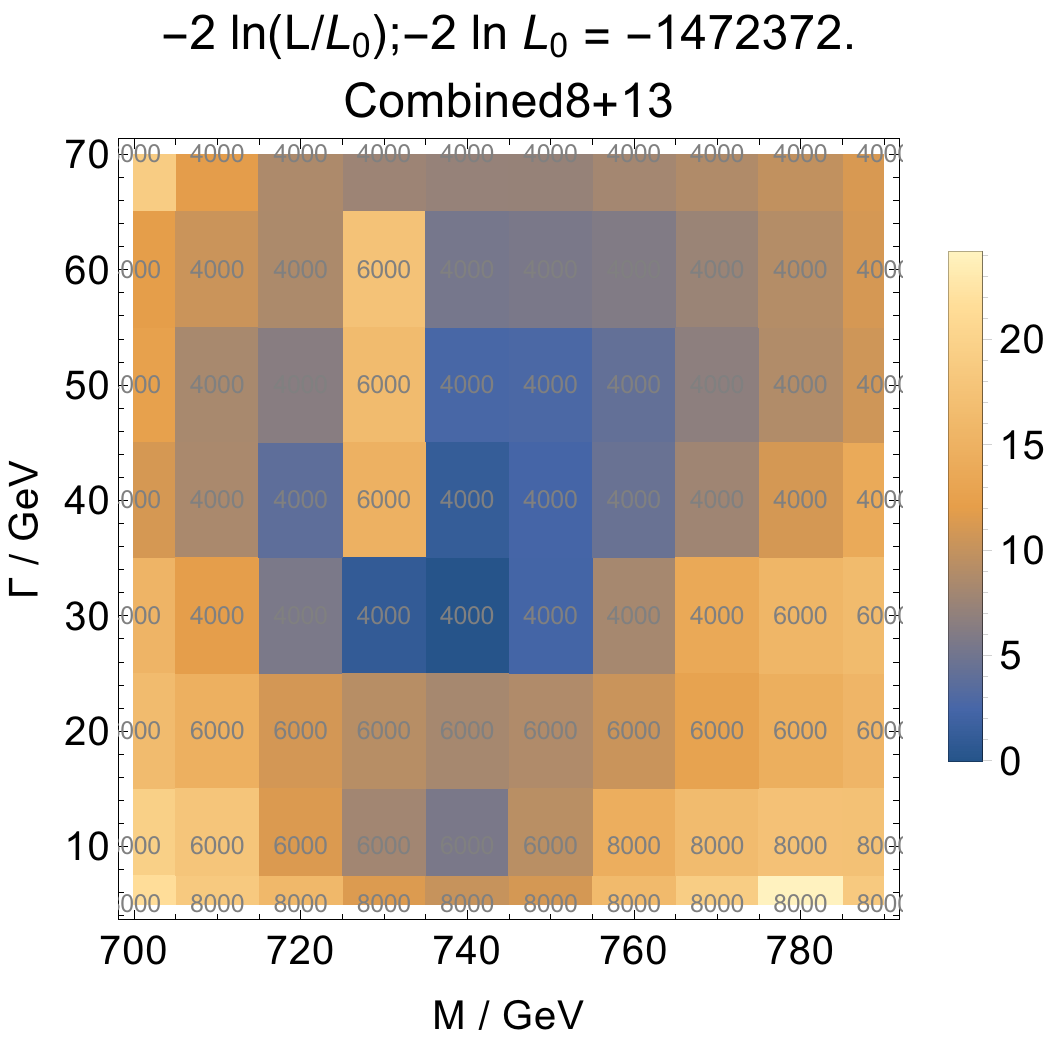}

\includegraphics[width=0.24\textwidth]{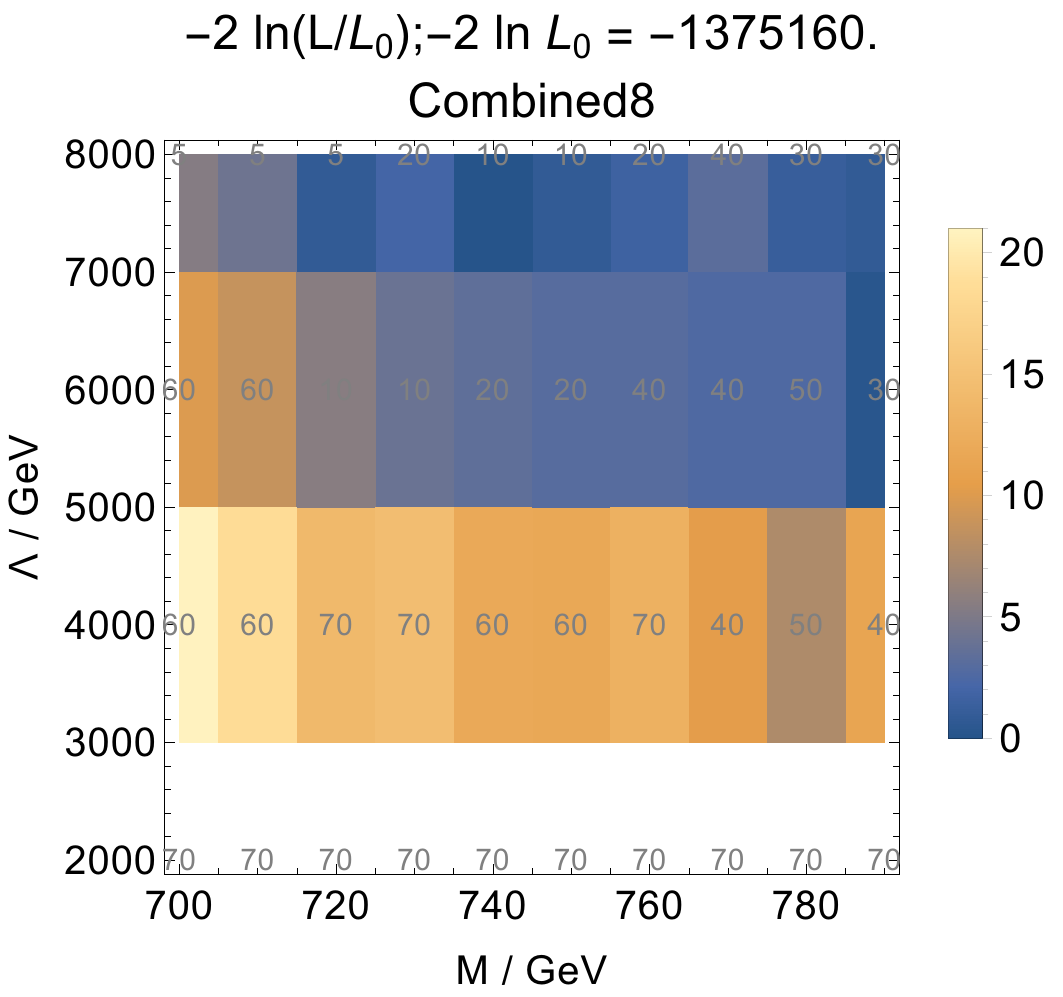}
\includegraphics[width=0.24\textwidth]{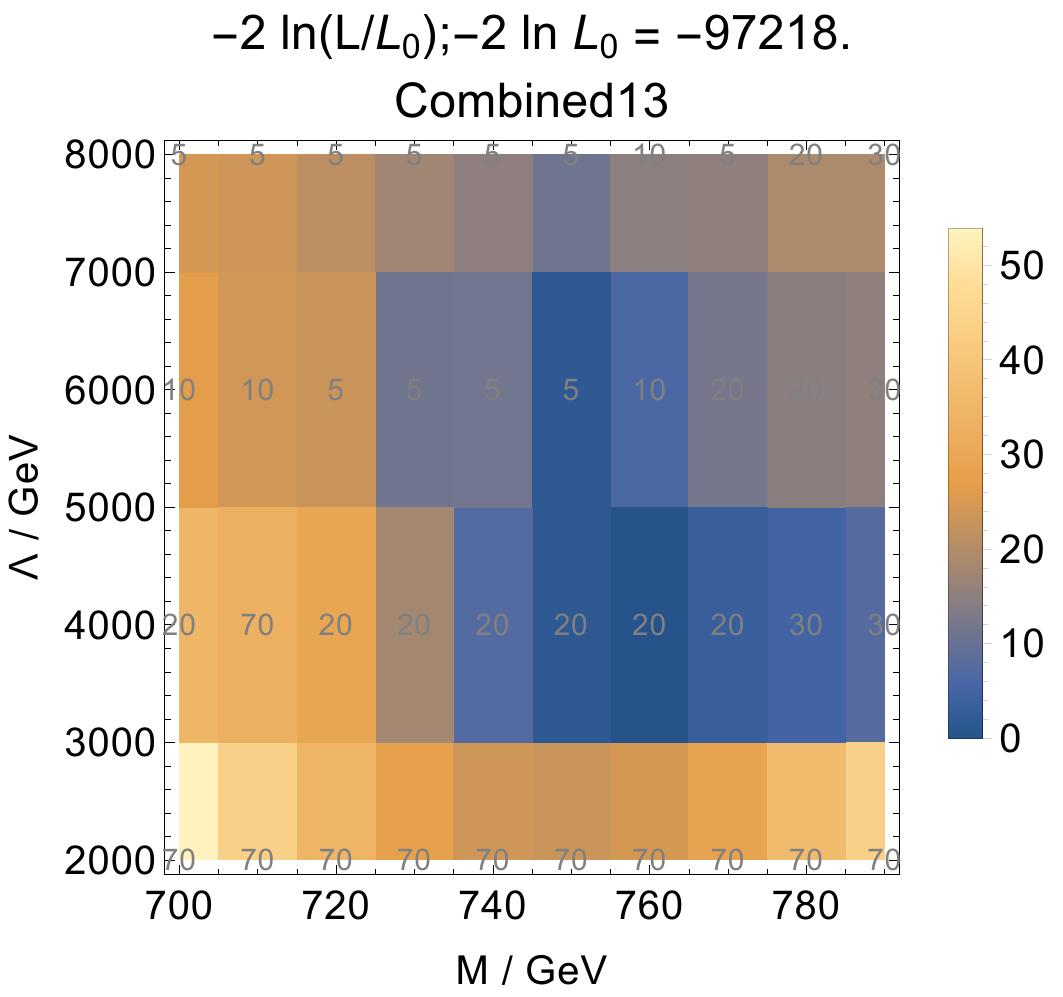}
\includegraphics[width=0.24\textwidth]{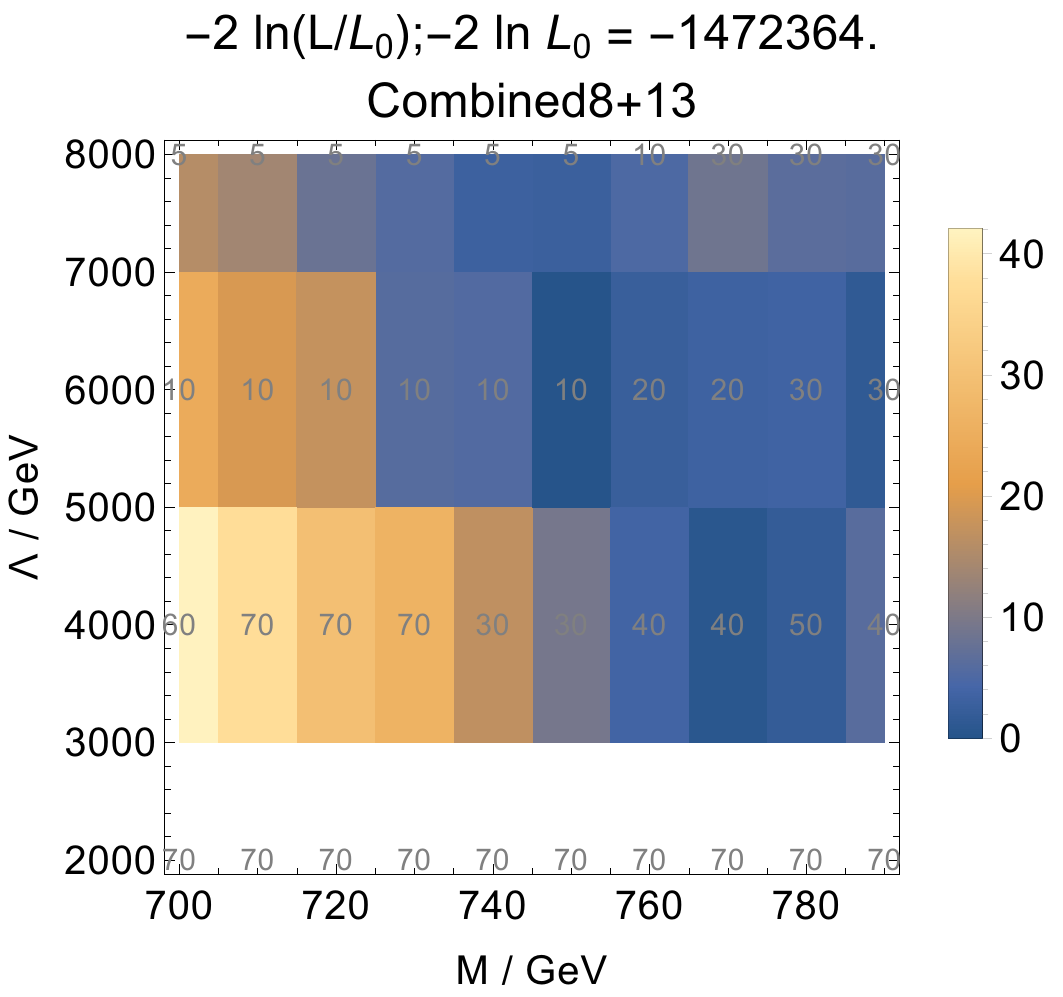}
\includegraphics[width=0.24\textwidth]{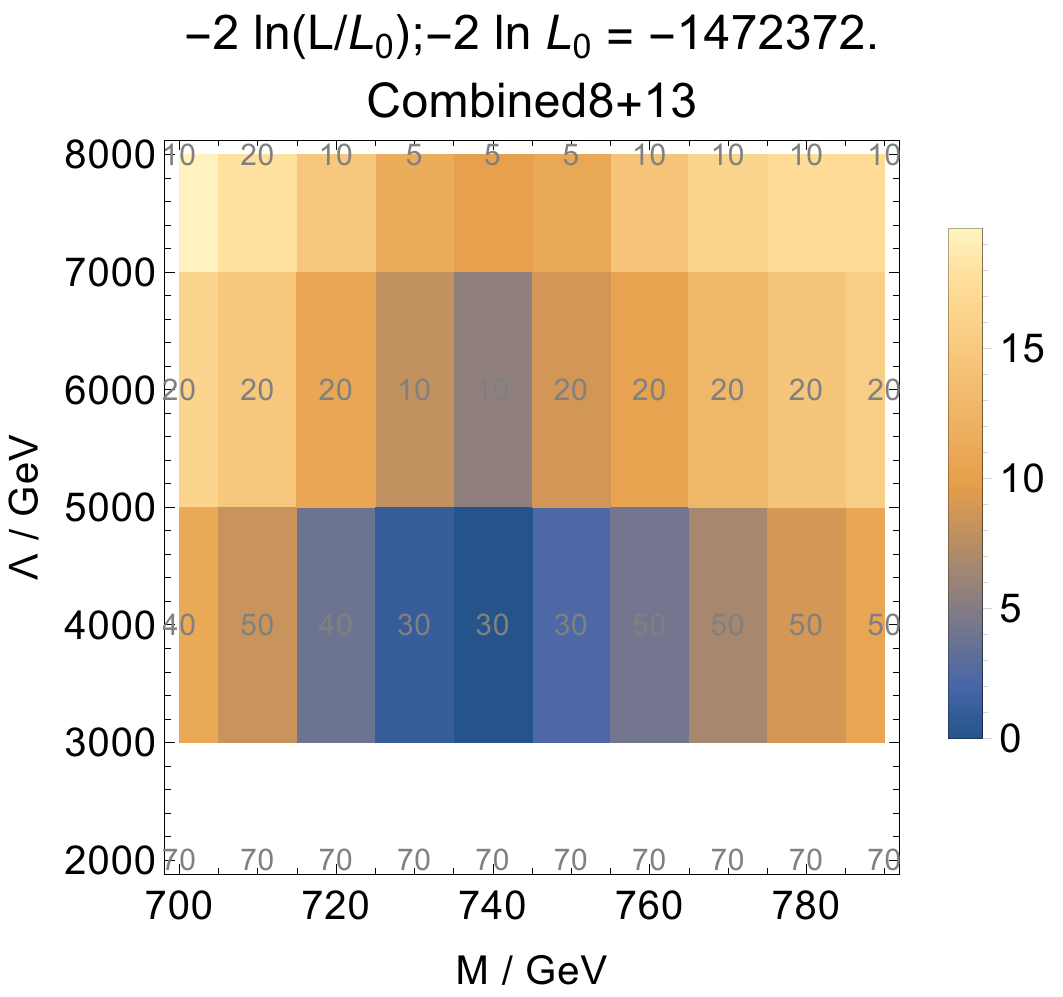}
\caption{Log likelihoods relative to the best fit points for the $qq$-initiated spin-2 signal hypothesis. Top row: Likelihoods as a function of signal mass $M$ and width $\Gamma$, profiling over the signal amplitude parameter $\Lambda$. The profiled value of $\Lambda$ is shown at each grid point. From left to right, the likelihoods are shown for the combined $\sqrt{s} = 8$ TeV data, combined $\sqrt{s} = 13$ TeV data, combined $\sqrt{s} = 8+13$ TeV data, and combined $\sqrt{s} = 8+13$ TeV data {\it neglecting interference effects}. Bottom row: Likelihoods as a function of signal mass $M$ and signal amplitude parameter $\Lambda$, profiling over the signal width $\Gamma$ (with the profiled value of $\Gamma$ shown at each grid point). \label{fig:tqq}}
\end{figure}

\begin{figure}
   \centering
\includegraphics[width=0.45\textwidth]{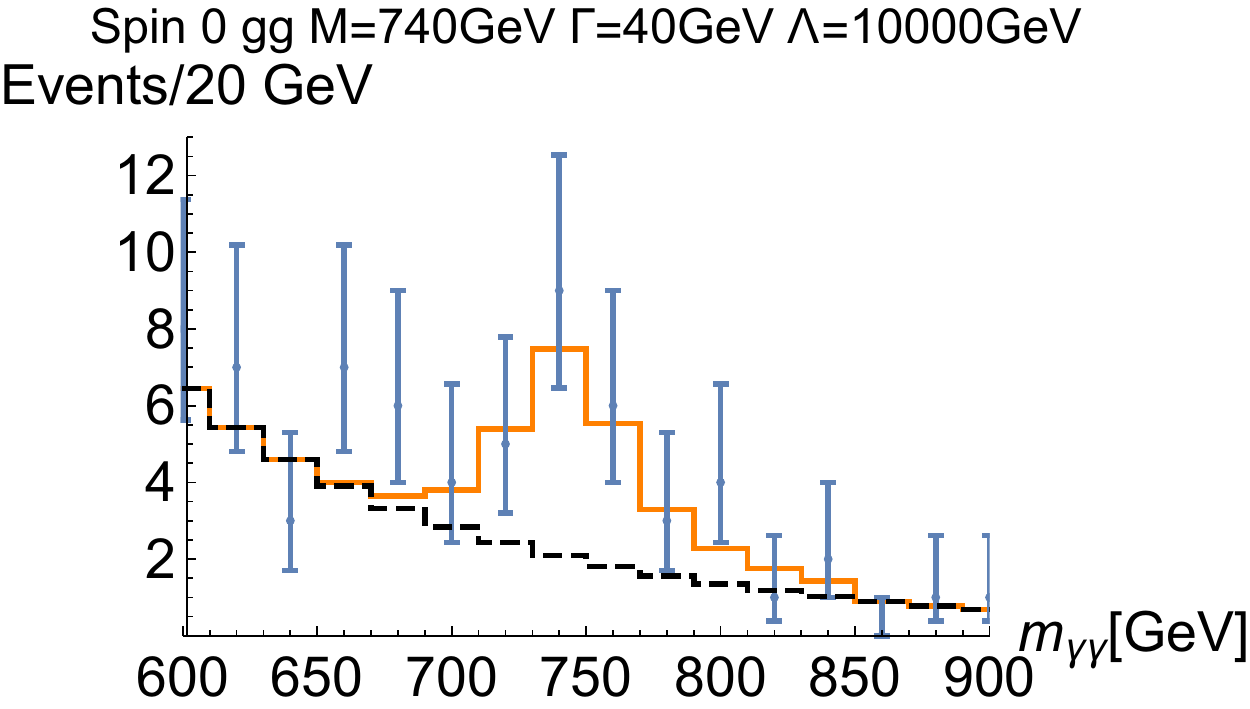}
\includegraphics[width=0.45\textwidth]{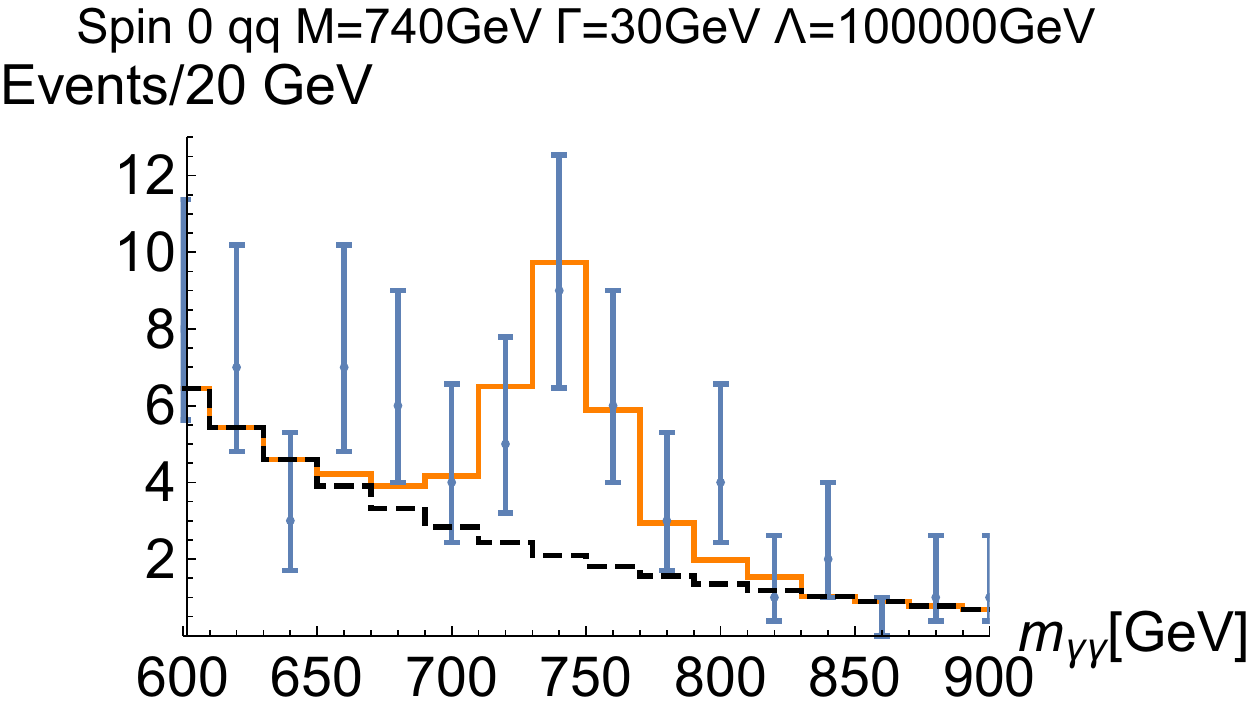}

\includegraphics[width=0.45\textwidth]{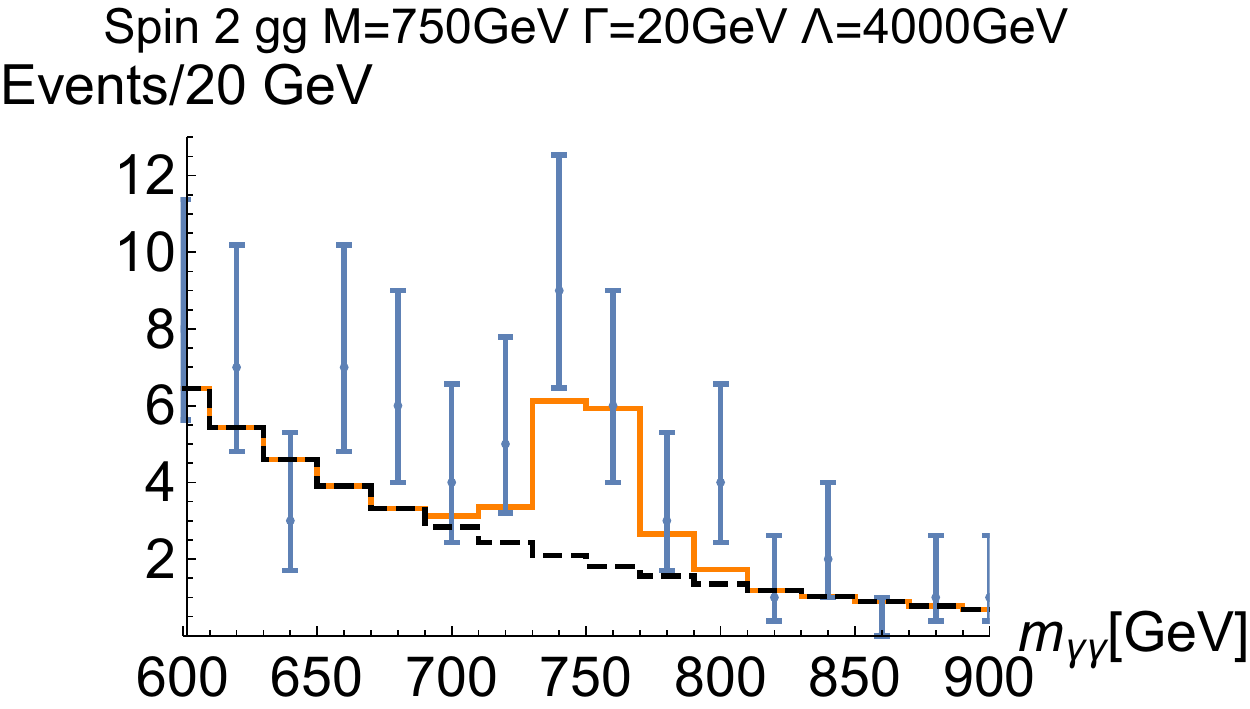}
\includegraphics[width=0.45\textwidth]{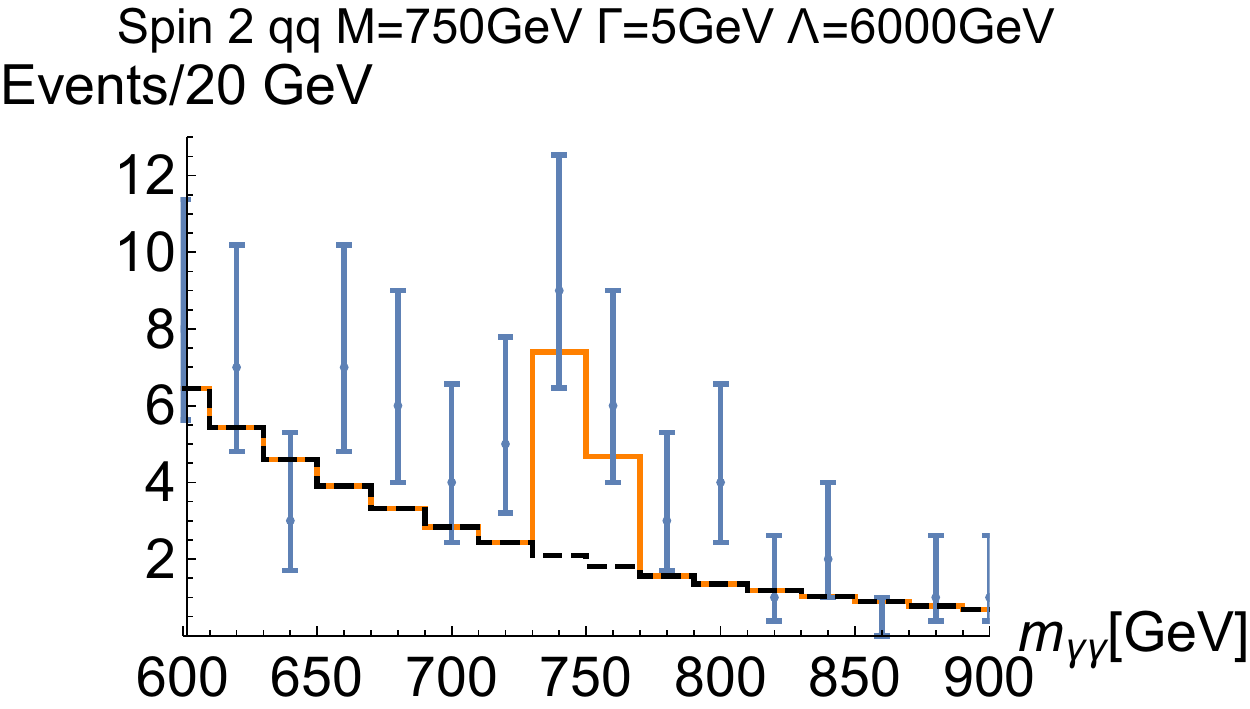}
\caption{The spectra of the points that best fit the ATLAS13Spin0 dataset, overlaid over the ATLAS13Spin0 data (blue points). The error bars on the data points are purely statistical. Top left: $gg$-initiated spin-0 scalar resonance. Top right: $q \bar q$-initiated spin-0 scalar resonance. Bottom left: $gg$-initiated spin-2 resonance. Bottom right: $q \bar q$-initiated spin-2 resonance. \label{fig:scottPlots}}
\end{figure}

We profile over each of the parameters $M$, $\Gamma$ and $\Lambda$ in turn to generate a series of two-dimensional plots of the difference in twice the log likelihood relative to the best fit point. We present the results for the $gg$-initiated scalar signal hypothesis in Fig.~\ref{fig:sgg}, including results using the combined $\sqrt{s} = 8$ TeV data; the combined $\sqrt{s} = 13$ TeV data; the full combination of $\sqrt{s} = 8, 13$ TeV data; and, for the purposes of comparison, the results of the full combination of $\sqrt{s} = 8, 13$ TeV data {\it neglecting interference effects.}  For the sake of simplicity we do not show the results for pseudoscalar signal hypotheses, as they are essentially indistinguishable from the scalar case in the absence of polarization data. The analogous information is shown for the $gg$-initiated spin-2 signal hypothesis in Fig.~\ref{fig:tgg}; for the $q \bar q$-initiated scalar signal hypothesis in Fig.~\ref{fig:sqq}; and for the $q \bar q$-initiated spin-2 signal hypothesis in Fig.~\ref{fig:tqq}. Finally, for the sake of illustration we show the binned spectra of the point, for each signal model, that best fits the ATLAS13Spin0 dataset in Fig.~\ref{fig:scottPlots}, overlaid with the data. 

As is apparent in Figs.~\ref{fig:sgg}-\ref{fig:tqq}, the excesses at $\sqrt{s} = 8$ TeV are modest and provide only mild preference for nonzero signal, with signal signficiance deriving primarily from $\sqrt{s} = 13$ TeV data. In general, the 8 TeV data prefers somewhat lower resonance masses compared to the 13 TeV data. Note that in both cases, the use of binned spectra lead us to slightly different best-fit values for the signal mass and width compared to those obtained by the ATLAS and CMS collaborations using unbinned data.

In the case of $gg$-initiated spin-0 signals, interference effects lead to a small dip preceding the signal peak (and therefore shift the apparent peak to higher values of $m_{\gamma \gamma}$). As is evident in Fig.~\ref{fig:sgg}, the inclusion of interference effects leads to a preference for slightly {\it lower} signal mass $M$ at large widths when compared to the fit neglecting interference effects. For the $gg$-initiated spin-2 signal, interference effects do not lead to a significant shift in the apparent peak position, so unsurprisingly the combined best-fit mass for a spin-2 signal is slightly {\it higher} than the spin-0 case, and comparable to the combined best-fit mass obtained by neglecting interference effects.

In the case of $q \bar q$-initiated spin-0 signals there is no interference between resonance and continuum, leading to a weak preference for finite width and combined best-fit mass comparable to the $gg$-initiated spin-2 scenario. In contrast, for $q \bar q$-initiated spin-2 signals the substantial resonance-continuum interference shifts the apparent signal peak to far lower values of $\mgamgam$ relative to the resonance mass $M$. As the width is increased for fixed $M$, the apparent peak migrates to lower values of $\mgamgam$, leading to the diagonal features in the likelihood apparent in Fig.~\ref{fig:tqq}. As a result, this means that the combined best-fit mass shifts considerably as a function of the width, ranging from $M = 750$ GeV for $\Gamma = 10$ GeV to $M = 770$ GeV for $\Gamma = 40$ GeV. In addition to shifting the position of the apparent resonance peak, resonance-continuum interference leads to a subsequent deficit. Ultimately, this leads to a preference for small width in the combined fit for a $q \bar q$-initiated spin-2 signal due to the lack of apparent deficits in the diphoton spectrum.

Taken together, the inclusion of interference effects in the interpretation of excesses near $M = 750$ GeV has a modest impact on the best-fit parameters for $gg$-initiated spin-0 signals (preferring a slightly lower mass $M$ due to the interference-induced shift of the $\mgamgam$ peak to higher values); little or no impact on $gg$-initiated spin-2 signals and $q \bar q$-initiated spin-0 signals; and a substantial impact on $q \bar q$-initiated spin-2 signals due to the distinctive peak-dip interference structure (preferring a narrow width due to the lack of nearby deficits, and tightly correlating the best-fit width and mass).

\subsection{2016 data \label{sec:2016}}

While excesses in the pre-2016 data set serve to illustrate the impact of resonance-continuum interference on signal interpretations, these particular excesses proved to be statistical fluctuations in light of 2016 data~\cite{CMS-PAS-EXO-16-027,ATLAS-CONF-2016-059}. Combined with previous data, both ATLAS and CMS measurements of the diphoton spectrum using 2016 data sizably reduce the overall significance of excesses around 750 GeV. For the sake of completeness, we repeat the above analysis including the 2016 data in our fit to the different lineshapes of the ($gg$- and $\bar{q}q$-initiated) S, PS and T models. 

We use the same method as \S\ref{sec:data}, and present the important properties of the three 2016 signal regions in Table~\ref{tab:data2016}. As before, we compare the fitted curves to our prediction for the background only component of the spectrum, computing $\mathcal{F} \equiv f^{\rm fit}(\mgamgam)/f_{\rm cont}^{\rm theory}(m_{\gamma \gamma})$. We find $\mathcal{F} \approx 1.25$ for all three 2016 datasets, and use this value to normalise our signal predictions accordingly.

\begin{table}
\begin{adjustbox}{width=\textwidth}
\begin{tabular}{ c | c | c | c | c | c | c}
Ref. & Dataset & Fit curve & Acceptance & $\sigma_\text{res}/\mgamgam$ & $L_\text{int} / \mathrm{fb}^{-1}$ & $C$ \\ \hline
\cite{ATLAS-CONF-2016-059} &
ATLAS2016SPIN0 &$f_\text{ATL}^{\rm fit}(;0.57,11.4,-2.88;)$ &\pbox{4cm}{$\abs{\eta_{1/2}} \in [0,2.37]$ \\ $E_{T,1} > 0.4 \mgamgam$ \\ $E_{T,2} > 0.3 \mgamgam$} &
$0.01$ & $12.2$ & $0.75$ \\ \hline
 
\multirow{2}{*}{\cite{CMS-PAS-EXO-16-027} } &
CMS2016EBEB & $f_\text{CMS}^{\rm fit}(;42.4,4.77,-0.76)$ &\pbox{4cm}{$\abs{\eta_{1/2}} \in [0,1.44]$ \\ $p_{T,1/2} > 75 \GeV$ \\ $\mgamgam > 230 \GeV$} & 0.01 & \multirow{2}{*}{ $12.9$ } & 0.81 \\ \cline{2-5} \cline{7-7}

 &
CMS2016EBEE & $f_\text{ATL}^{\rm fit}(;25434,18.7,-0.68)$ &\pbox{4cm}{$\abs{\eta_{1}} \in [0,1.44]$ \\ $\abs{\eta_{2}} \in [1.57,2.5]$\\ $p_{T,1/2} > 75 \GeV$ \\ $\mgamgam > 320 \GeV$} & 0.015 & & 0.72
 \\ 
\end{tabular}
\end{adjustbox}
\caption{ATLAS and CMS diphoton spectrum measurements based on 2016 data, including the best-fit values for the background curves $f_{\rm ATL}^{\rm fit}$ and $f_{\rm CMS}^{\rm fit}$; the geometric acceptance; the diphoton invariant mass resolution $\sigma_\text{res}/\mgamgam$; the integrated luminosity $L_{\rm int}$ in fb$^{-1}$; and the efficiency factor $C$ for each data set. In the ``Fit curve'' column, the entries are of the form $f_{\rm ATL}^{\rm fit}(;N,b,a_0;)$ and $f_{\rm CMS}^{\rm fit}(;N,a,b)$, corresponding to the parameters appearing in (\ref{eq:ATLASfunc}) and (\ref{eq:CMSfunc}), respectively. In the ``Acceptance'' column, the subscripts `1' and `2' respectively refer to the leading and subleading photon in $p_T$.   \label{tab:data2016}}
\end{table}

\begin{figure}
   \centering
\includegraphics[width=0.24\textwidth]{Fit_spin0gg_masswidth_Combined8+13.pdf}
\includegraphics[width=0.24\textwidth]{Fit_spin0qq_masswidth_Combined8+13.pdf}
\includegraphics[width=0.24\textwidth]{Fit_spin2gg_masswidth_Combined8+13.pdf}
\includegraphics[width=0.24\textwidth]{Fit_spin2qq_masswidth_Combined8+13.pdf}

\includegraphics[width=0.24\textwidth]{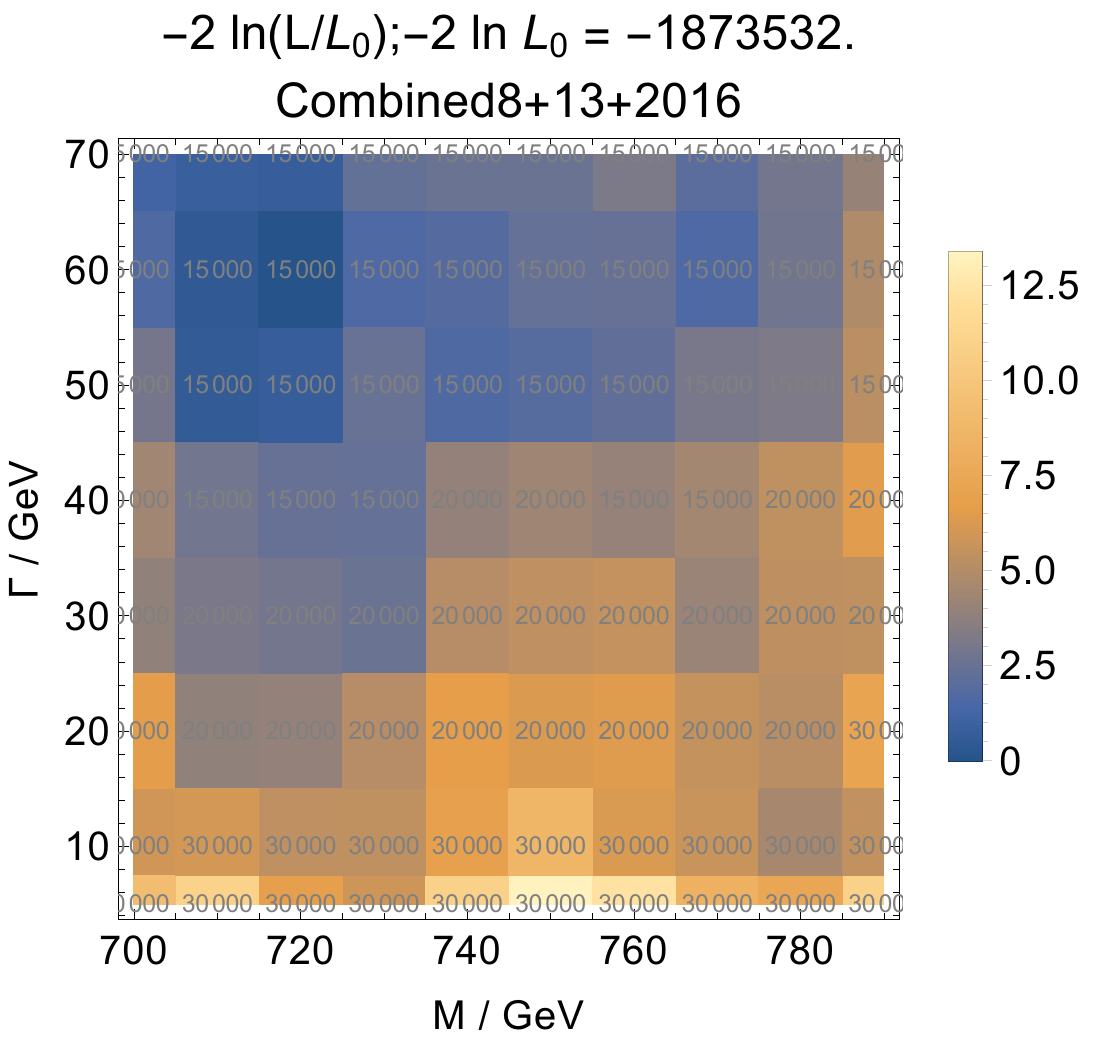}
\includegraphics[width=0.24\textwidth]{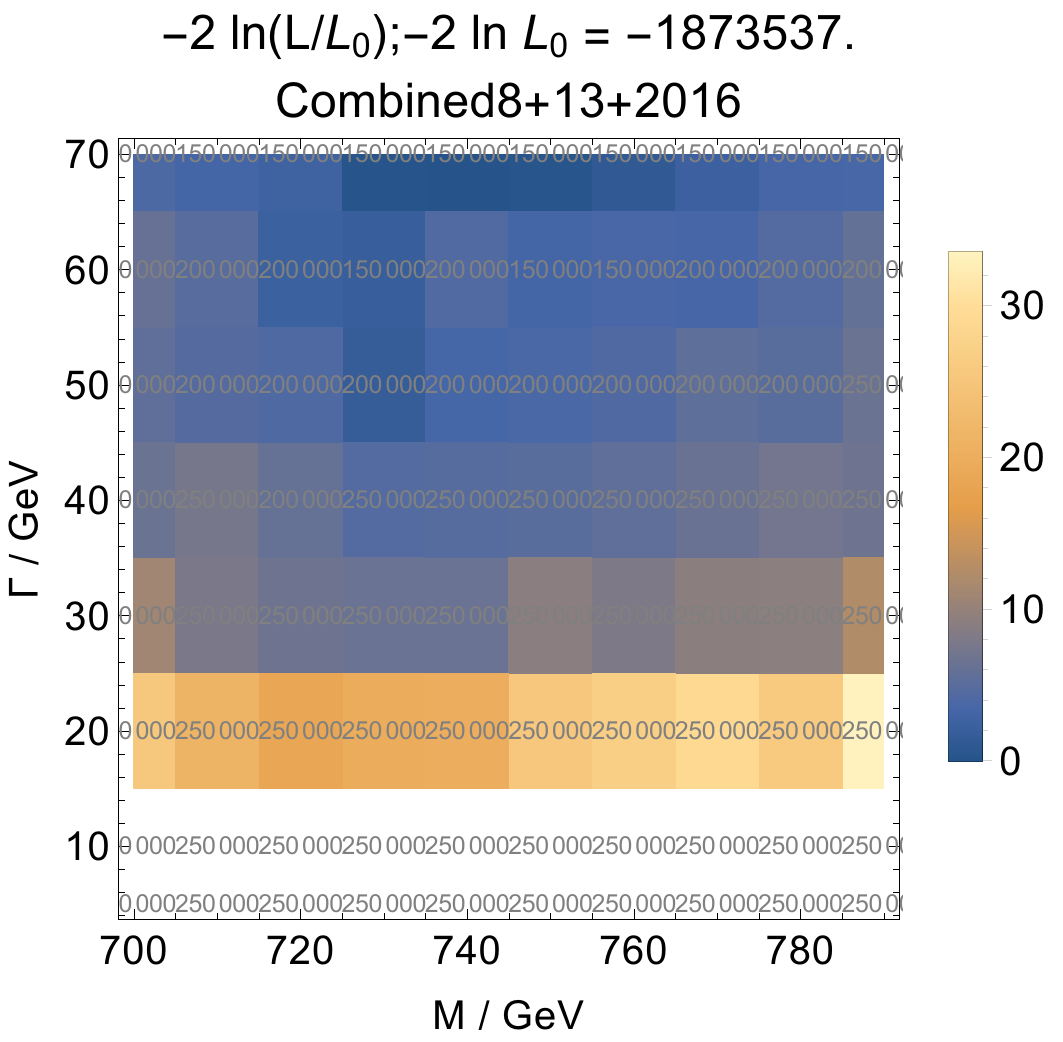}
\includegraphics[width=0.24\textwidth]{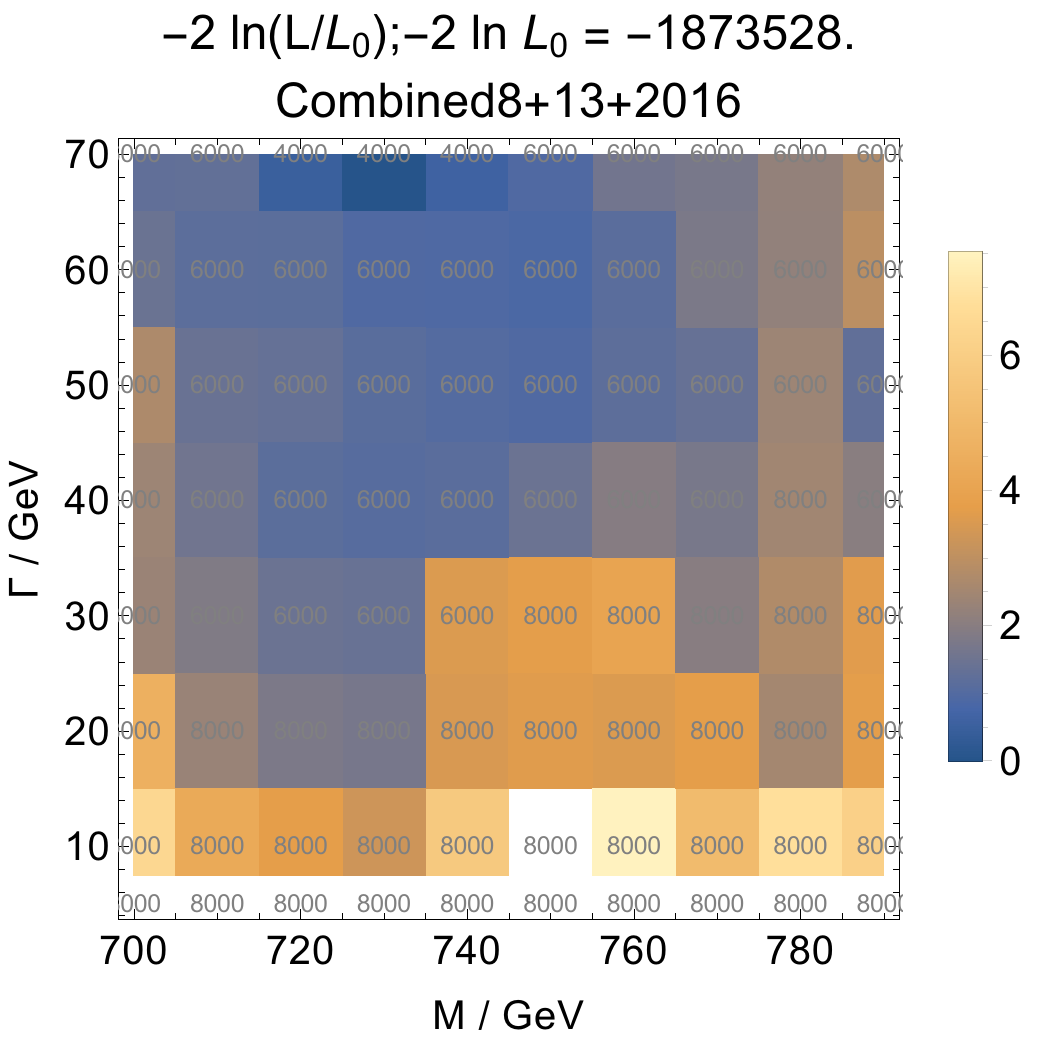}
\includegraphics[width=0.24\textwidth]{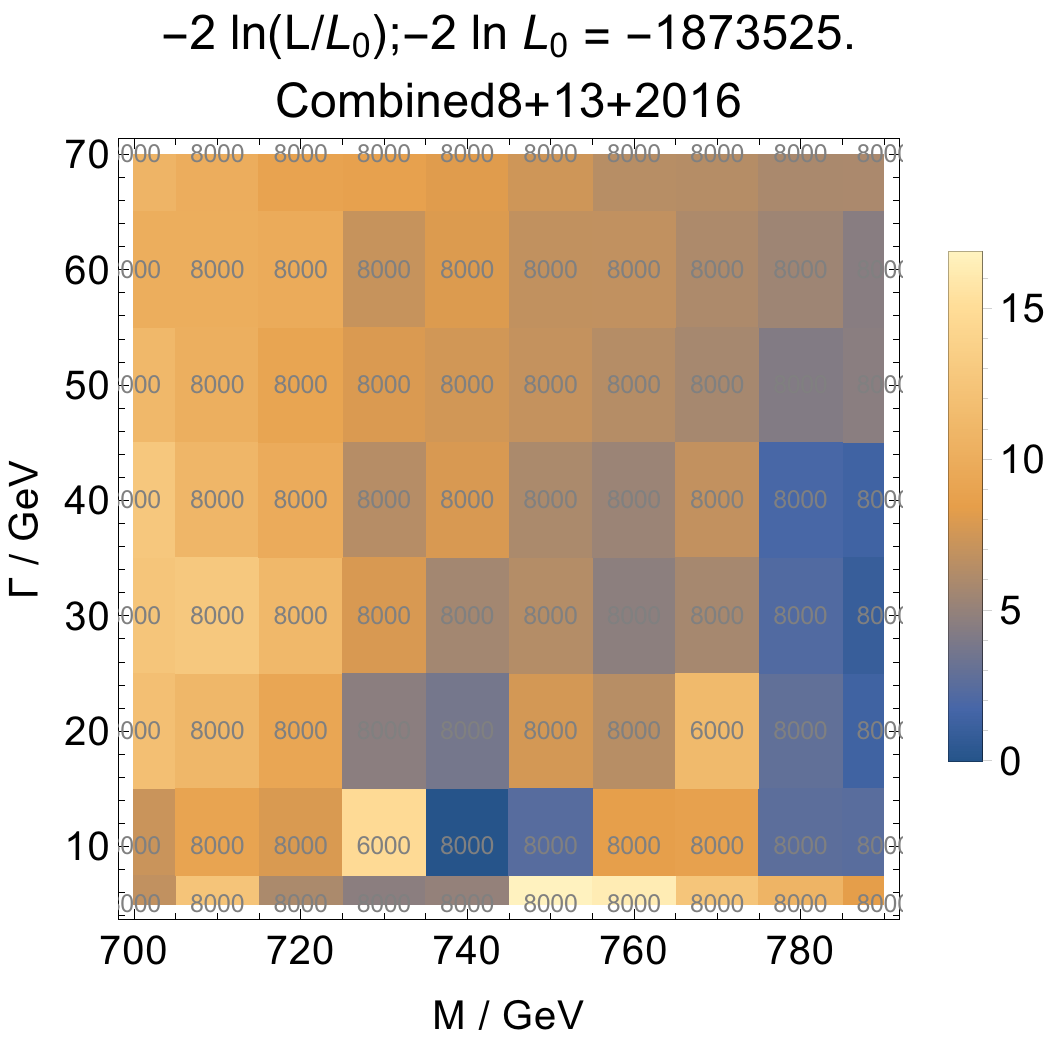}
\caption{From left to right: log likelihoods relative to the best fit points for the $gg$-initiated spin-0, $q\bar{q}$-initiated spin-0, $gg$-initiated spin 2 and $q\bar{q}$-initiated spin-2 signal hypotheses. Top row: Likelihoods based on pre-2016 data (at $\sqrt{s}=8+13$ TeV) as a function of signal mass $M$ and width $\Gamma$, profiling over the signal amplitude parameter $\Lambda$. The profiled value of $\Lambda$ is shown at each grid point. Bottom row: the effect of including the 2016 data on the respective signal hypotheses.\label{fig:2016fitmasswidth}}
\end{figure}

\begin{figure}
   \centering
\includegraphics[width=0.24\textwidth]{Fit_spin0gg_massamp_Combined8+13.pdf}
\includegraphics[width=0.24\textwidth]{Fit_spin0qq_massamp_Combined8+13.pdf}
\includegraphics[width=0.24\textwidth]{Fit_spin2gg_massamp_Combined8+13.pdf}
\includegraphics[width=0.24\textwidth]{Fit_spin2qq_massamp_Combined8+13.pdf}

\includegraphics[width=0.24\textwidth]{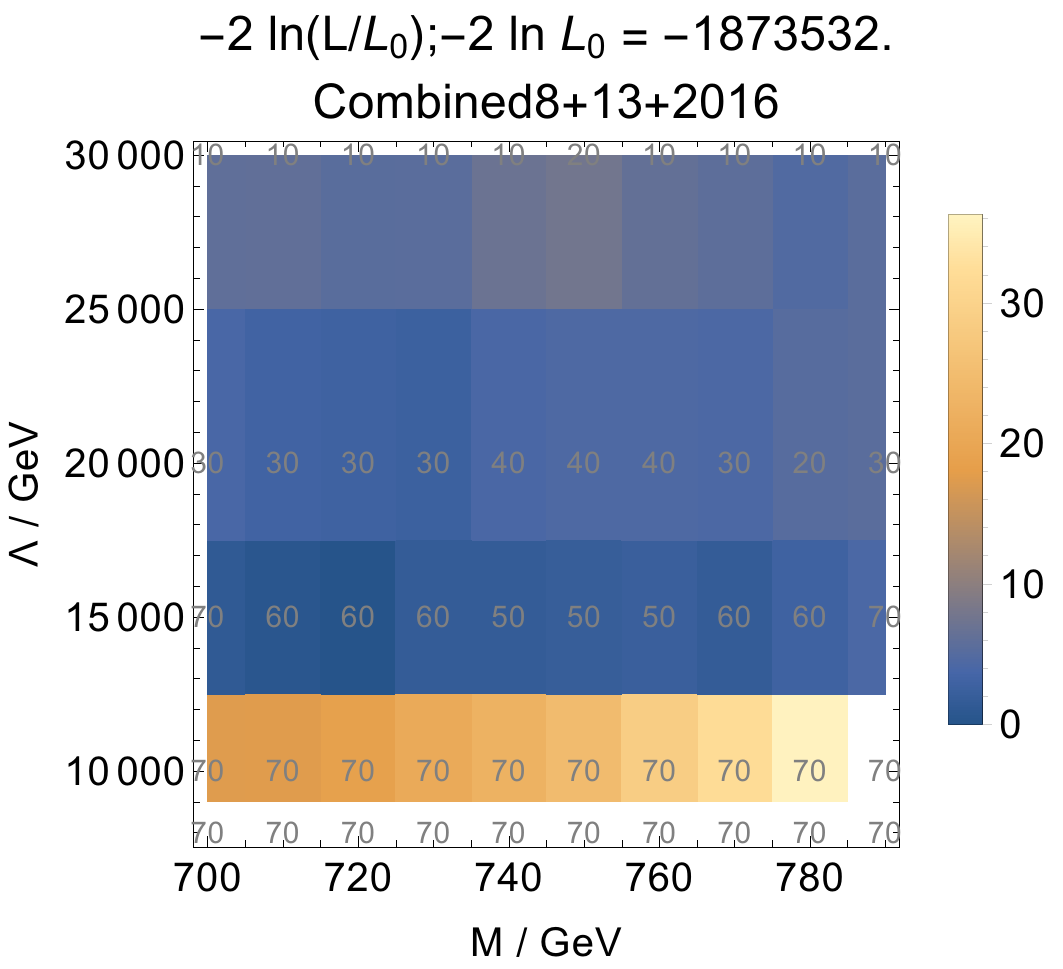}
\includegraphics[width=0.24\textwidth]{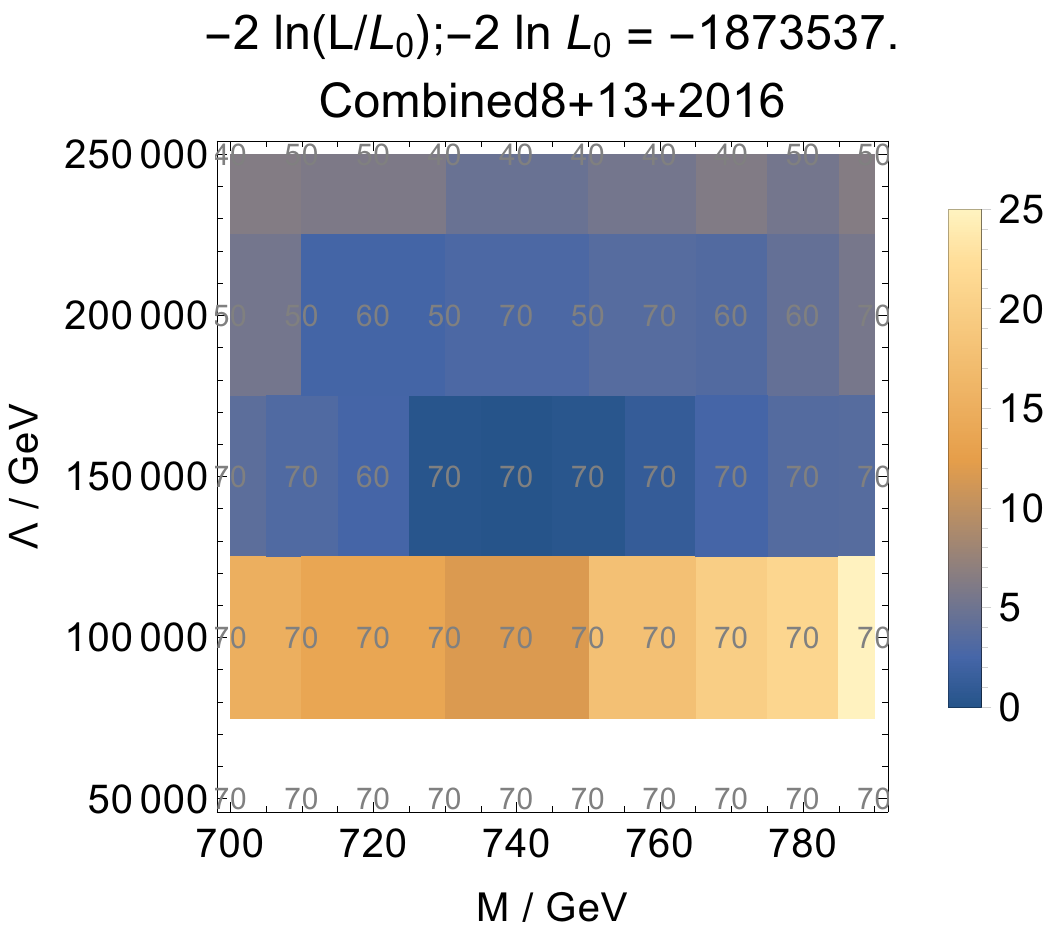}
\includegraphics[width=0.24\textwidth]{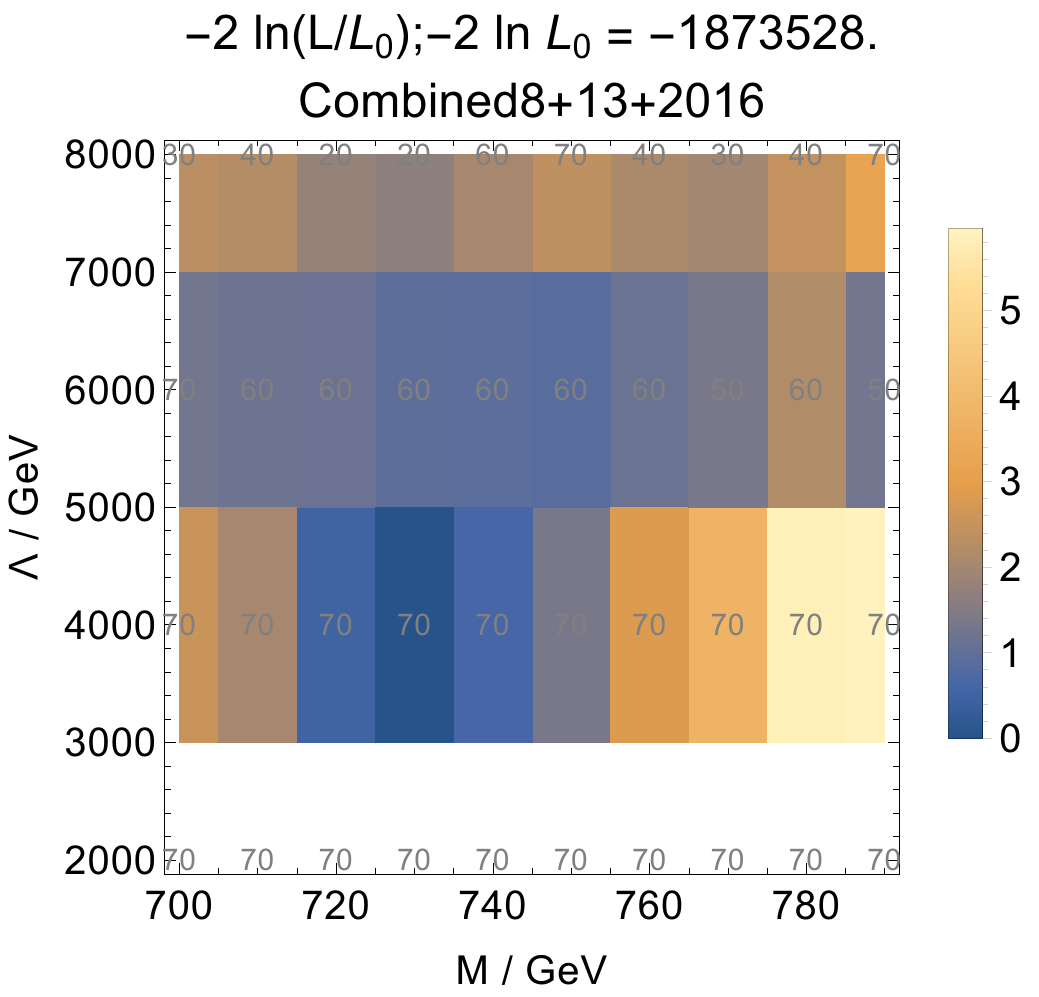}
\includegraphics[width=0.24\textwidth]{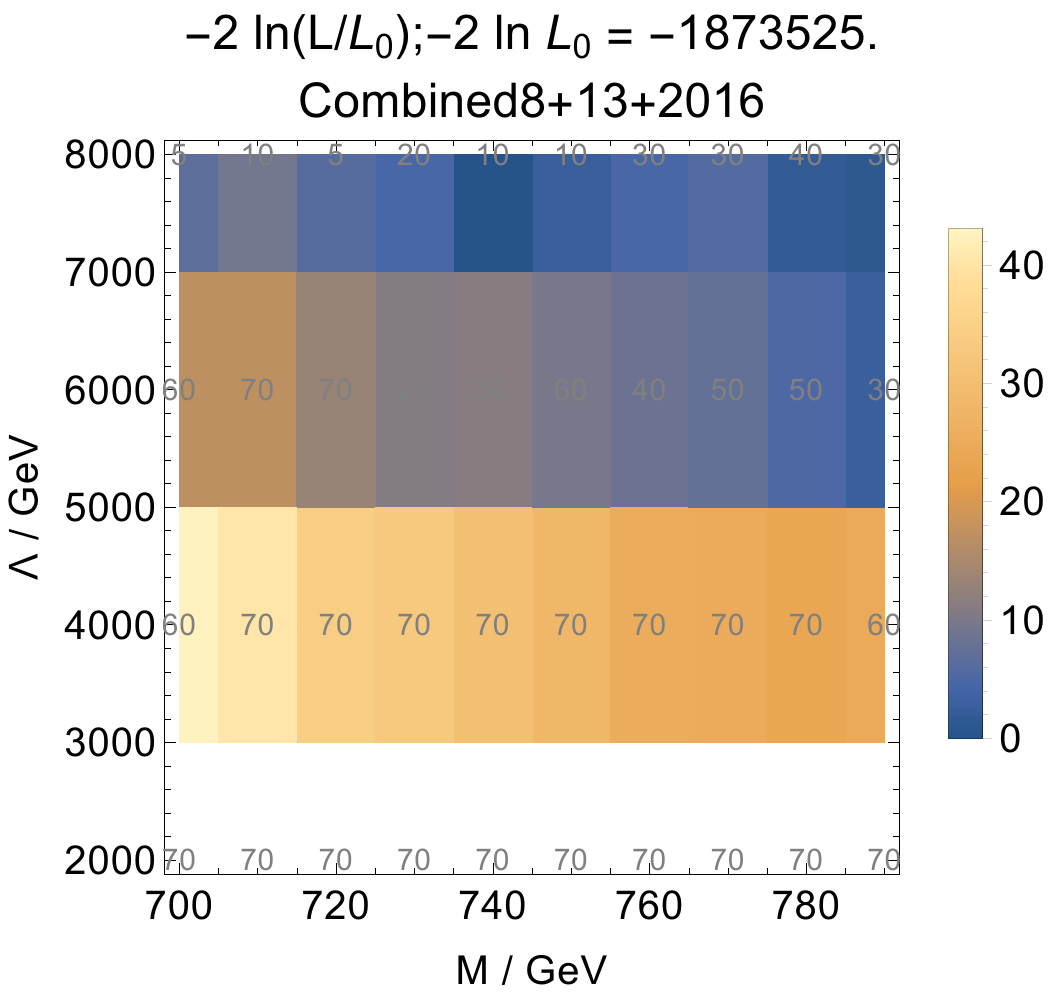}
\caption{From left to right: log likelihoods relative to the best fit points for the $gg$-initiated spin-0, $q\bar{q}$-initiated spin-0, $gg$-initiated spin 2 and $q\bar{q}$-initiated spin-2 signal hypotheses. Top row: Likelihoods based on pre-2016 data (at $\sqrt{s}=8+13$ TeV) as a function of signal mass $M$ and signal amplitude parameter $\Lambda$, profiling over the width $\Gamma$. The profiled value of $\Gamma$ is shown at each grid point. Bottom row: the effect of including the 2016 data on the respective signal hypotheses.\label{fig:2016fitmassamp}}
\end{figure}

The effect of including the 2016 data on the various signal fits is shown in Figs.~\ref{fig:2016fitmasswidth} and~\ref{fig:2016fitmassamp}. The overall significance relative to background reduces signifcantly ($\Delta \chi^2 \sim 9$ in the case of the spin-0 scalar $gg$-initiated model), and the preference switches to $q\bar{q}$-initiated models, which, on account of parton luminosities, predict smaller features in the largely smooth 2016 data. However, the differences in maximum likelihood between spin 0 and spin 2 remain unaffected.

\section{Diphoton Valleys} \label{sec:valley}

While much attention has focused on the interpretation of excesses in the diphoton mass spectrum, in light of the variety of signal shapes afforded by the models of \S\ref{sec:signal} it is worthwhile to consider the implications of possible {\it deficits} in the spectrum. Given that these effects are most apparent in the case of $q \bar q$-initiated spin-2 resonances, in this Section we consider this signal hypothesis as an interpretation for unorthodox `excesses' in the measured diphoton spectra elsewhere in the measured diphoton spectra. 

We consider two distinctive scenarios. In the first scenario we continue to focus on real values of the amplitude coefficient $A$, where interference effects lead to a peak-dip structure atop a falling continuum background. In our analysis of possible excesses near $\mgamgam = 750$ GeV this led to a preference for small width for $q \bar q$-initiated spin-2 signals, given the lack of apparent deficits in this part of the invariant mass spectrum. However, it is worth considering whether such peak-dip structures might appear elsewhere in the ATLAS and CMS diphoton spectra. 

\begin{figure}
\centering
\includegraphics[height=4.5cm]{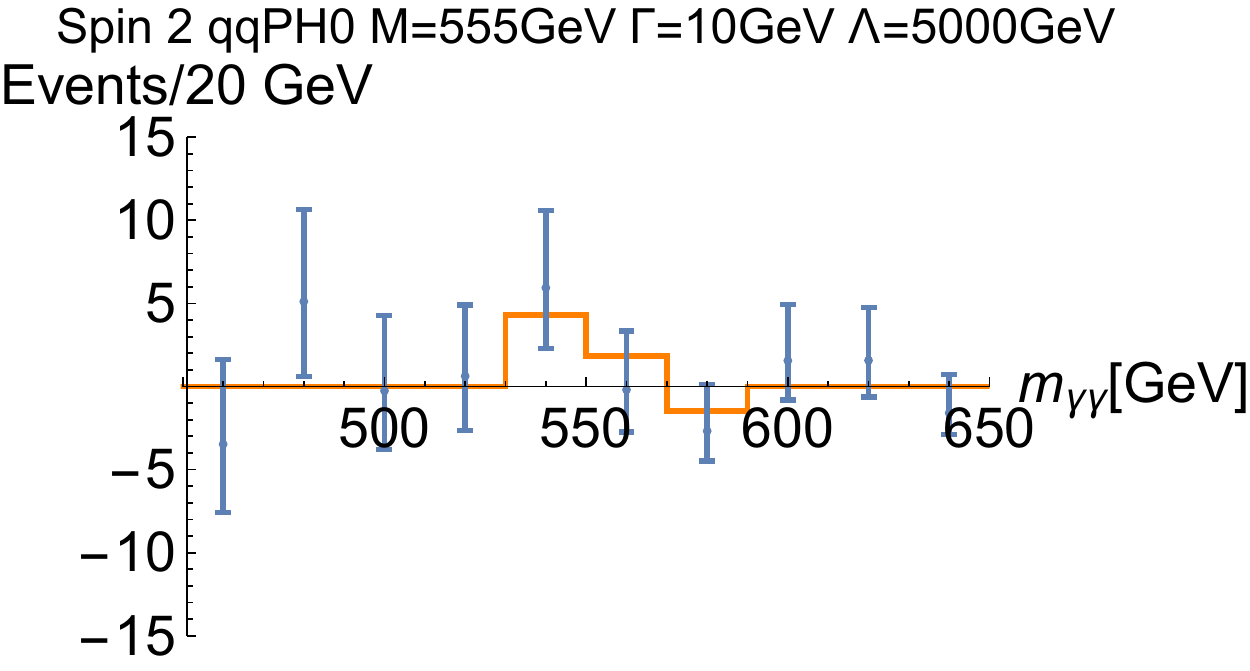}
\caption{The background-subtracted spectrum of the $q \bar q$-initiated spin-2 model point that best fits the ATLAS13Spin0 dataset (blue points) around $550 \GeV$. The error bars on the data points are purely statistical. \label{fig:scottPlotDip}}
\end{figure}

\begin{figure}
\centering

\includegraphics[width=0.24\textwidth]{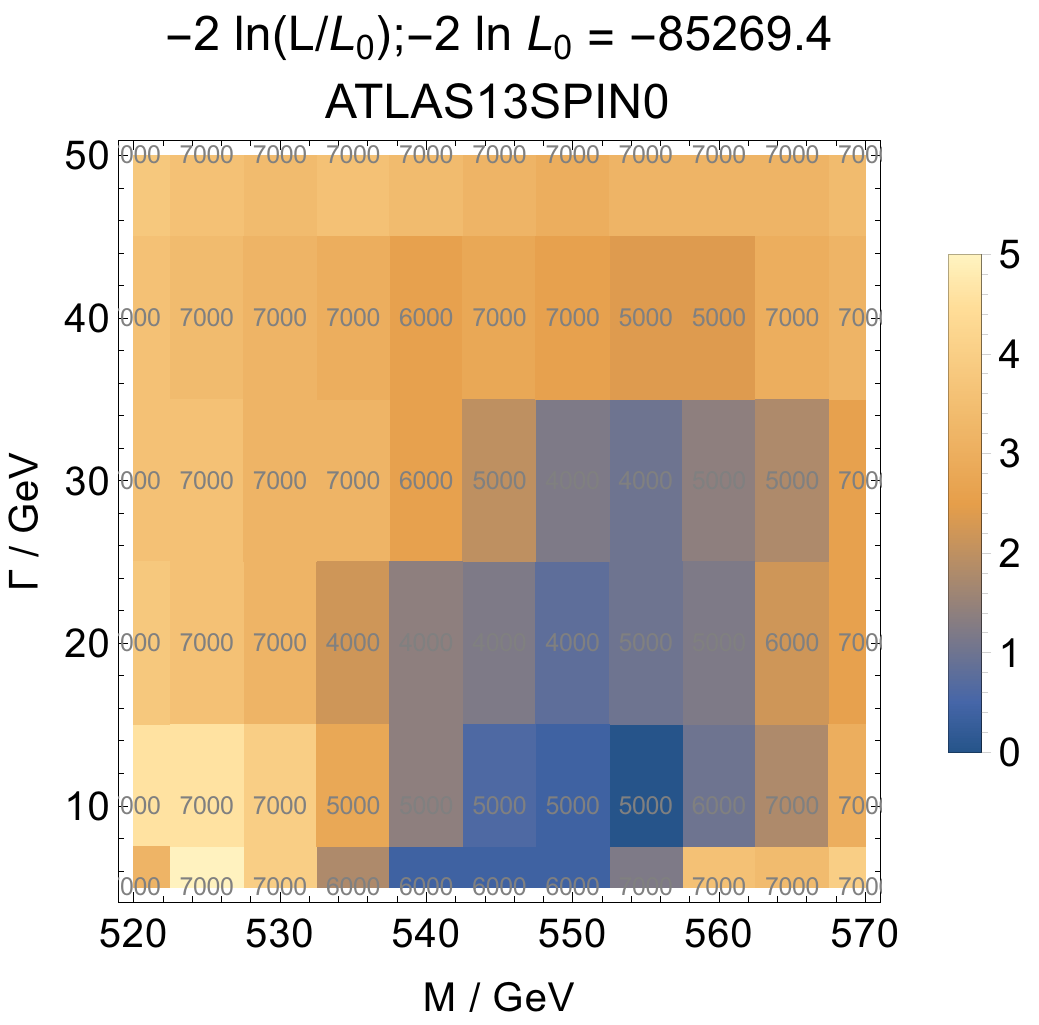}
\includegraphics[width=0.24\textwidth]{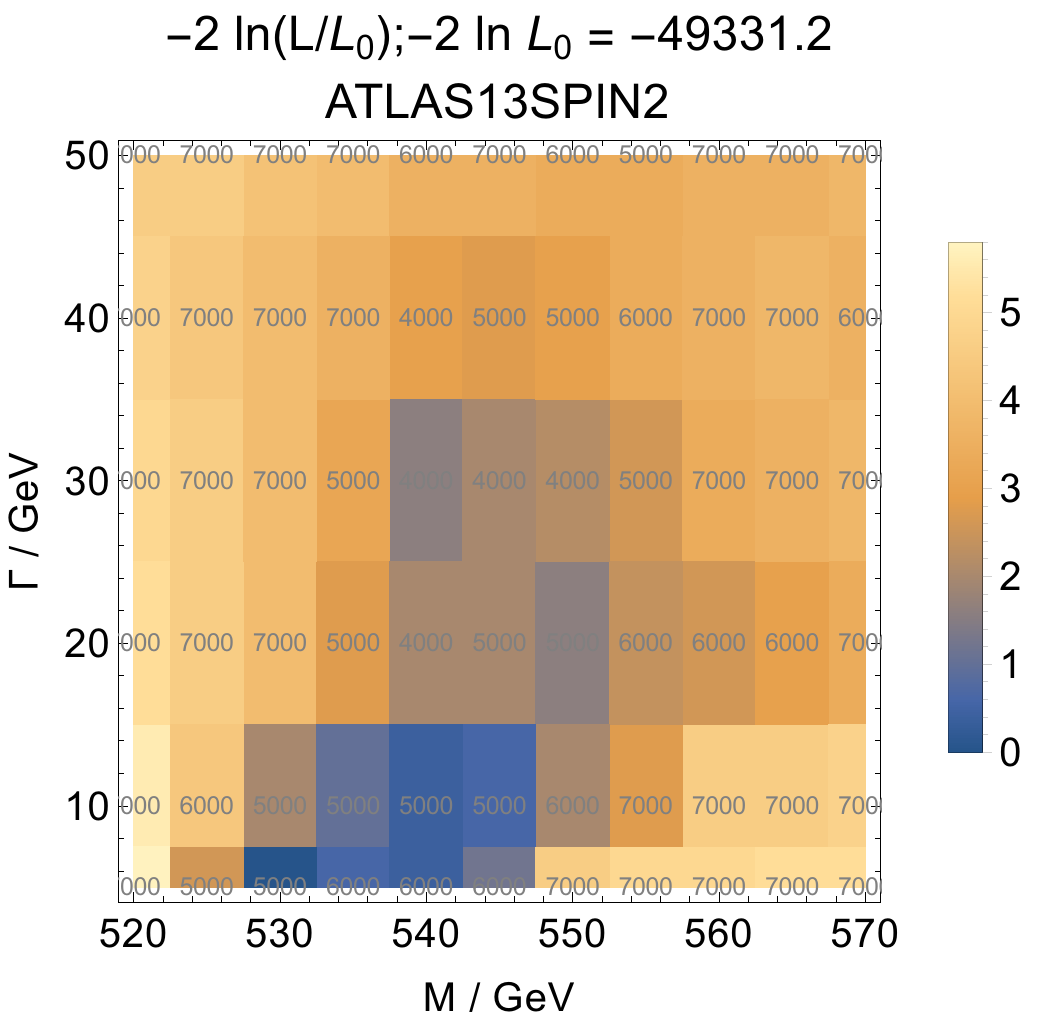}
\includegraphics[width=0.24\textwidth]{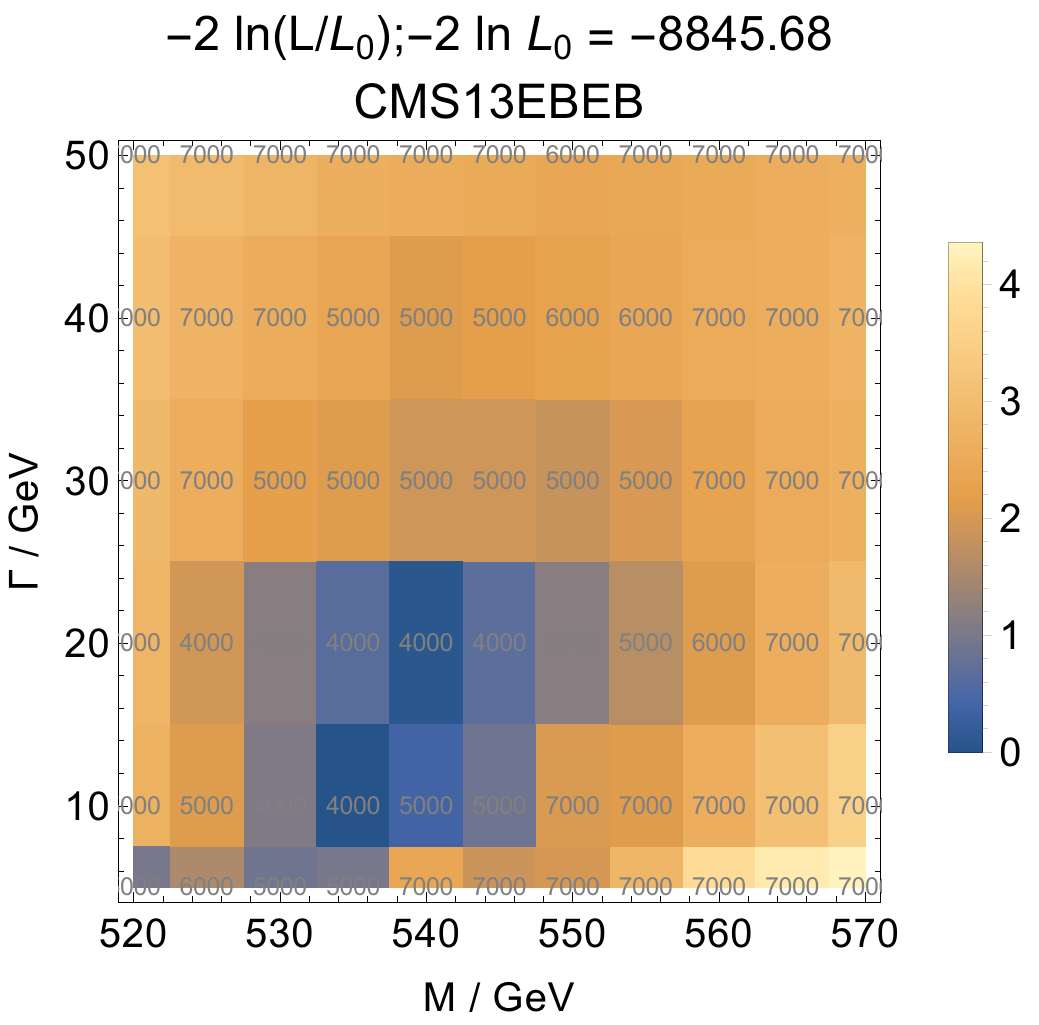}
\includegraphics[width=0.24\textwidth]{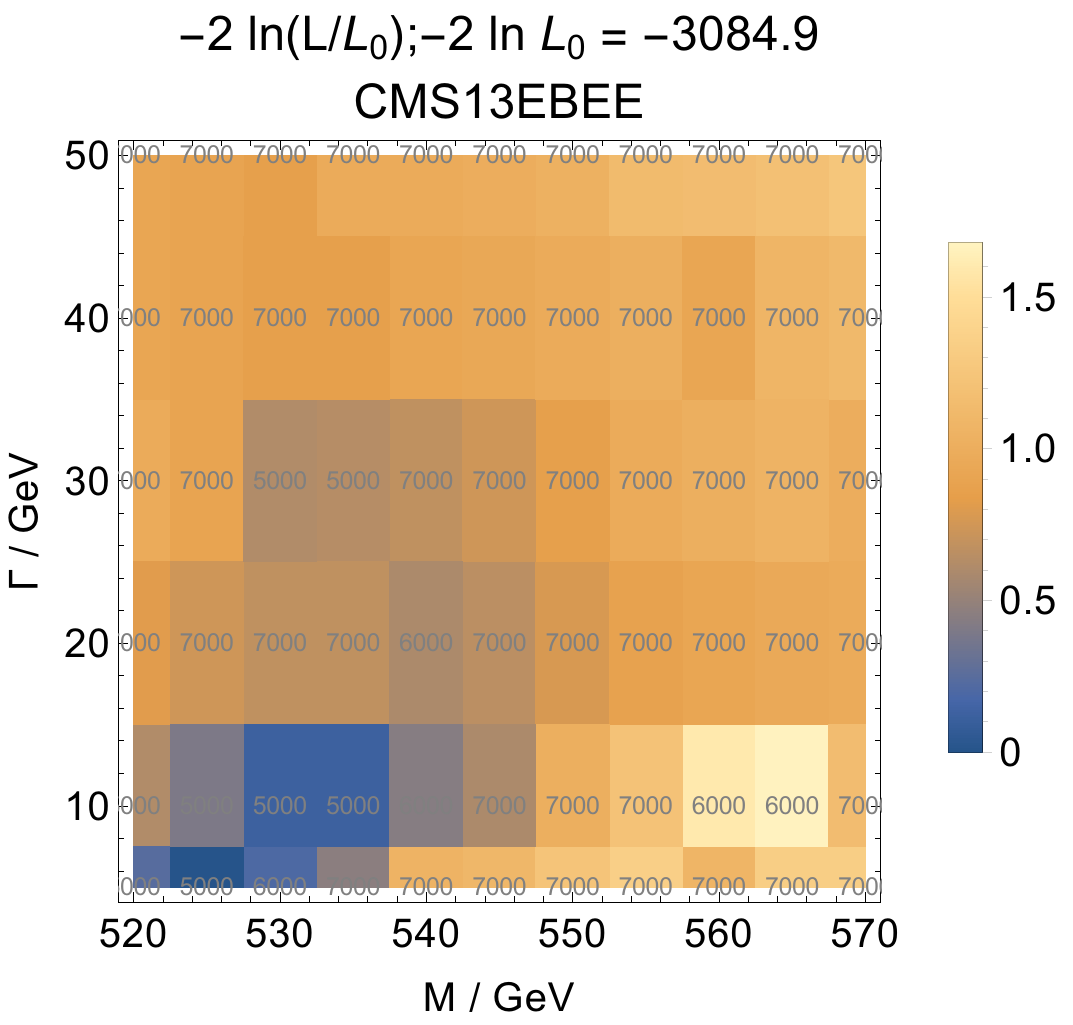}

\includegraphics[width=0.24\textwidth]{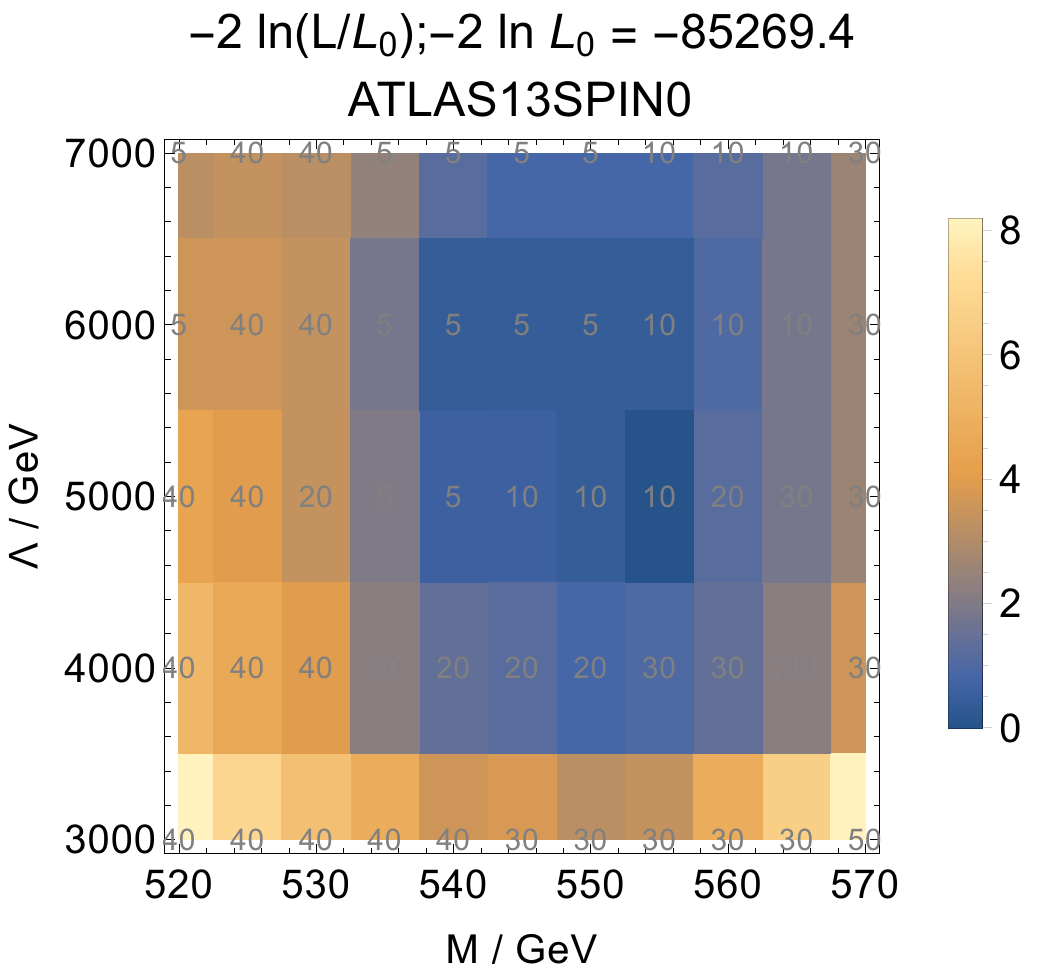} 
\includegraphics[width=0.24\textwidth]{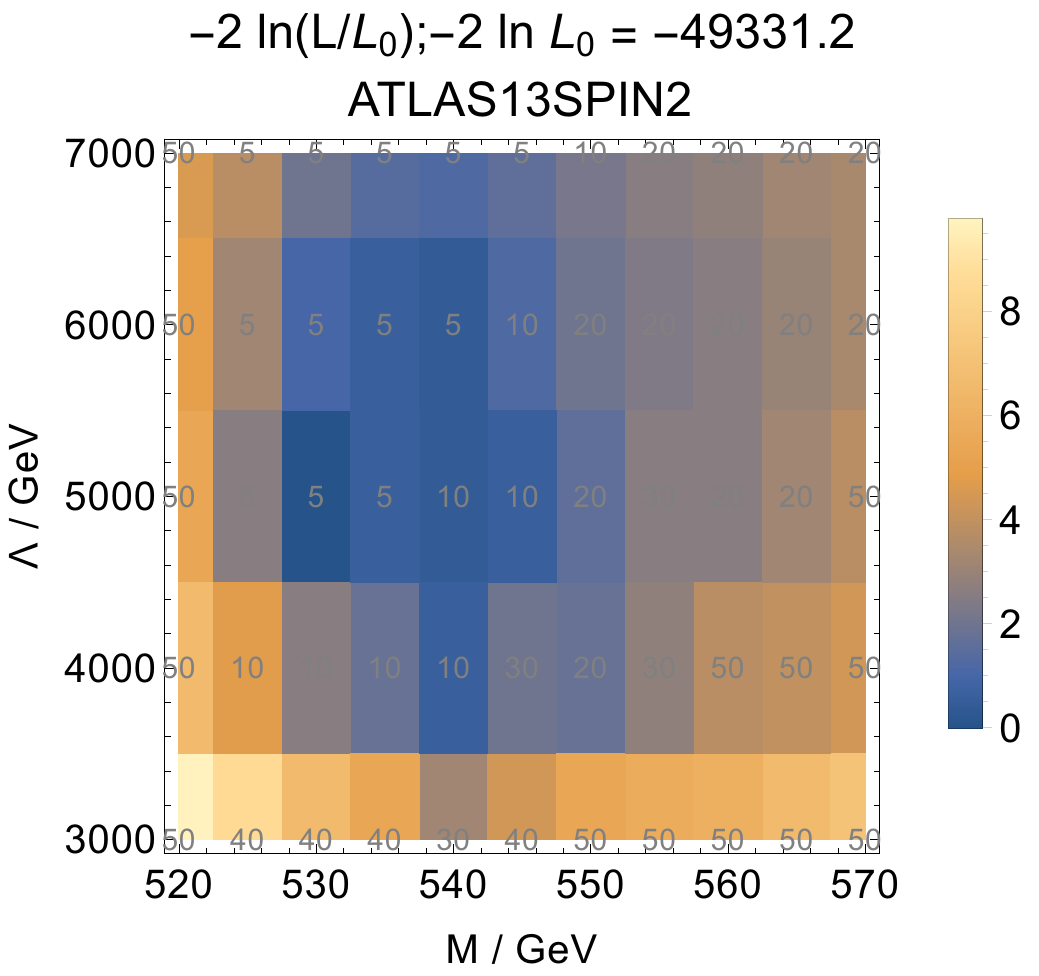}
\includegraphics[width=0.24\textwidth]{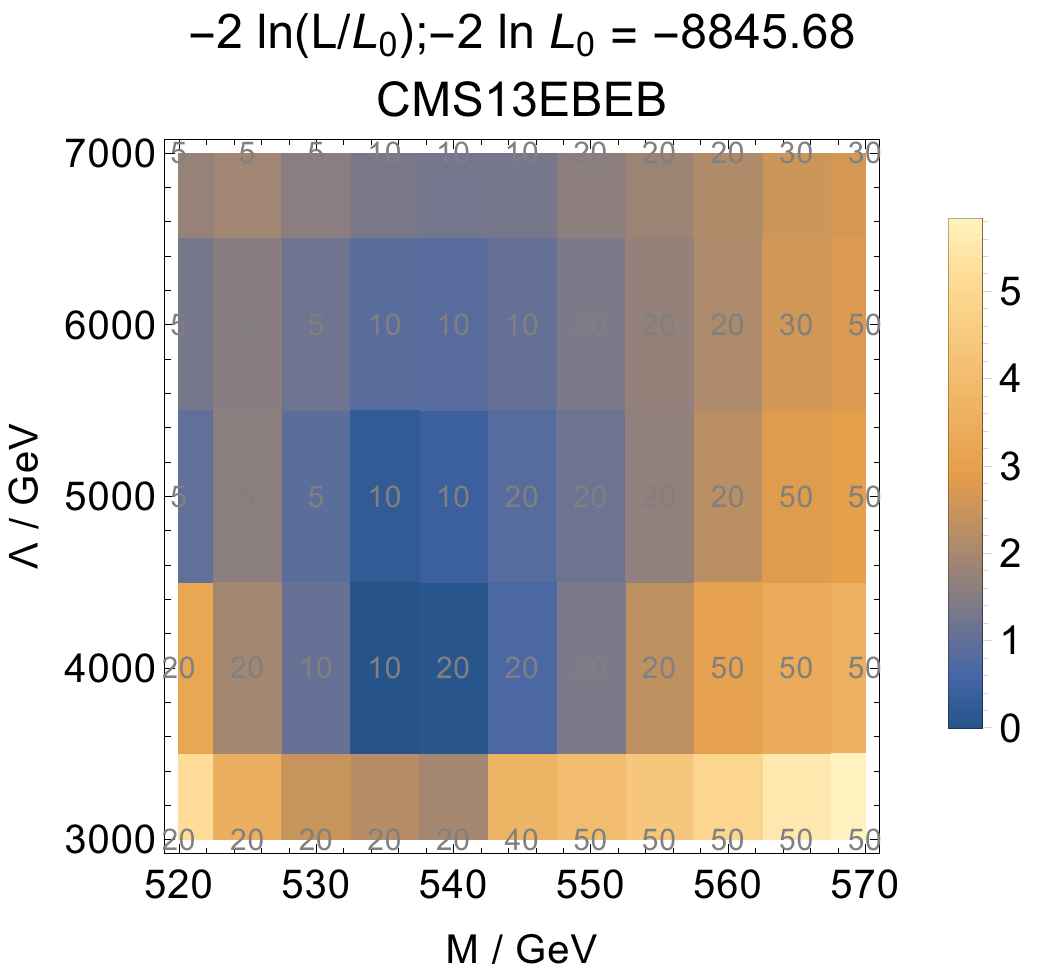}
\includegraphics[width=0.24\textwidth]{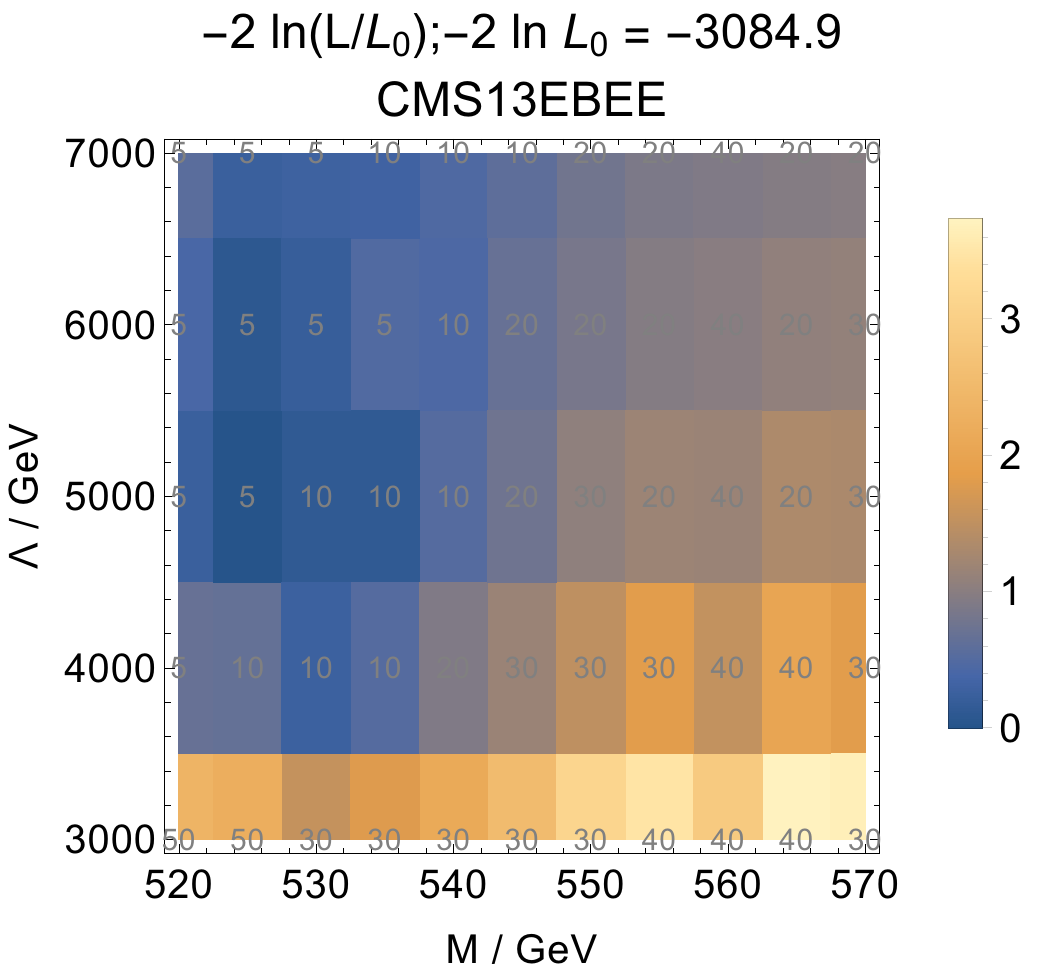}

\caption{Log likelihoods relative to the best fit points for the $q \bar q$-initiated spin-2 signal hypothesis near 550 GeV. Top row: Likelihoods as a function of signal mass $M$ and width $\Gamma$, profiling over the signal amplitude parameter $\Lambda$. The profiled value of $\Lambda$ is shown at each grid point. From left to right, the likelihoods are shown for the 13 \TeV~ datasets ATLAS13Spin0, ATLAS13Spin2, CMS13EBEB and CMS13EBEE. Bottom row: Likelihoods as a function of signal mass $M$ and signal amplitude parameter $\Lambda$, profiling over the signal width $\Gamma$ (with the profiled value of $\Gamma$ shown at each grid point).
\label{fig:dipHunt}}
\end{figure}

For simplicity and purely illustrative purposes we focus on the pre-2016 $\sqrt{s} = 13$ TeV data set, where a peak-dip structure is apparent in both ATLAS and CMS diphoton spectra near $\mgamgam = 550$ GeV.  Figure~\ref{fig:dipHunt} shows the result of fitting to the 13 \TeV ~spectra around $\mgamgam = 550 \, \GeV$ using the fitting procedure detailed in \S\ref{sec:data}. Both ATLAS and CMS see a (slight) peak-dip structure at similiar invariant masses, which act in the `Combined13' fit to improve the chi-squared by around $\Delta \chi^2 \sim 6$ at the best fit point (whose spectrum is displayed in Fig.~\ref{fig:scottPlotDip}), compared to the background only hypothesis. While not particularly significant and not shared by the $\sqrt{s} = 8$ TeV dataset, it serves to illustrate the valuable point that new physics may first appear in the diphoton spectrum in the form of peak-dip structures, rather than the pure peaks currently considered by the ATLAS and CMS collaborations.

\begin{figure}
\centering
\includegraphics[height=5cm]{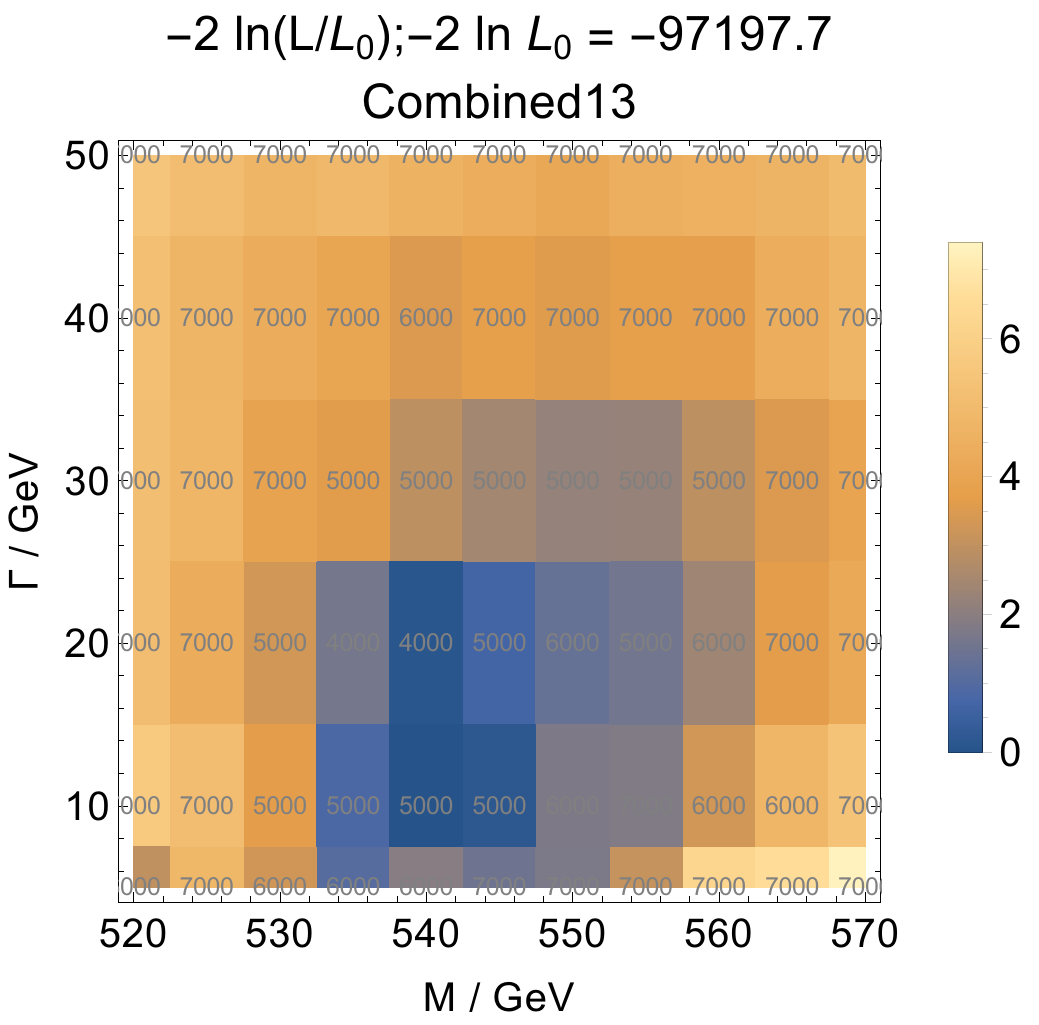}
\includegraphics[height=5cm]{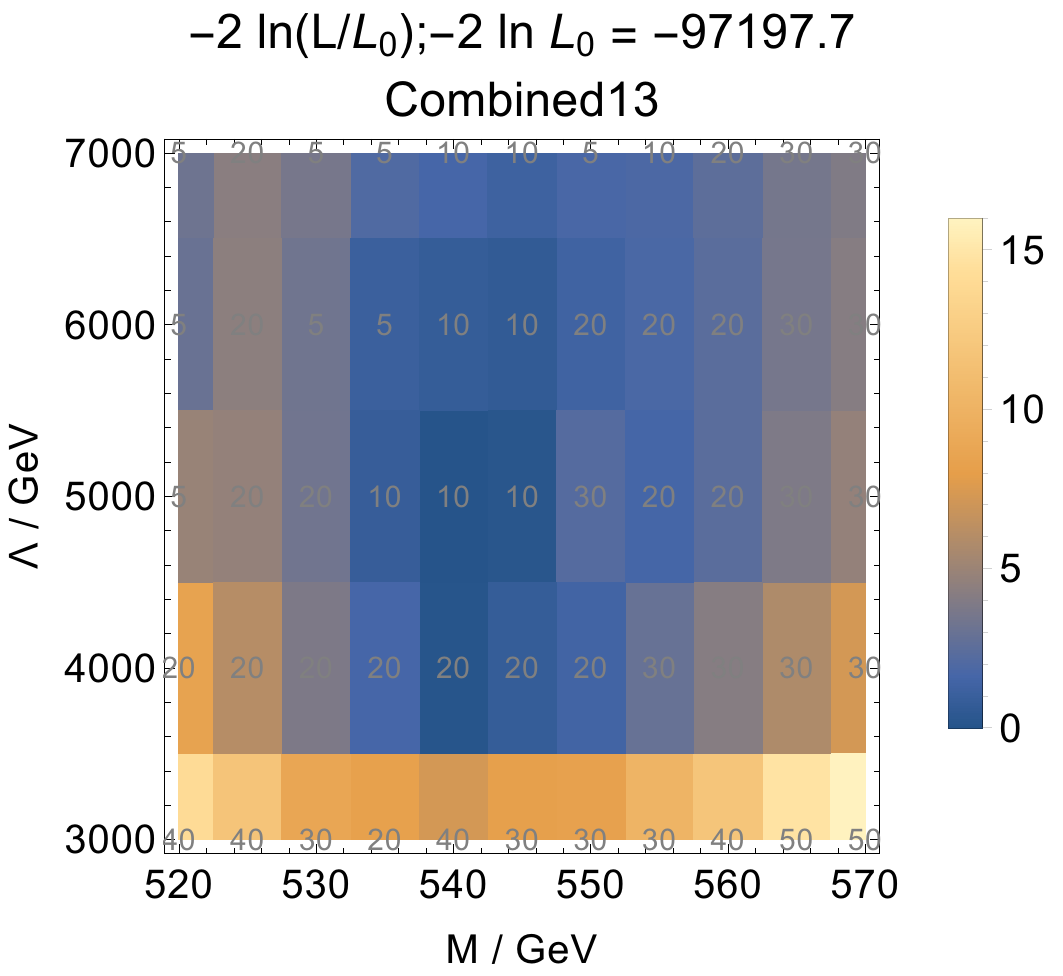}

\caption{Log likelihoods relative to the best fit points for the $q \bar q$-initiated spin-2 signal hypothesis near 550 GeV for the combined pre-2016 $\sqrt{s} = 13$ TeV data set. Left: Combined likelihood as a function of signal mass $M$ and width $\Gamma$, profiling over the signal amplitude parameter $\Lambda$. The profiled value of $\Lambda$ is shown at each grid point. Right: Combined likelihood as a function of signal mass $M$ and signal amplitude parameter $\Lambda$, profiling over the signal width $\Gamma$.
\label{fig:dipHuntcombo}}
\end{figure}

\begin{figure}
\begin{center}
\includegraphics[width=0.55\textwidth]{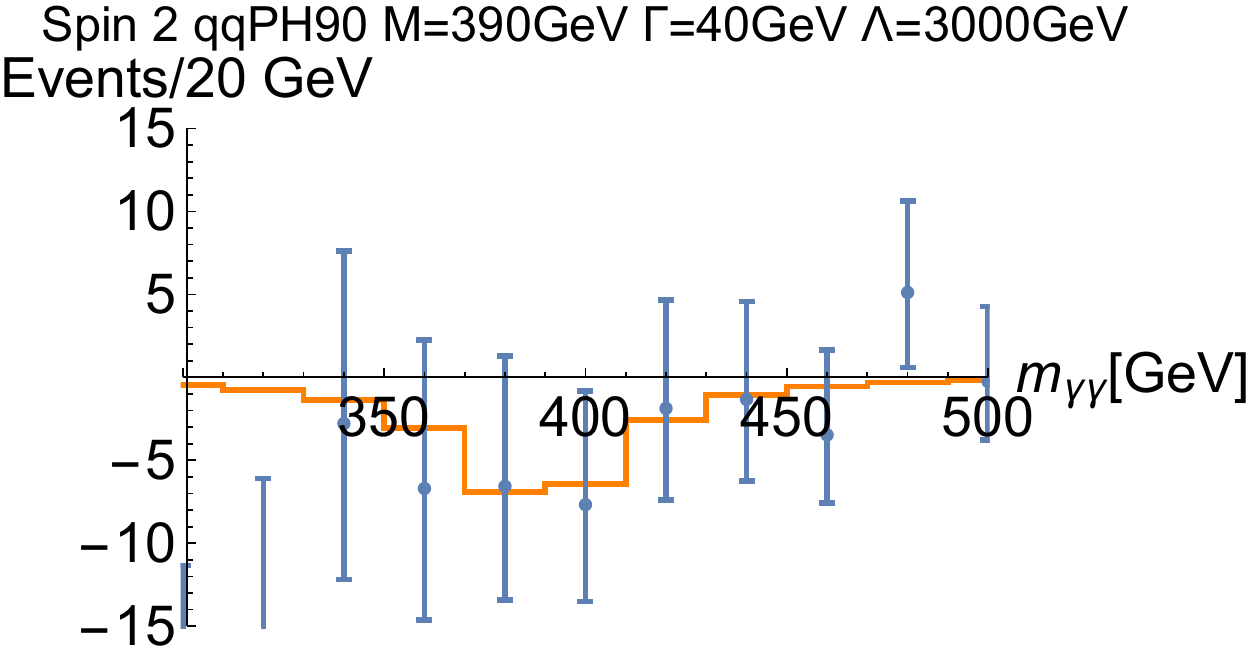}
\end{center}
\caption{The background-subtracted spectrum of the $q \bar q$-initiated spin-2 model point that best fits the ATLAS13Spin0 dataset (blue points) around $400 \GeV$, assuming a 90$^{\circ}$ phase in the signal amplitude. The error bars on the data points are purely statistical.\label{fig:scottPlotDip2}}
\end{figure}

\begin{figure}
\centering

\includegraphics[width=0.24\textwidth]{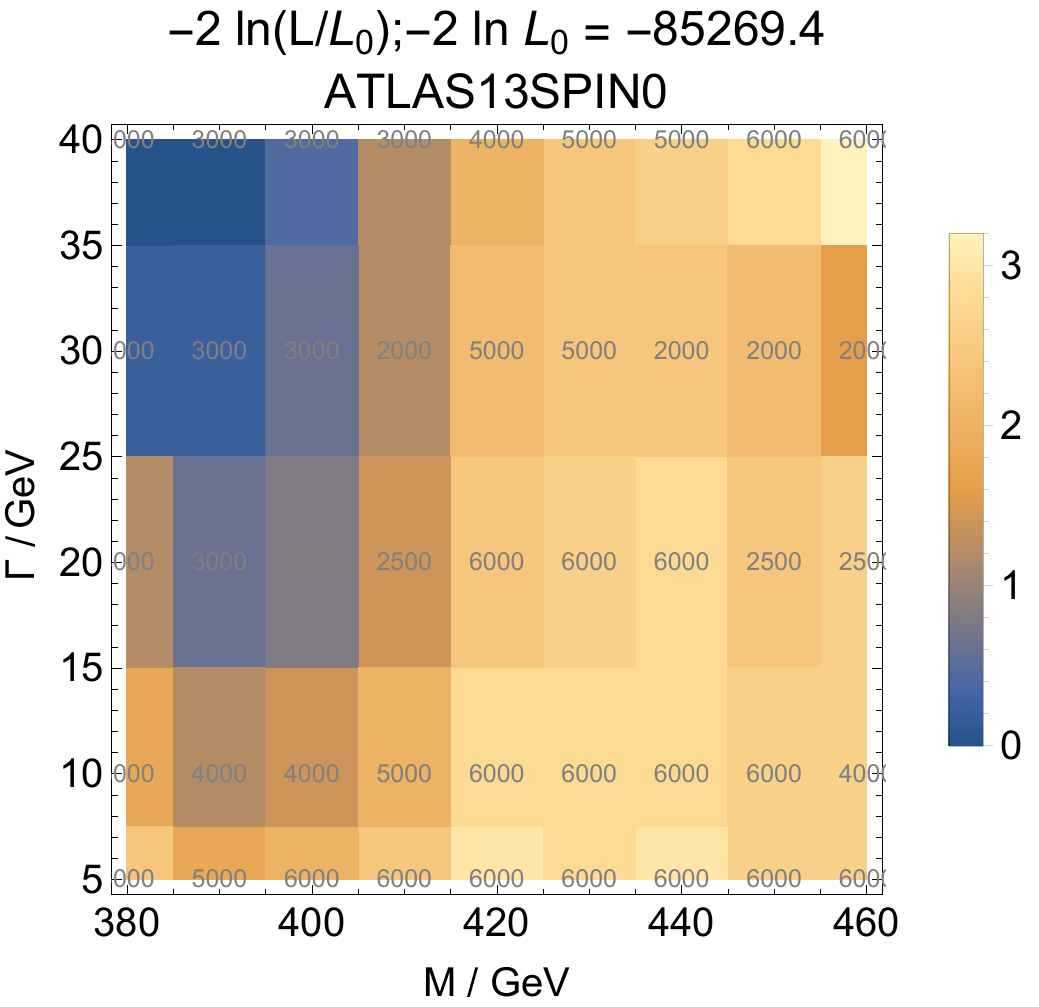}
\includegraphics[width=0.24\textwidth]{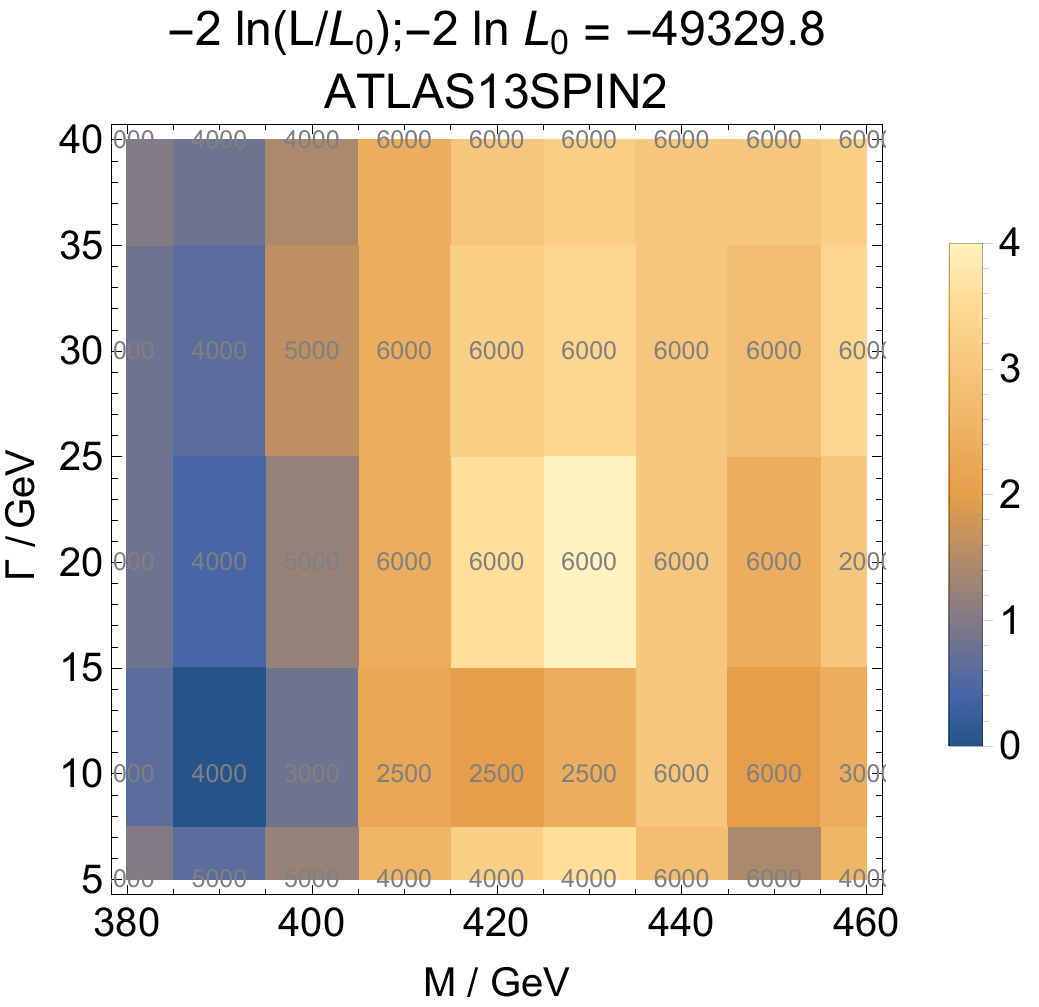}
\includegraphics[width=0.24\textwidth]{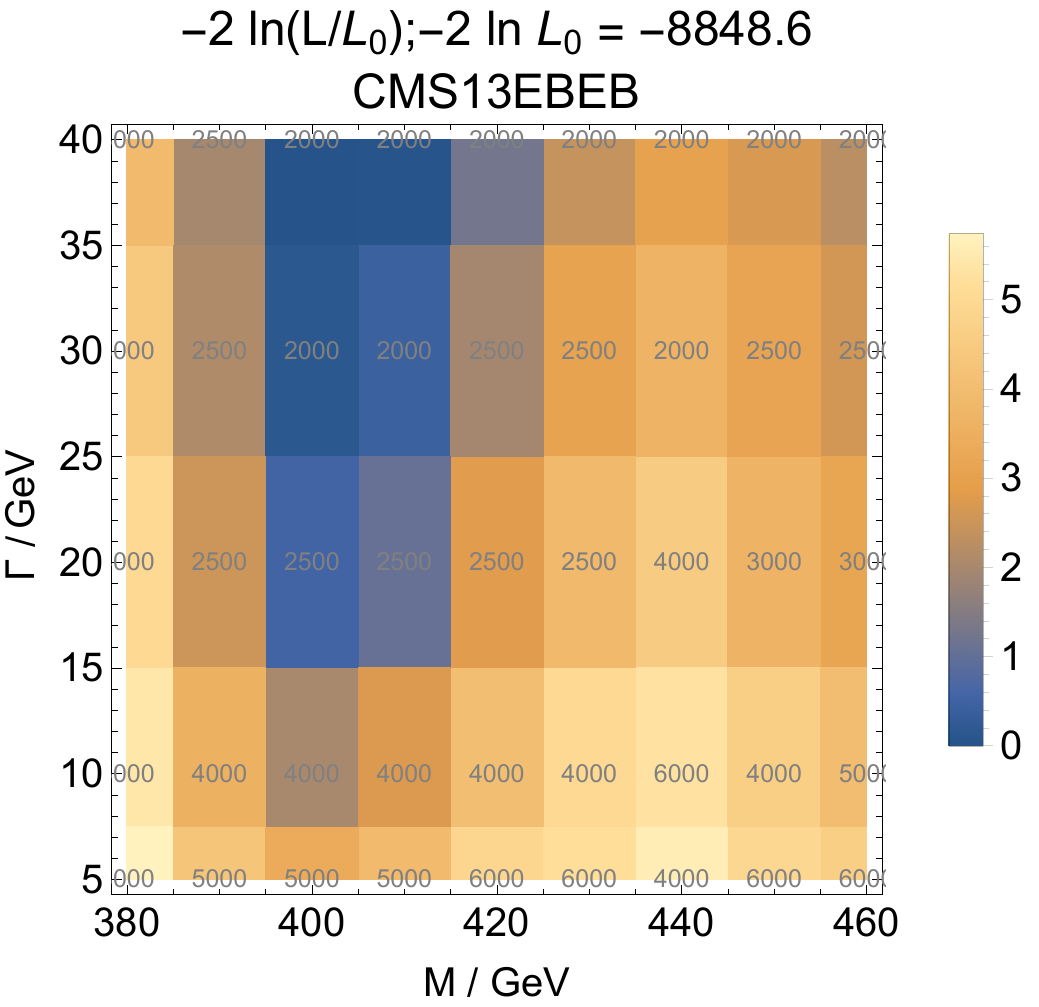}
\includegraphics[width=0.24\textwidth]{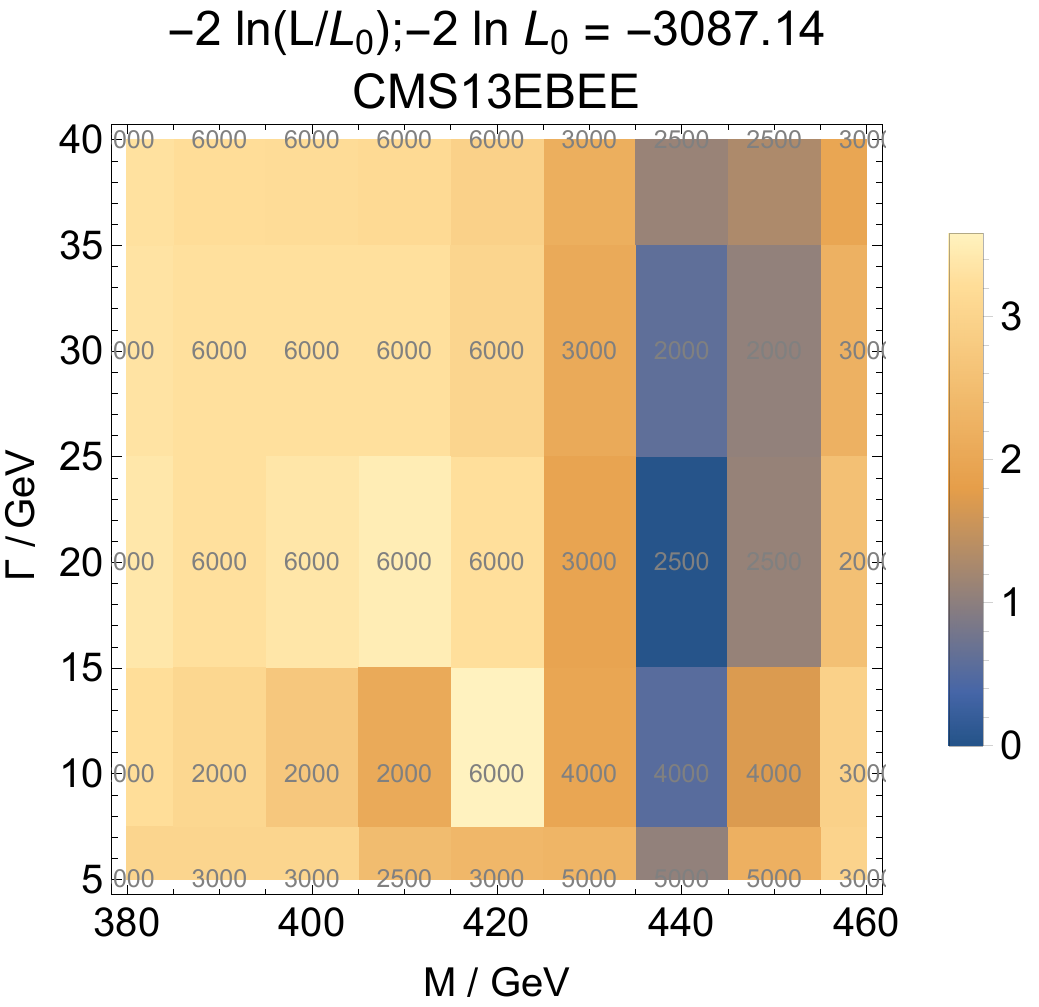}

\includegraphics[width=0.24\textwidth]{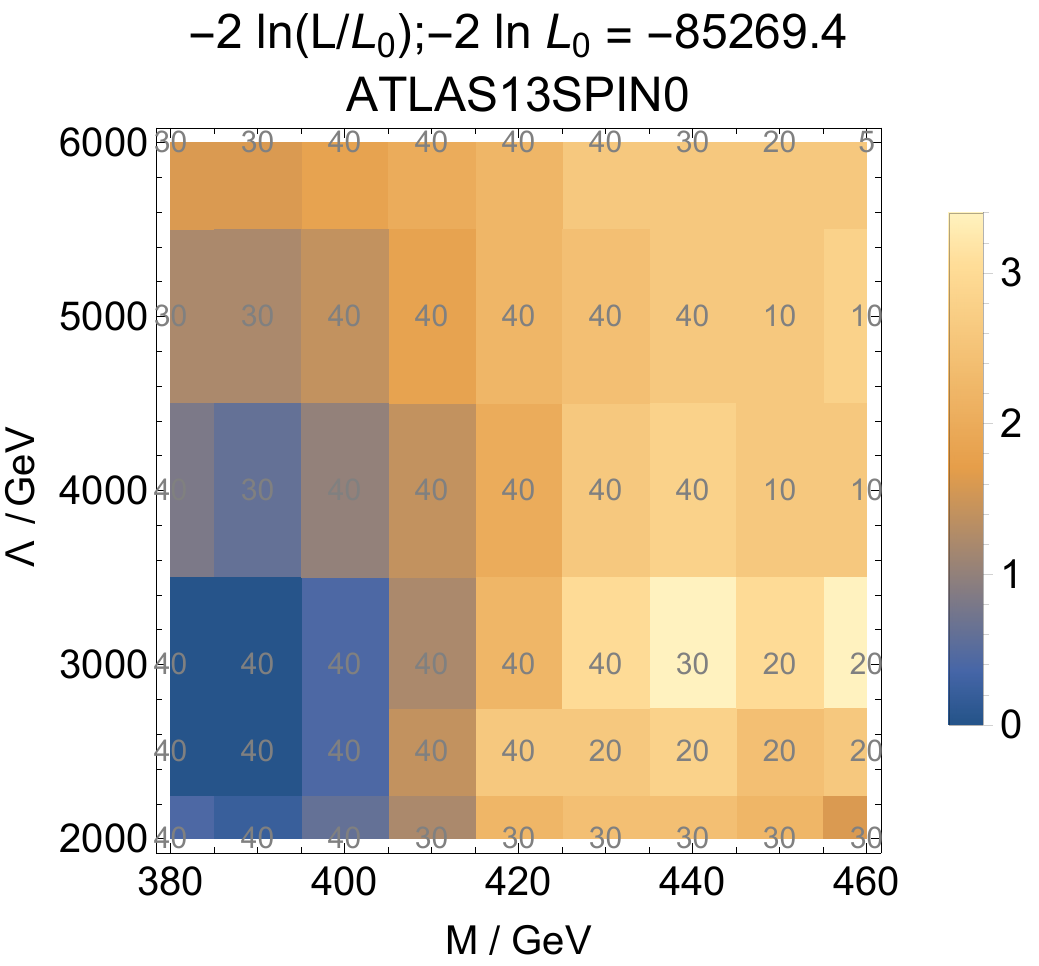} 
\includegraphics[width=0.24\textwidth]{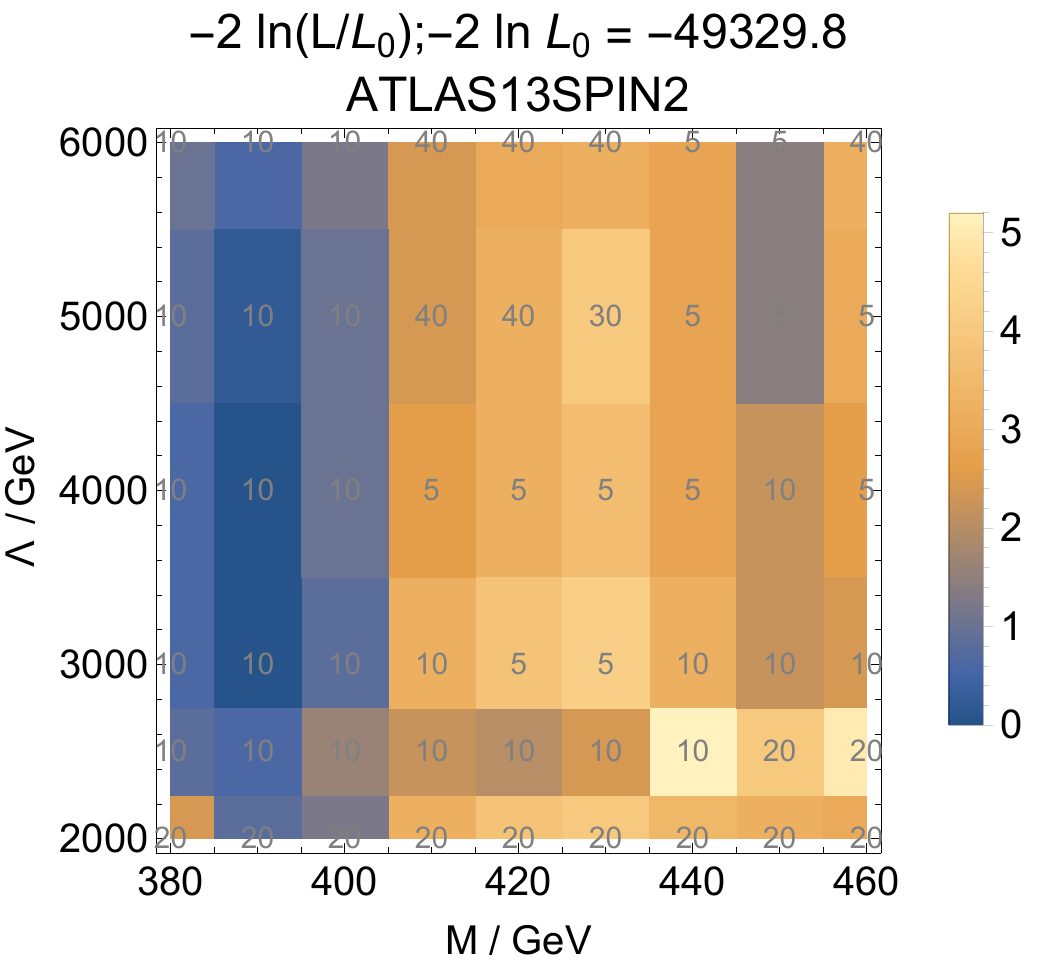}
\includegraphics[width=0.24\textwidth]{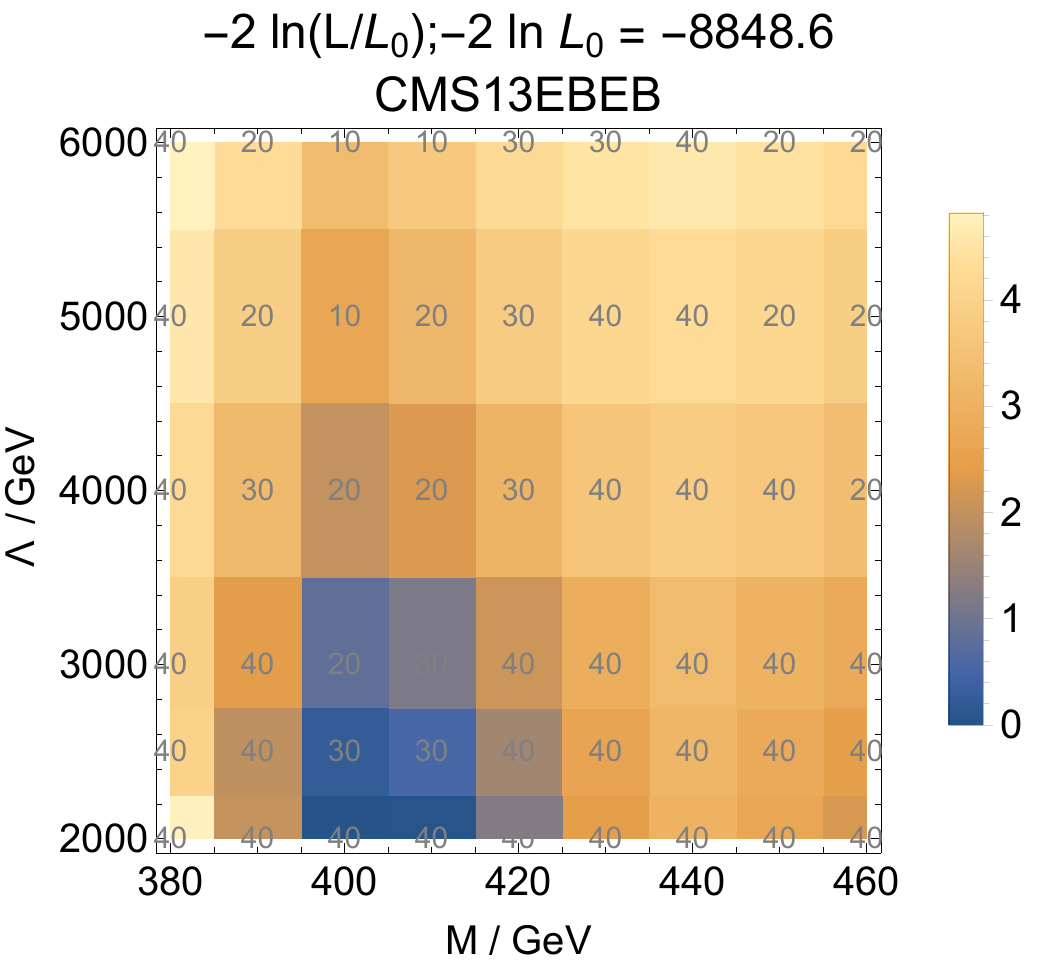}
\includegraphics[width=0.24\textwidth]{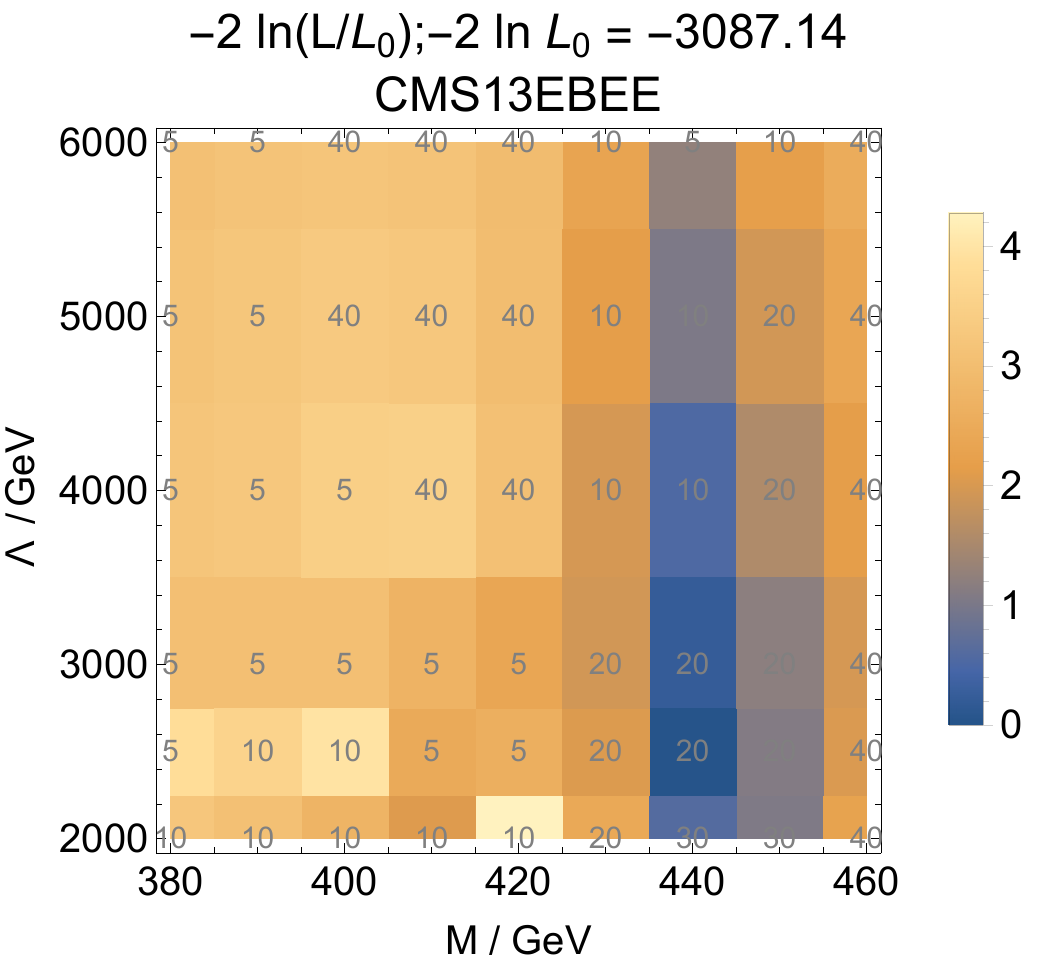}

\caption{Log likelihoods relative to the best fit points for the $q \bar q$-initiated spin-2 signal hypothesis near 400 GeV, assuming a 90$^{\circ}$ phase in the signal amplitude. Top row: Likelihoods as a function of signal mass $M$ and width $\Gamma$, profiling over the signal amplitude parameter $\Lambda$. The profiled value of $\Lambda$ is shown at each grid point. From left to right, the likelihoods are shown for the 13 \TeV~ datasets ATLAS13Spin0, ATLAS13Spin2, CMS13EBEB and CMS13EBEE. Bottom row: Likelihoods as a function of signal mass $M$ and signal amplitude parameter $\Lambda$, profiling over the signal width $\Gamma$ (with the profiled value of $\Gamma$ shown at each grid point).\label{fig:dipHunt2}}
\end{figure}

\begin{figure}
\centering
\includegraphics[height=5cm]{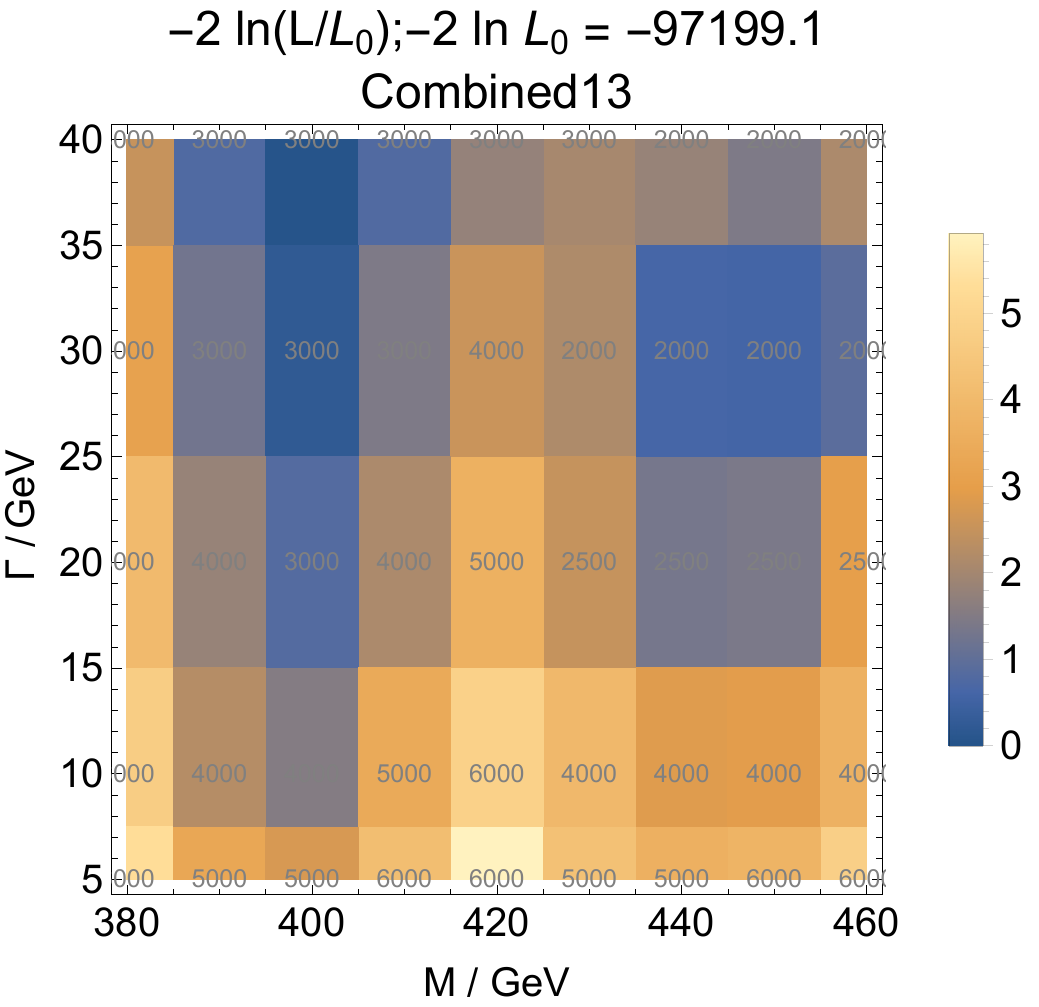}
\includegraphics[height=5cm]{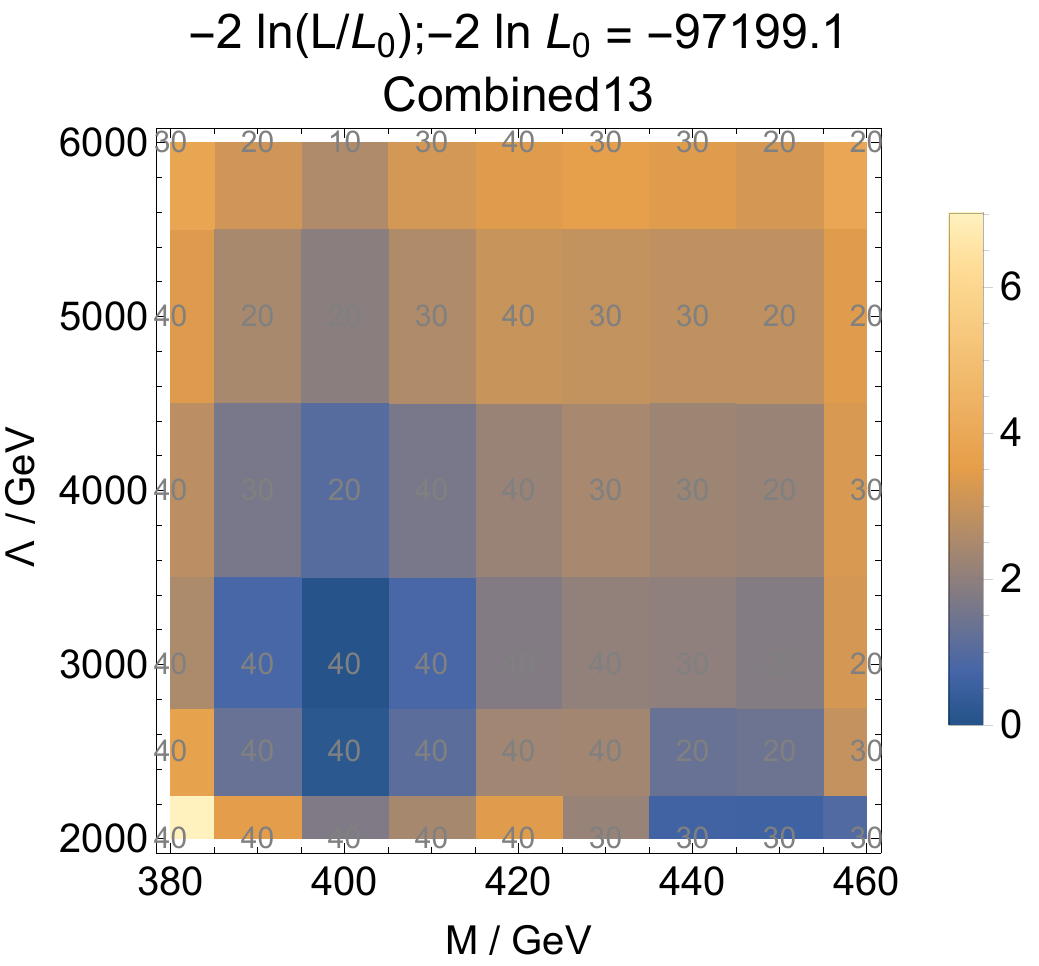}

\caption{Log likelihoods relative to the best fit points for the $q \bar q$-initiated spin-2 signal hypothesis near 400 GeV, assuming a 90$^{\circ}$ phase in the signal amplitude, for the combined pre-2016 $\sqrt{s} = 13$ TeV data set. Left: Combined likelihood as a function of signal mass $M$ and width $\Gamma$, profiling over the signal amplitude parameter $\Lambda$. The profiled value of $\Lambda$ is shown at each grid point. Right: Combined likelihood as a function of signal mass $M$ and signal amplitude parameter $\Lambda$, profiling over the signal width $\Gamma$. \label{fig:dipHuntcombo2}}
\end{figure}

We next consider a scenario where $A$ is complex with a 90$^\circ$ phase, in which case resonance-continuum interference leads to a pure {\it deficit} with respect to the background diphoton distribution. While such a phase does not appear in the on-shell decays of a spin-2 resonance coupling directly to quarks and photons through purely local operators, it is in principle possible when the coupling to quarks and/or photons arises through a loop of particles lighter than half the resonance mass.\footnote{This naturally raises the prospect of on-shell decays of the resonance directly to the mediating particles, an interesting possibility beyond the scope of the present work.}

Again for simplicity and purely illustrative purposes we focus on the pre-2016 $\sqrt{s} = 13$ TeV data set, where a pure deficit is apparent in both ATLAS and CMS diphoton spectra near $\mgamgam = 400$ GeV. Figure~\ref{fig:dipHunt2} shows the result of fitting this signal hypothesis to the 13 \TeV ~spectra around $\mgamgam = 400 \, \GeV$, again using the procedure detailed in \S\ref{sec:data}. Both ATLAS and CMS see a modest deficit at similiar invariant masses, which act in the `Combined13' fit to improve the chi-squared by around $\Delta \chi^2 \sim 5$ at the best fit point compared to the background only hypothesis; the corresponding spectrum is displayed in Fig.~\ref{fig:scottPlotDip2}. While again not particularly significant compared and not shared by the $\sqrt{s} = 8$ TeV data, this illustrates the potential for correlated deficits in diphoton spectra to serve as a sign of new physics.

\section{Conclusions} \label{sec:conc}

In this work we have investigated the effects of resonance-continuum interference on the diphoton spectrum in the presence of spin-0 and spin-2 resonances produced via $q \bar q$ or $gg$ initial states. We have demonstrated these effects in data at the level of a binned likelihood analysis using ATLAS and CMS data at $\sqrt{s} = 8, 13$ TeV, examining the impact of resonance-continuum interference on the interpretation of statistical fluctuations near $\mgamgam = 750$ GeV as well as elsewhere in the diphoton spectrum.

 With the exception of $q \bar q$-initiated spin-0 resonances, resonance-continuum interference leads to significant changes in the best-fit mass and width when fitting signal hypothesis to the diphoton invariant mass spectrum. The largest effects are observed for $q \bar q$-initiated spin-2 resonances, where resonance-continuum interference can shift the best-fit masses and widths by tens of GeV. In the case of fluctuations near 750 GeV in pre-2016 data, it leads to a preference for negligible width for a spin-2 particle produced via $q \bar q$ given the absence of adjacent deficits in the spectrum. 
 
The substantial interference effects for a spin-2 resonance also admit significant peak-dip structures or even pure deficits in the diphoton spectrum. While deficits improve limit-setting when resonance-continuum interference is neglected, once interference is taken into account it raises the suggestive possibility of searching for ``signal-like'' deficits in the diphoton spectrum. We have illustrated this possibility using peak-dip structures in LHC diphoton spectra near $\mgamgam = 550$ GeV and pure deficits near $\mgamgam = 400$ GeV.

Our results highlight the importance of accounting for resonance-continuum interference in fitting signals to the diphoton spectrum. They both illustrate the value of incorporating resonance-continuum interference into LHC searches for new physics in the diphoton spectrum, and also indicate the potential value of systematically searching for {\it deficits} in the diphoton spectrum as a sign of new physics.

\section*{Acknowledgments}

We thank Scott Thomas for insightful discussions. We gratefully acknowledge the hospitality of the Kavli Institute for Theoretical Physics, supported in part by the National Science Foundation under Grant No. NSF PHY11-25915. The work of NC was supported in part by the Department of Energy under the grant DE-SC0014129. DS acknowledges the support of the Science and Technology Facilities Council and Emmanuel College, Cambridge. SR and DS are grateful for the support of the Betty and Gordon Moore Foundation during their stay at the KITP.

\bibliography{interferencerefs}
\bibliographystyle{utphys}

\end{document}